%% file: ms.tex
\documentclass[apj]{emulateapj}
\usepackage{apjfonts}
\usepackage{graphicx}

\slugcomment{Erratum to this paper is attached in Appendix.}
\shorttitle{A Revised Effective Temperature Scale for the {\it Kepler} Input Catalog}
\shortauthors{Pinsonneault et~al.}

\begin{document}
\title{A Revised Effective Temperature Scale for the {\it Kepler} Input Catalog}

\author{
Marc H.\ Pinsonneault\altaffilmark{1},
Deokkeun An\altaffilmark{2},
Joanna Molenda-\.Zakowicz\altaffilmark{3},\\
William J.\ Chaplin\altaffilmark{4},
Travis S.\ Metcalfe\altaffilmark{5},
Hans Bruntt\altaffilmark{6}
}

\altaffiltext{1}{Department of Astronomy, the Ohio State University,
Columbus, OH, 43210 USA}
\altaffiltext{2}{Department of Science Education, Ewha Womans University,
Seoul 120-750, Republic of Korea; deokkeun@ewha.ac.kr.}
\altaffiltext{3}{Astronomical Institute, University of Wroc{\l}aw, ul. Kopernika 11,
51-622 Wroc{\l}aw, Poland}
\altaffiltext{4}{School of Physics and Astronomy, University of Birmingham,
Edgbaston, Birmingham, B15 2TT, UK}
\altaffiltext{5}{High Altitude Observatory,
National Center for Atmospheric Research, Boulder, CO 80307, USA}
\altaffiltext{6}{Department of Physics and Astronomy, Aarhus University,
DK-8000 Aarhus C, Denmark}

\begin{abstract}
We present a catalog of revised effective temperatures for stars
observed in long-cadence mode in the {\it Kepler} Input Catalog (KIC).  We
use SDSS $griz$ filters tied to the fundamental temperature scale.  Polynomials for
$griz$ color-temperature relations are presented, along with correction
terms for surface gravity effects, metallicity, and statistical
corrections for binary companions or blending.  We compare our temperature scale
to the published infrared flux method (IRFM) scale for $V_TJK_s$ in both
open clusters and the {\it Kepler} fields.  We find good agreement overall,
with some deviations between $(J\, -\, K_s)$-based temperatures from the IRFM and
both SDSS filter and other diagnostic IRFM color-temperature relationships
above $6000$~K.
For field dwarfs we find a mean shift towards hotter
temperatures relative to the KIC, of order $215$~K,
in the regime where the IRFM scale is well-defined ($4000$~K to $6500$~K).
This change is of comparable magnitude in both color systems and in spectroscopy
for stars with $T_{\rm eff}$ below $6000$~K. Systematic differences between 
temperature estimators appear for hotter stars, and we define
corrections to put the SDSS temperatures on the IRFM scale for
them.  When the theoretical 
dependence on gravity is accounted for we find a similar 
temperature scale offset between the fundamental and KIC scales for 
giants.  We demonstrate that statistical corrections to color-based 
temperatures from binaries are significant.  Typical errors, mostly 
from uncertainties in extinction, are of order $100$~K.  Implications 
for other applications of the KIC are discussed.
\end {abstract}

\keywords{stars: fundamental parameters}

\section {Introduction}

One of the most powerful applications of stellar multi-color photometry
is the ability to precisely infer crucial global
properties. Photometric techniques are especially efficient for
characterizing large samples and providing basic constraints for more
detailed spectroscopic studies.  Modern surveys frequently used
filters designed for the Sloan Digital Sky Survey \citep[SDSS;][]{dr8},
however, while traditional correlations between color and effective
temperature ($T_{\rm eff}$), metallicity ([Fe/H]), and surface gravity
($\log{g}$) have employed other filter sets,
typically on the Johnson-Cousins system.
In \citet[][hereafter A09]{an:09a}
we used SDSS photometry of a solar-metallicity cluster M67 \citep{an:08} to define
a photometric $ugriz$--$T_{\rm eff}$ relation, 
and checked the metallicity scale
using star clusters over a wide range of metallicity. This scale was applied to
the Virgo overdensity in the halo by \citet{an:09b}.
The approach used is similar in spirit
to earlier work in the Johnson-Cousins filter system   
\citep{pinsono:03,pinsono:04,an:07a,an:07b}; the latter effort
used the color-temperature relationships of
\citet{lejeune:97,lejeune:98} with
empirical corrections based on cluster studies.

A revised color-temperature-metallicity relationship for late-type
stars has recently been published by \citet[][hereafter C10]{casagrande:10};
it is based on the infrared flux method (IRFM).  There are a number of advantages
of this approach, as discussed in C10, but there is a lack of native
SDSS data in the stars used to define the calibration itself.
Fortunately, the color-temperature relationships in C10 are defined for
$JHK_s$ colors in the Two Micron All Sky Survey \citep[2MASS;][]{skrutskie:06},
and the {\it Kepler} mission provides a large body of
high quality $griz$ photometry for stars in the 2MASS catalog \citep{brown:11}.

In this paper we use $griz$ data in the {\it Kepler} Input Catalog (KIC)
in conjunction with 2MASS to compare the effective temperature scale for
the $griz$ colors to the IRFM scale. For this initial paper we concentrate on
the mean relationships between the two systems for the average
metallicity of the field sample, taking advantage of the weak
metallicity dependence of the color-$T_{\rm eff}$ relationships that we have
chosen.  In a follow-up paper we add information from spectroscopic
metallicity and $\log{g}$ determinations to compare empirical photometric
relationships involving these quantities to the theoretical
relationships used in the current work.  Unresolved binaries and
extinction errors can be severe problems for photometric temperature
estimates, and another goal of this work is to quantify their
importance.

Another important matter, which we uncovered in the course of our
research, concerns systematic errors in the $griz$ photometry in the KIC.
For large photometric data sets, it can be difficult to assess such errors.
Fortunately, we can also compare photometry used in the KIC with photometry
in the same fields from the SDSS; the latter is important for numerous
applications of data derived from the {\it Kepler} mission. We will
demonstrate that there are significant systematic differences between
the two, and derive corrections to minimize these effects.

We therefore begin with a discussion of our method in Section~\ref{sec:method}.
Along with a description on the sample selection in the KIC (\S~\ref{sec:sample}),
we compare the SDSS and KIC photometry and derive corrected KIC magnitudes and colors
(\S~\ref{sec:phot}). A basis model isochrone in the SDSS colors is presented
(\S~\ref{sec:model}), and a method of determining photometric $T_{\rm eff}$
from $griz$ is described (\S~\ref{sec:teff}). Both the IRFM/$VJHK_s$ and SDSS/$griz$
temperature scales are compared to the KIC dwarf temperatures in Section~\ref{sec:main},
where a $\sim200$~K offset is found in the KIC with respect to both IRFM and SDSS
temperature scales. We also present a method of correcting the dwarf temperature
scale for giants (\S~\ref{sec:giant}). For well-studied open clusters, we find a good
agreement overall between SDSS and IRFM, but find some systematic deviations between
IRFM ($J\, -\, K_s$)-based temperatures from the IRFM and both SDSS filter and other
diagnostic IRFM color-temperature relations (\S~\ref{sec:cluster}). We provide a
formula to put SDSS $T_{\rm eff}$ on the consistent scale with IRFM. These findings
are confirmed using spectroscopic temperature determinations (\S~\ref{sec:spec}).
We also discuss the impact of unresolved binaries and uncertainties in the extinction
estimates (\S~\ref{sec:binary} and \S~\ref{sec:error}). Our revised catalog is presented
in Section~\ref{sec:catalog}, where we provide a recipe for estimating $T_{\rm eff}$
for interested readers, if the application of our technique is desired to the entire
KIC sample in general. We discuss the implications of our new fundamental $T_{\rm eff}$
scale in Section~\ref{sec:summary}.

\section {Method}\label{sec:method}

Our basic data come from the long-cadence sample in the KIC. From this we
extracted a primary sample of dwarfs in the temperature range where our
calibrations are best constrained; our procedure is given in Section~\ref{sec:sample}.
We uncovered some offsets between KIC and native SDSS photometry, and describe
correction terms in Section~\ref{sec:phot}.  Our methods for deriving
color-temperature relationships in $griz$ are described in Sections~\ref{sec:model}
and \ref{sec:teff}.

\subsection {Sample}\label{sec:sample}

We took $griz$ photometry from the KIC \citep{brown:11}; photometric uncertainties
were taken as $0.01$~mag in $gri$ and $0.03$~mag in $z$. Errors were taken
from the quadrature sum of uncertainties in the individual filters.
$JHK_s$ photometry was taken from the All Sky Data Release of the 2MASS
Point Source Catalog \citep[PSC;][]{skrutskie:06}\footnote{See http://www.ipac.caltech.edu/2mass/.},
and checked against complementary information in the KIC itself.

\begin{figure}
\epsscale{1.05}
\plotone{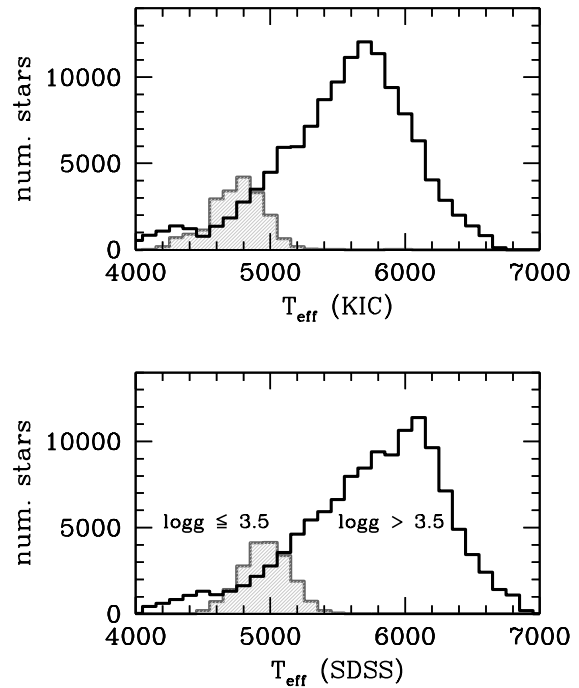}
\caption{Long-cadence data from the KIC ({\it top}) and with our
revised SDSS-based effective temperature scale ({\it bottom}).
Data are binned in $100$~K increment. Dwarfs with KIC $\log{g} > 3.5$
(open histogram) are separated from giants with lower $\log{g}$
(shaded histogram).
\label{fig:KICsample}}
\end{figure}
  
For our sample we chose long-cadence targets in the KIC; our initial source had $161,994$ candidates.
We selected stars with $griz$ photometry detected in all of the bandpasses.
This sample was nearly complete in the
2MASS catalog.  We excluded a small number of sources with 2MASS
photometry quality flags not equal to {\tt AAA} ($N=3,602$) and stars with colors
outside the range of validity of either the IRFM or SDSS scales
($N=11,830$), leaving us with a main sample of
$146,562$ stars. We then further restricted our sample by excluding
stars with $\log{g}$ estimates
below $3.5$~dex in the KIC ($N=19,663$) for a dwarf comparison sample
of $126,899$.  We illustrate the distribution of stars in the sample in
$100$~K bins in Figure~\ref{fig:KICsample}, both in the initial
catalog (top panel) and the revised one in this paper (bottom panel).  We did not
use the giants in our comparison of the dwarf-based temperature scale
(Section~\ref{sec:main}),
but we do employ theoretical $\log{g}$ corrections to the photometric
temperatures for the purposes of the main catalog (see Sections~\ref{sec:giant} and
\ref{sec:table}).

\subsection{Recalibration of the KIC Photometry}\label{sec:phot}

We adopted three primary color indices ($g\, -\, r$, $g\, -\, i$, and $g\, -\, z$)
as our temperature indicators for the SDSS filter system.  A preliminary
comparison of colors yielded surprising internal differences and
trends as a function of mean $T_{\rm eff}$ in the 
relative temperatures inferred
from these color indices (see below). Because the A09 color-color
trends were calibrated using SDSS photometry of M67, this
reflects a zero-point difference between the KIC and SDSS photometry in the
color-color plane.  It is not likely that this difference is caused by
extinction or stellar population differences because all three colors
have similar sensitivities to extinction and metallicity.  Initially,
we suspected problems with the SDSS calibration \citep[see][for a discussion
of zero-point uncertainties]{an:08}. However, the
differences seen were outside of the error bounds for the SDSS
photometry.  For a fraction of the targets (about $2\%$) 
the temperatures inferred from different color sources
(SDSS versus IRFM from 2MASS colors)
are also discordant by more than three standard deviations, in
some cases by thousands of degrees in $T_{\rm eff}$.  We examine both phenomena below.

\begin{figure}
\centering
\includegraphics[scale=0.34]{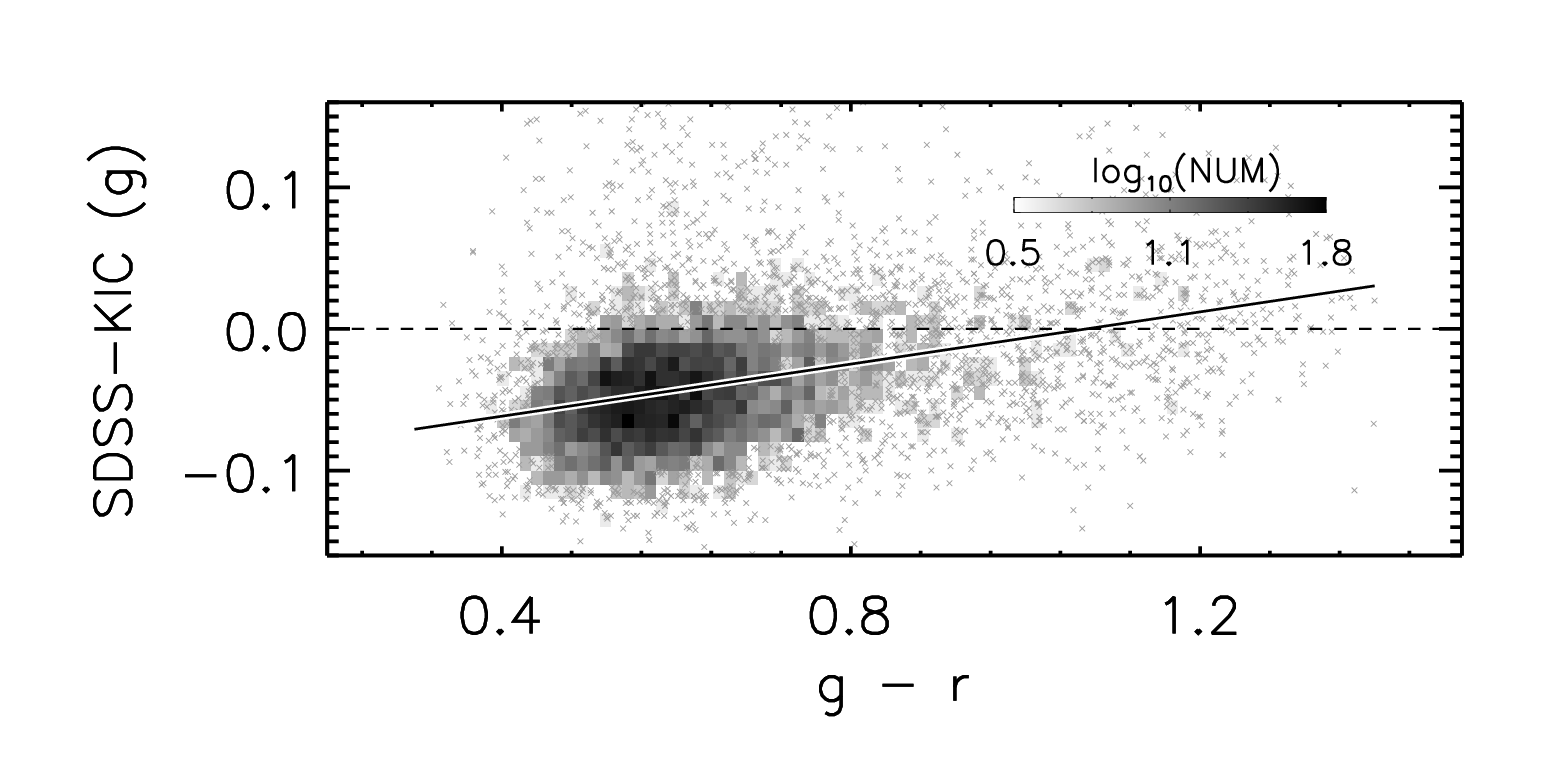}
\includegraphics[scale=0.34]{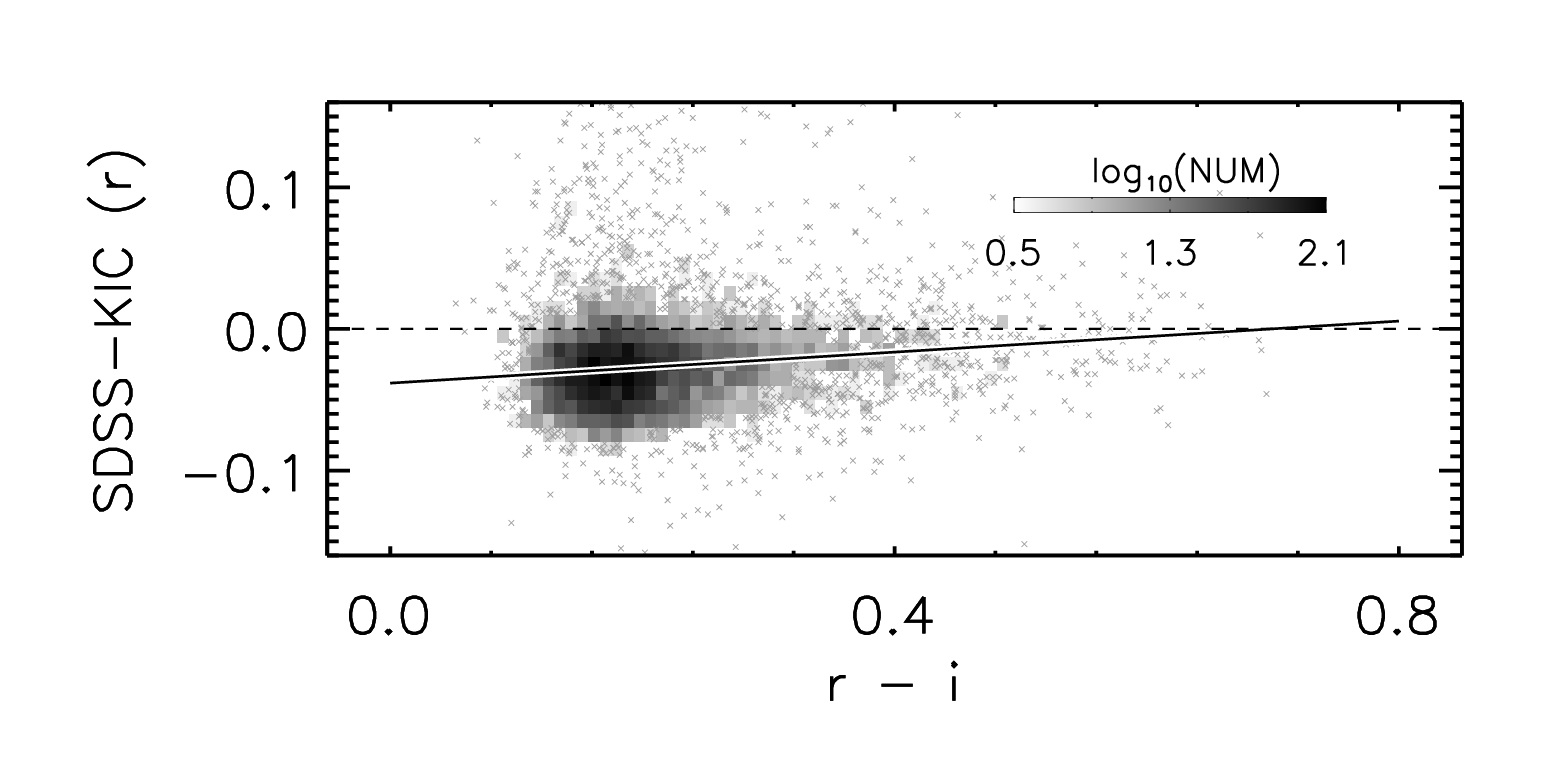}
\includegraphics[scale=0.34]{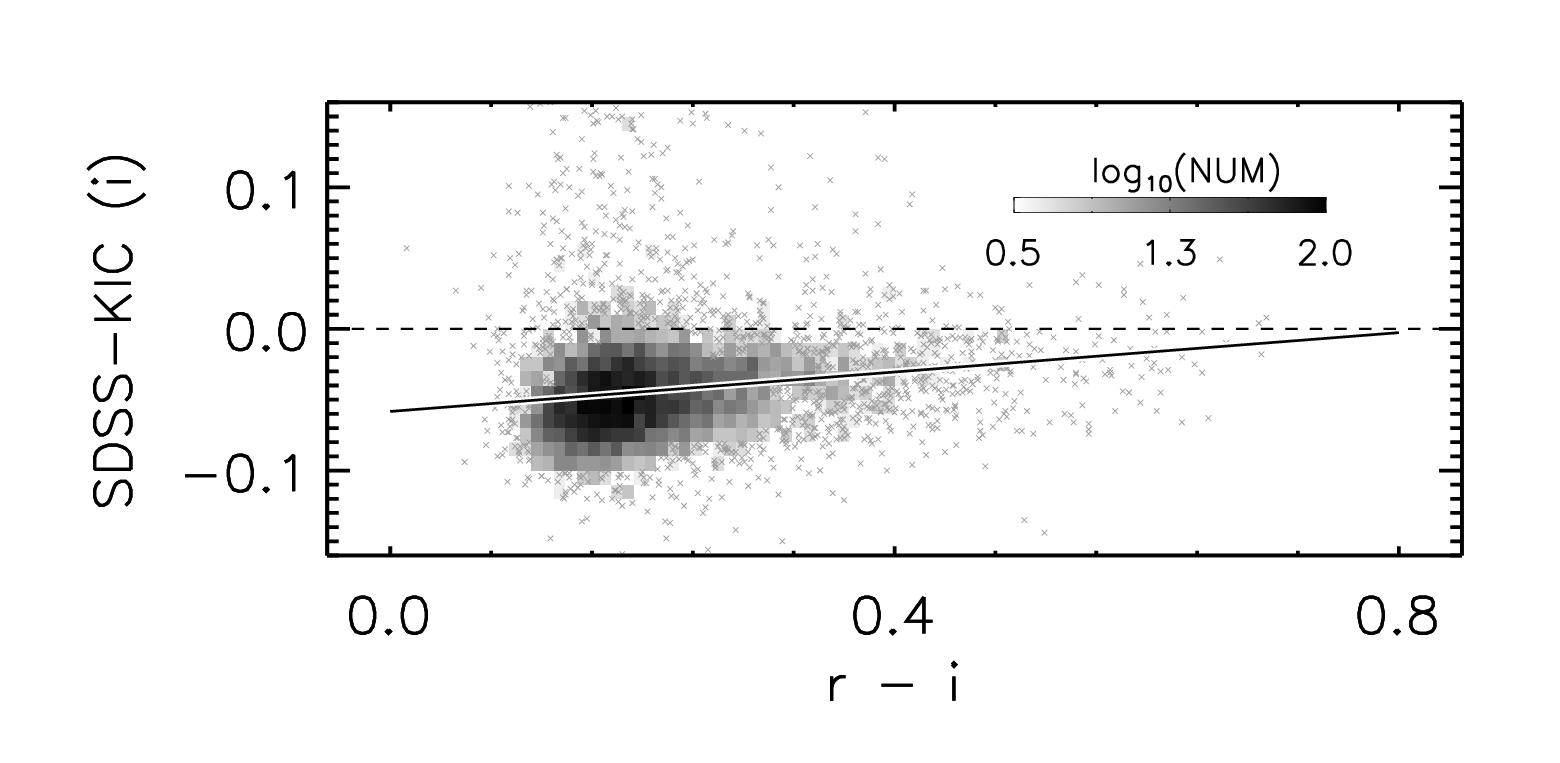}
\includegraphics[scale=0.34]{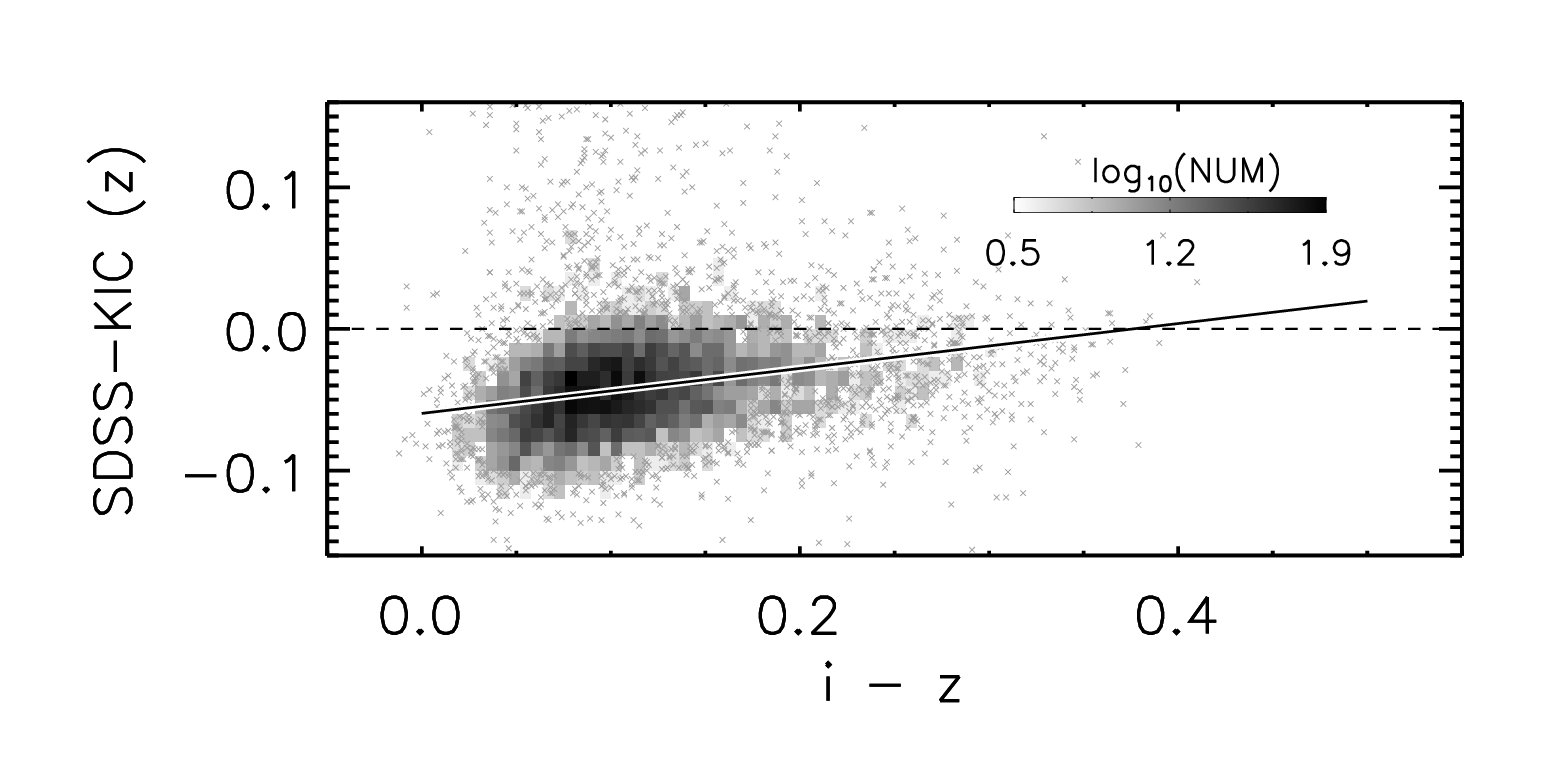}
\caption{Photometry comparisons between SDSS (DR8) and the KIC in the sense
of the former minus the latter. Comparisons are shown in $griz$ from top to
bottom panels. Solid lines are a linear fit to the residuals.
\label{fig:photcomp}}
\end{figure}

About 10\% of the stars in the {\it Kepler} field are covered in the most
recent data release (DR8) of the SDSS imaging survey \citep{dr8}. There is
an overlap in the two photometric sets at $14 \la r \la 18$. We compare
photometry for stars in common in Figure~\ref{fig:photcomp}. With a $1\arcsec$
search radius, we found that the median differences (in the sense of the SDSS
minus KIC), after rejecting stars with differences greater than $0.2$~mag
on both sides, are $\Delta g=-0.040$, $\Delta r=-0.028$, $\Delta i=-0.045$,
and $\Delta z=-0.042$.

Inspection of Figure~\ref{fig:photcomp} shows that these differences are
also functions of color. Solid lines are a linear fit to the data after
an iterative $3\sigma$ rejection. The linear transformation equations are
as follows:
\begin{eqnarray}
g_{\rm SDSS} = g_{\rm KIC} +0.0921 (g\, -\, r)_{\rm KIC} -0.0985,\\
r_{\rm SDSS} = r_{\rm KIC} +0.0548 (r\, -\, i)_{\rm KIC} -0.0383,\\
i_{\rm SDSS} = i_{\rm KIC} +0.0696 (r\, -\, i)_{\rm KIC} -0.0583,\\
z_{\rm SDSS} = z_{\rm KIC} +0.1587 (i\, -\, z)_{\rm KIC} -0.0597,
\label{eq:phot}
\end{eqnarray}
where the subscripts indicate either SDSS or KIC photometry.

It is possible that the SDSS photometry in the {\it Kepler} field has
some zero-point shifts with respect to the main SDSS survey database;
the SDSS photometry pipeline can fail to work properly
if the source density is too high. To check this, we compared
the KIC and the DAOPHOT crowded-field photometry \citep{an:08} in the
NGC~6791 field. Although the sample size is smaller, a comparison of
DAOPHOT and KIC photometry in the NGC~6791 field yields systematic offsets
in the same sense as the field mean in all band passes.
Given that the cluster fiducial sequence from DAOPHOT photometry
matches that from an independent study \citep{clem:08} relatively well
\citep{an:08}, it is unlikely that the offsets seen in
Figure~\ref{fig:photcomp} are due to zero-point issues in the SDSS photometry.

We also checked the standard star photometry in \citet{brown:11}, which is
originally from the SDSS DR1 photometry for 284 stars outside of the {\it Kepler} field.
We compared with the SDSS DR8 photometry, but did not find the aforementioned
trends outside of those expected from random photometric errors: the mean
differences (SDSS minus KIC values) were $-0.009$, $-0.004$, $+0.004$, and
$+0.012$~mag in $griz$, respectively, with an error in the mean of the order of
$0.001$~mag. Therefore, revisions of the standard magnitude system (SDSS DR1
versus DR8) do not appear to be the explanation either. We also investigated
the  possibility of a zero-point difference between the faint and bright stars in
the KIC, which had different exposures.
However, we found that magnitude offsets with respect to SDSS are similar
for both samples, and that the internal dispersion of the KIC temperature
estimates is essentially the same with each other.
Regardless of the origin, the differences between the SDSS and KIC
photometry are present in the overlap sample, and we therefore adjusted
the mean photometry to be on the most recent SDSS scale.

Inspection of Figure~\ref{fig:photcomp} also reveals another problem in the
KIC photometry: a sub-population of stars are much brighter in the KIC than
in the SDSS even after photometric zero-point shifts have been accounted for.
We do believe that unresolved background stars explains the occasional cases
where different colors predict very different temperatures.
In Figure~\ref{fig:photcomp} there are many data points  that have 
KIC magnitudes brighter than the SDSS ones. We attribute these stars to blended sources
in the KIC.  The mean FWHM of SDSS images is $1.4\arcsec$, while that of
KIC photometry is $2.4\arcsec$.

To check on this possibility, we cross-checked $13,284$ stars in common between
DR8 of the SDSS and our KIC sample. $312$ stars had a resolved SDSS source within
$2.4\arcsec$, while $20$ have two or more such blended sources. $2.5\%$ of the stars
would therefore have resolved blends between the resolution of the two surveys.
If we assume that the space density of blends is constant, we can use the density
of blends to estimate the fraction present even in the higher resolution SDSS sample.
When this effect is accounted for, we would expect $3.8\%$ of the KIC sources to
have a blended star within the resolution limit of the KIC. The average such star
was $2.85$~mag fainter than the KIC target, sufficient to cause a significant anomaly
in the inferred color-temperature relationships. A comparable fraction of the
catalog is likely to have similar issues.  A significant contribution from background
stars would in general combine light from stars with different temperatures.
As a result, one would expect different color-temperature relations to predict
discordant values. We therefore assess the internal consistency of the photometric
temperatures as a quality control check in our revised catalog to
identify possible blends (Section~\ref{sec:teff}).

To identify blended sources in the KIC, we further performed a test using the
separation index \citep{stetson:03}, which is defined as the logarithmic ratio of
the surface brightness of a star to the summed brightness from all neighboring stars
\citep[see also][]{an:08}. However, we found that applying the separation index to
the KIC does not necessarily provide unique information for assessing the effects of
the source blending.

\subsection{Base Model Isochrone}\label{sec:model}

We adopted stellar isochrones in A09 for the estimation of photometric
temperatures. Interior models were computed using YREC, and theoretical
color-$T_{\rm eff}$ relations were derived from the MARCS stellar atmospheres
model: see A09 and \citet{an:09b} for details. These model colors were then calibrated using
observed M67 sequences as in our earlier work in the Johnson-Cousins system
\citep{pinsono:03,pinsono:04,an:07a,an:07b}. The empirical color corrections
in $ugriz$ were defined using M67 at its solar metallicity, and
a linear ramp in [Fe/H] was adopted so that the color corrections become
zero at or below [Fe/H]$<-0.8$. Detailed test on the empirical color
corrections will be presented elsewhere (An et al. 2012, in preparation).

As a base case of this work, we adopted the mean metallicity recorded in
the KIC of [Fe/H]$=-0.2$. This metallicity is comparable to, or slightly below, 
that in the solar neighborhood.  For example, the Geneva-Copenhagen Survey
\citep{nord:04} has a mean [Fe/H] of $-0.14$~dex with a dispersion of $0.19$~dex;
a recent revision by \citet{casa:11} raises the mean [Fe/H] to $-0.07$~dex,
which is a fair reflection of the systematic uncertainties. The bulk of the
KIC dwarfs are about $100$~pc above the galactic plane, and thus would be
expected to have somewhat lower metallicity. In the following analysis, we
assumed [Fe/H]$=-0.2$ when using $griz$- or IRFM color-$T_{\rm eff}$ relationships,
unless otherwise stated.

\input{tab1.tex}

Table~\ref{tab:isochrone} shows our base model isochrone at [Fe/H]$=-0.2$ and
the age of $1$~Gyr. All colors are color-calibrated as described above.
Note that the isochrone calibration is defined for the main-sequence only;
the relevant corrections for the lower gravities of evolved stars are described
separately in Section~\ref{sec:giant}.
The SDSS photometry did not cover the main-sequence turn-off region of M67
because of the brightness limit in the SDSS imaging survey at $r \sim 14$~mag.
As a result, the M67-based $griz$ color calibration is strictly
valid at $4000 \leq T_{\rm eff} \leq 6000$~K (see Figure~$17$ in A09).

The choice of $1$~Gyr age in our base model isochrone has a negligible
effect on the color-$T_{\rm eff}$ relations. The difference between $1$~Gyr
and $12$~Gyr isochrones is only less than $5$~K near main-sequence turn-off.
However, younger age of the models enables the determination of photometric
$T_{\rm eff}$ over a wider range of colors at the hot $T_{\rm eff}$ end.

\input{tab2.tex}

From Table~\ref{tab:isochrone} we derived polynomial color-$T_{\rm eff}$ relations
of our base model for convenience of use. The following relationship was used
over the temperature range $4080$~K $\leq T_{\rm eff} {\rm (YREC)} < 7000~K$:
\begin{equation}
   5040/T_{\rm eff} = a_0+a_1 x+a_2 x^2+a_3 x^3+a_4 x^4+a_5 x^5
\label{eq:teff}
\end{equation}
where $x$ represents $g\, -\, r$, $g\, -\, i$, or $g\, -\, z$, and $a_0$--$a_5$
are coefficients for each color index as listed in Table~\ref{tab:poly}.
Difference in $T_{\rm eff}$ inferred from these polynomial equations
compared to those found
in Table~\ref{tab:isochrone} from interpolation in the full tables
are at or below the $6$~K level.

\input{tab3.tex}

In Table~\ref{tab:feh} we provide the metallicity sensitivity of the
color-$T_{\rm eff}$ relations in the model isochrones at several [Fe/H].
To generate this table, we compared $1$~Gyr old isochrones at individual
[Fe/H] with our fiducial model (Table~\ref{tab:isochrone}) at [Fe/H]$=-0.2$
for each color index, and estimated the $T_{\rm eff}$ difference at
a given color (individual models minus the fiducial isochrone).
We include the sensitivity to metallicity predicted by atmospheres models,
but do not include an additional empirical correction below [Fe/H]$=-0.8$
because the cluster data did not require one.
The $T_{\rm eff}$ at a fixed color generally becomes cooler at a lower [Fe/H].
We use the metallicity corrections in the comparisons with spectroscopic
$T_{\rm eff}$ where we have reliable [Fe/H] measurements (see Section~\ref{sec:spec}),
but do not apply corrections to the KIC sample (see Sections~\ref{sec:dwarf}
and \ref{sec:table}).

\subsection{Photometric $T_{\rm eff}$ Estimation}\label{sec:teff}

The stellar parameters for the KIC were generated using a
Bayesian method \citep[see][for a discussion]{brown:11}. We adopt a less
ambitious approach focused on KIC stars identified as dwarfs. The
three key assumptions in our work are that we define $T_{\rm eff}$ at a
reference [Fe/H] and the model $\log{g}$ (Table~\ref{tab:isochrone}),
and that we adopt the map-based E$(B\, -\, V)$
in the KIC as a prior.  Within this framework we can then derive independent
temperature estimates from the $griz$ photometry and infer the
random $T_{\rm eff}$ errors. Uncertainties in the extinction, the 
impact on the colors of unresolved binaries,
and population (metallicity and $\log{g}$) differences can then be
treated as error sources.  In the latter case, we can compute
correction terms to be used if there is an independent method of
measurement.  This approach is not the same as the one that we have
employed in earlier studies, so a brief justification is in order.

The traditional approach to photometric parameter estimation is to
take advantage of the fact that different filter combinations respond
to changes in metallicity and extinction. If one has the proper
template metallicity and extinction, for example, the answers from the
various colors will agree within photometric errors;
if not, the pattern of differences can be used to solve for them
\citep[see][]{an:07a,an:07b}.

The particular problem for the KIC is that the available color
combinations in $griz$ are rather insensitive to both over the narrow
metallicity range and the modest mean extinctions ($0 \la E(B\, -\, V)
\la 0.2$) in the field \citep[see][for a discussion of $griz$-based
estimates]{an:09b}. In other words, all the color combinations in $griz$
produce similar metallicity sensitivities of color-$T_{\rm eff}$ relations.
Therefore, even though the absolute change of photometric $T_{\rm eff}$
can be significant by the error in the adopted metallicity, it is
difficult to infer photometric metallicities based on the available filter
combinations in $griz$ alone.

The temperature estimates in \citet{lejeune:97,lejeune:98},
which were used as the basis color-$T_{\rm eff}$ relations in our prior
color calibration in the Johnson-Cousins system \citep{an:07a,an:07b},
are insensitive to $\log{g}$ near the main sequence, and the IRFM scale
in C10 does not include an explicit $\log{g}$ dependence for the temperatures.
As a result, we believe that the most fruitful approach is to define
a benchmark temperature estimate. If additional color information or
spectroscopic [Fe/H] data become available, the relevant corrections can
be applied, and we present methods below to do so (Section~\ref{sec:main}).

The KIC gravities for cool stars are precise enough to separate
dwarfs (KIC $\log{g} > 3.5$) from giants (KIC $\log{g} \leq 3.5$) and
to be used as a basis for corrections to the temperatures for giant stars
(Sections~\ref{sec:giant} and \ref{sec:table}).
The KIC metallicities are more problematic, and we do
not use them for temperature corrections. Instead the metallicity sensitivity
is included as an error source in our effective temperature estimates.

We adopted the map-based KIC catalog extinction estimates ($A_V$)
and the \citet{cardelli:89} standard extinction curves with
$A_V = 3.1 E(B\, -\, V)$. Extinction coefficients in $griz$ were derived
in A09:
$A_g=1.196 A_V$ , $A_r=0.874 A_V$ , $A_i=0.672 A_V$, and
$A_z=0.488 A_V$. We further took $A_J=0.282 A_V$, $A_H=0.180 A_V$,
$A_{K_s}=0.117 A_V$, and $A_{V_T}=1.050 A_V$, where $V_T$ represents
the Tycho $V$ passband \citep{an:07a}.

For a given extinction-corrected set of $griz$ magnitudes, we searched the
best-fitting stellar template in the model isochrone for each star in the KIC.
The mean $T_{\rm eff}$ was obtained by simultaneously fitting the models
in $griz$, assuming $0.01$~mag error in $gri$ and $0.03$~mag error in $z$.
We also estimated $T_{\rm eff}$ using the same model isochrone, but
based on data from each of our fundamental color indices ($g\, -\, r$,
$g\, -\, i$, and $g\, -\, z$), which is simply a photometric $T_{\rm eff}$
estimation from a single color-$T_{\rm eff}$ relation. Its purpose is
to readily identify and quantify the internal consistency of our primary
temperature determination from the multi-color-$T_{\rm eff}$ space.

\begin{figure}
\epsscale{1.05}
\plotone{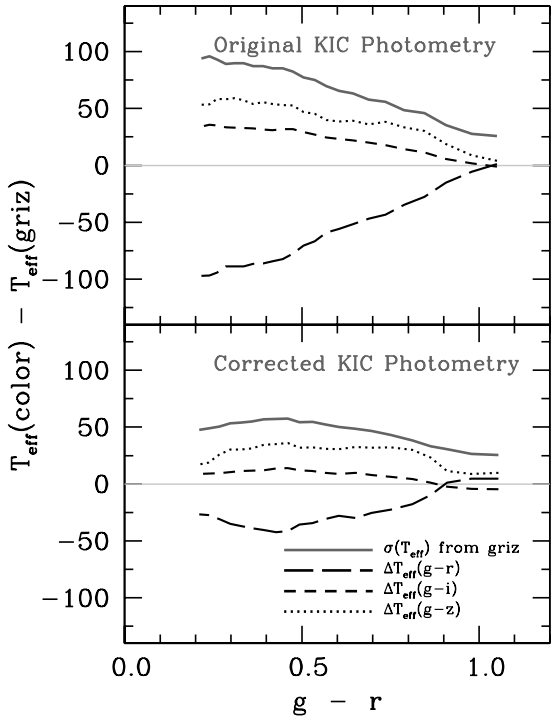}
\caption{Internal dispersion in $T_{\rm eff}$ estimates (solid line) and
differences between the mean $griz$-based $T_{\rm eff}$ and that inferred
from $g\, -\, r$ (long-dashed), $g\, -\, i$ (short-dashed), and $g\, -\, z$ (dotted).
Original KIC photometry is used in the top panel, while corrected KIC photometry
is used in the bottom panel.
\label{fig:SDSSdispersion}}
\end{figure}

In Figure~\ref{fig:SDSSdispersion} we plot the internal dispersion and the
mean trends of $T_{\rm eff}$ from a given color index with respect to the
average error-weighted temperature from $griz$ for all of the dwarfs in our
sample. The top panel shows the case of the original KIC data, and the bottom
panel shows the one for the corrected KIC photometry.
The magnitude corrections described in Section~\ref{sec:phot} were motivated
by concordance between SDSS and the KIC. Nevertheless, the results when
using the recalibrated KIC photometry as temperature indicators were
extremely encouraging.

Although the internal agreement is not complete, the remaining differences
in the bottom panel of Figure~\ref{fig:SDSSdispersion} are comparable to
the zero-point uncertainties discussed in \citet{an:08}.
We view this as strong supporting evidence for the physical reality of
the magnitude corrections illustrated in Figure~\ref{fig:photcomp}. We
therefore recommend that the zero-points of the KIC photometry be modified
according to equations~1--4. In the remainder of the paper, we use magnitudes
and colors adjusted using these equations.

\section{Revised $T_{\rm eff}$ Scale for the KIC}\label{sec:main}

We begin by evaluating the $T_{\rm eff}$ inferred from the IRFM and the SDSS
systems for dwarfs (KIC $\log{g} > 3.5$).  We then use open clusters and
comparisons with high-resolution spectroscopy to establish agreement between
the two scales, indicating the need for correction to the KIC effective temperatures.
We then evaluate the impact of binaries, surface gravity, and metallicity
on the colors.  We provide statistical corrections to the temperatures caused
by unresolved binary companions, as well as corrections
for $\log{g}$ and metallicity.  We then perform a global error analysis
including extinction uncertainties and the mild metallicity dependence of our
color-temperature relationships.  The latter is treated as a temperature
error source because we evaluate all KIC stars at a mean 
reference metallicity ([Fe/H]$=-0.2$).

\subsection{Temperature scale comparisons for dwarfs}\label{sec:dwarf}

We have three native temperature scales to compare: the one in the
KIC, our isochrone-based scale from $griz$ (hereafter SDSS or
$griz$-based scale unless otherwise stated), and one from the
($J\, -\, K_s$)-based IRFM.
Below we compare the mean differences between them and compare
the dispersions to those expected from random error sources alone.
We find an offset between the KIC and the other two scales.  The IRFM
and SDSS scales are closer, but some systematic differences
between them are also identified.  In this section, we examine various
effects that could be responsible for these differences, and finish with
an overall evaluation of the error budget.

We computed IRFM and SDSS $T_{\rm eff}$ estimates assuming [Fe/H]$=-0.2$.
In terms of the temperature zero-point, adopting this metallicity led to
mean shifts of $+20$~K in $J\, -\, K_s$, and $-40$~K in the $griz$-based
$T_{\rm eff}$ estimate, relative to those which would have been obtained
with solar abundance. In other words, changes in the adopted {\it mean}
metallicity would cause zero-point shifts of $\sim60$~K in the overall
$T_{\rm eff}$ scale comparison. On the other hand, a scatter around the
mean metallicity in the {\it Kepler} field is another source of error that
would make the observed $T_{\rm eff}$ comparison broader. We discuss this
in Section~\ref{sec:error} along with other sources of uncertainties.

In the comparisons below we repeatedly clipped the samples, rejecting stars
with temperature estimates more than three standard deviations from the mean,
until we achieved convergence.  This typically involved excluding about 1 \%
of the sample. Such stars represent cases where the extinction corrections
break down or where the relative colors differ drastically from those expected
for single unblended stars.

Random errors were taken from the photometric errors alone and
yield a minimum error in temperature.  For the SDSS colors we also
computed the internal dispersion in the three temperature estimates
from $g\, -\, r$, $g\, -\, i$, and $g\, -\, z$, and used the larger of
either this dispersion or the one induced by
photometric errors as a random uncertainty.  
Median random errors for the SDSS and 
IRFM temperatures were $40$~K and $171$~K, respectively.
These estimates are consistent with expectations from the observed
dispersions of the colors (see Figure~\ref{fig:binaryerror}).
We then compared stars at fixed KIC temperature 
and computed the average $T_{\rm eff}$ difference between those inferred
from the IRFM, those inferred from $griz$, and the scale in the KIC itself.
For a limited subset of stars, we also had {\it Tycho} photometry, and
computed temperatures from $V_T\, -\, K_s$. This sample is small,
so we used it as a secondary temperature diagnostic.

\begin{figure}
\epsscale{1.25}
\plotone{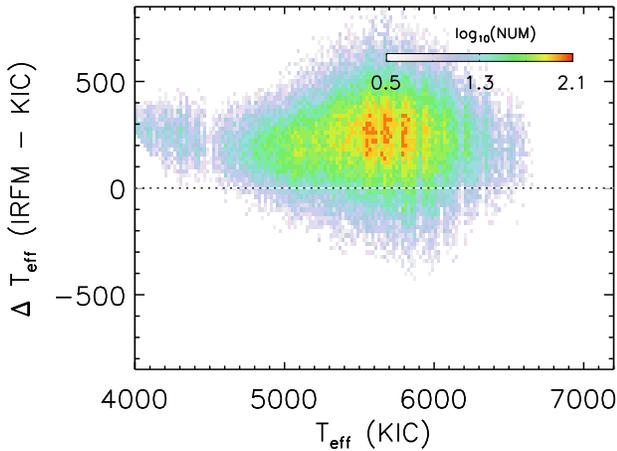}
\caption{Comparisons of the temperatures inferred from IRFM ($J\, -\, K_s$) as a function
of KIC $T_{\rm eff}$. The color coding indicates the logarithmic number density
of stars with a temperature and temperature difference at the indicated point (see legend).
\label{fig:compteffirfmkic}}
\end{figure}

\begin{figure}
\epsscale{1.25}
\plotone{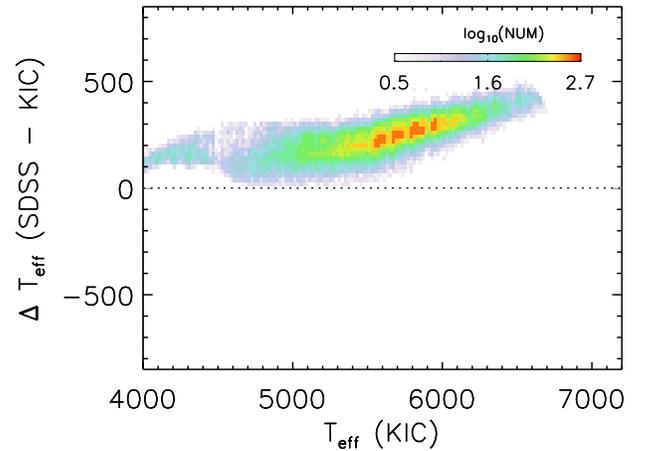}
\caption{Same as in Figure~\ref{fig:compteffirfmkic}, but from $griz$ colors.
\label{fig:compteffkicyrec}}
\end{figure}

In Figures~\ref{fig:compteffirfmkic}--\ref{fig:compteffkicyrec} we illustrate
the differences between KIC and the IRFM and SDSS, respectively. For the IRFM
scale in Figure~\ref{fig:compteffirfmkic}, we compare $T_{\rm eff}$ from
$J\, -\, K_s$. In Figure~\ref{fig:compteffkicyrec} we compare the mean
SDSS temperatures inferred from $griz$ to that in the KIC. In both cases
we see a significant zero-point shift, indicating a discrepancy between the
fundamental effective temperature scale and that adopted by the KIC.

\begin{figure*}
\centering
\includegraphics[scale=0.62]{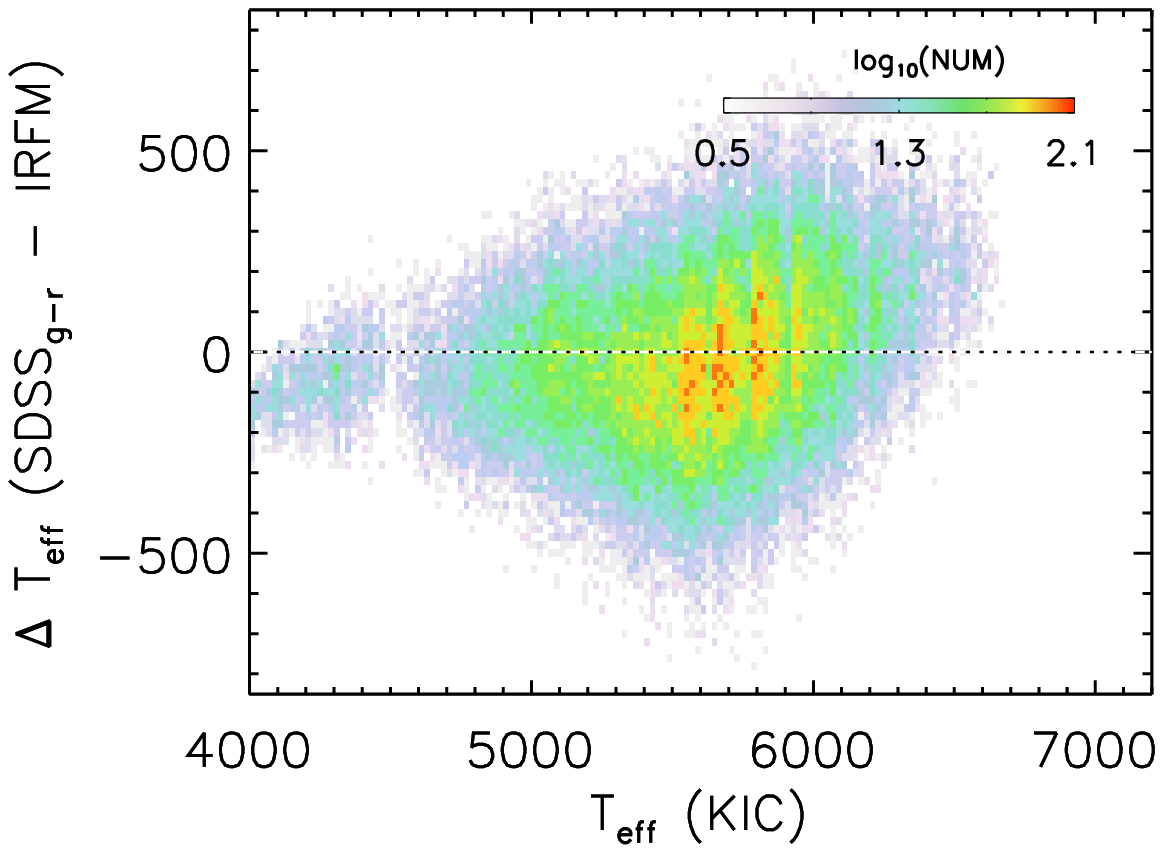}
\includegraphics[scale=0.62]{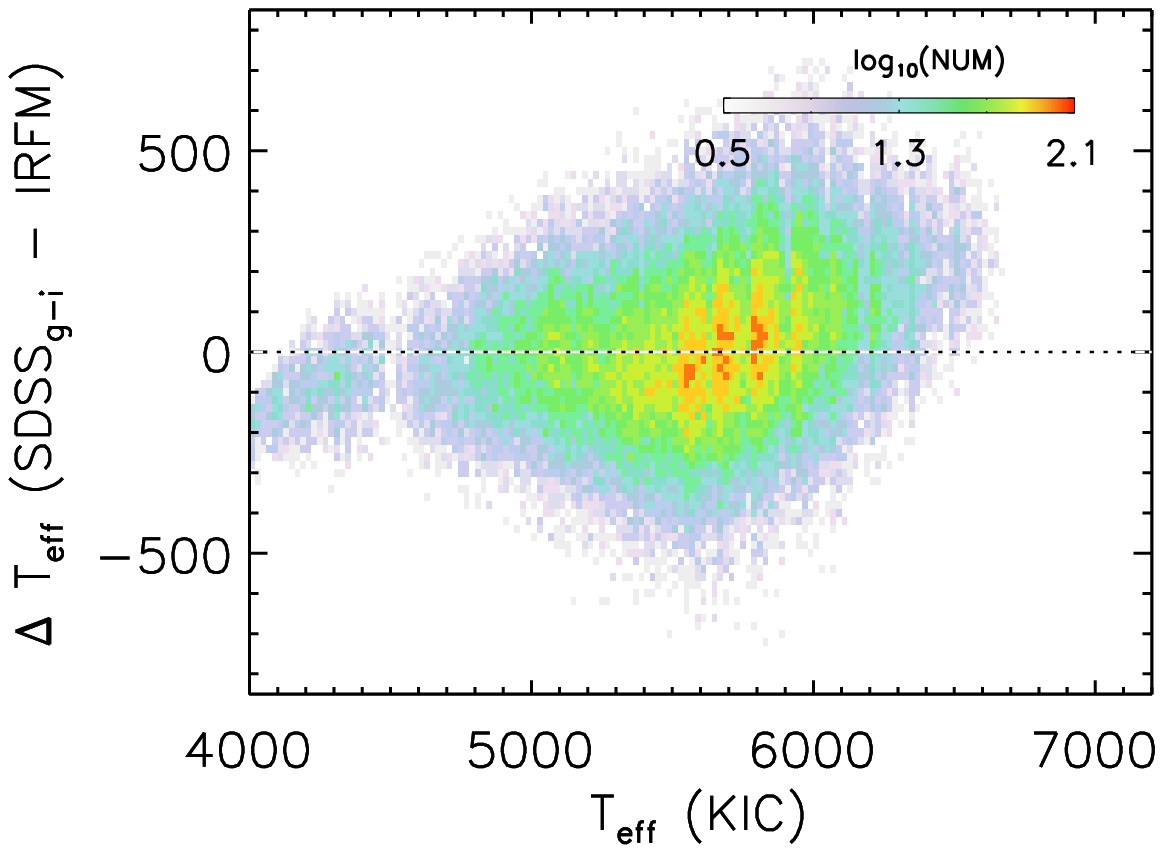}
\includegraphics[scale=0.62]{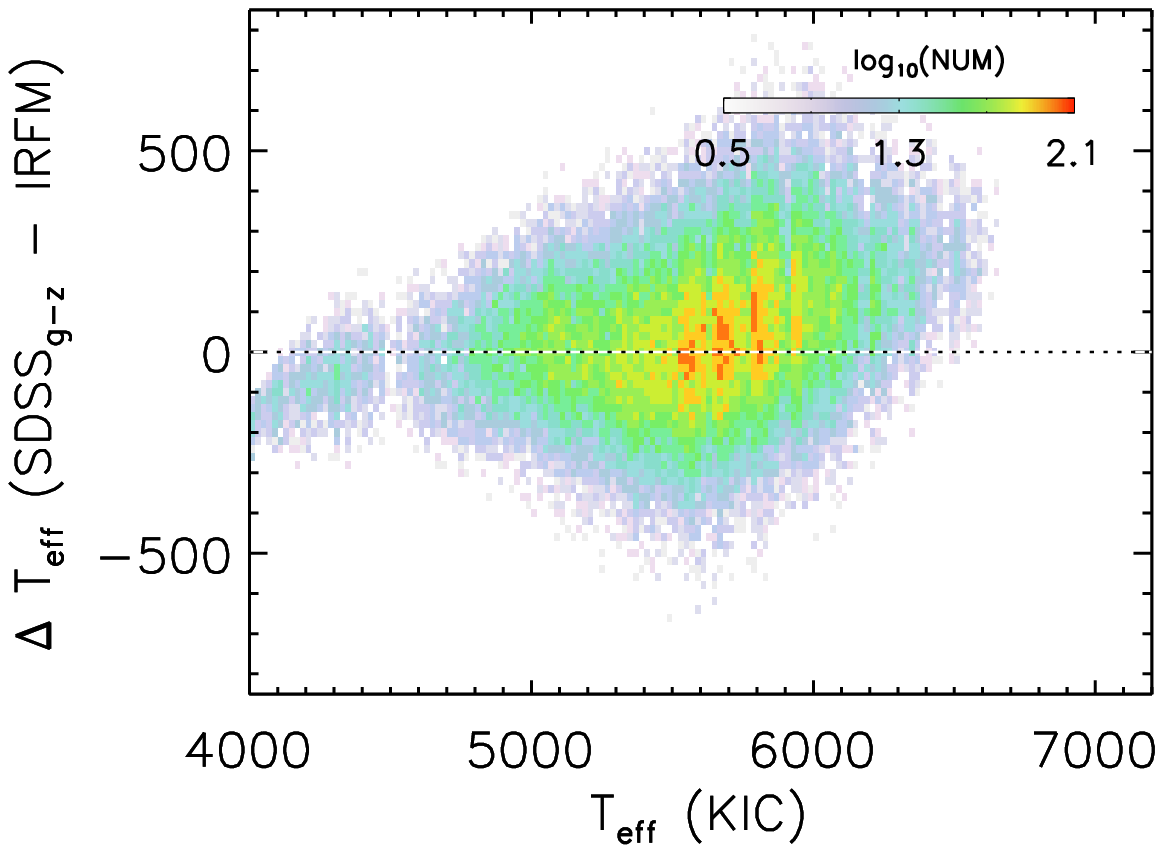}
\includegraphics[scale=0.62]{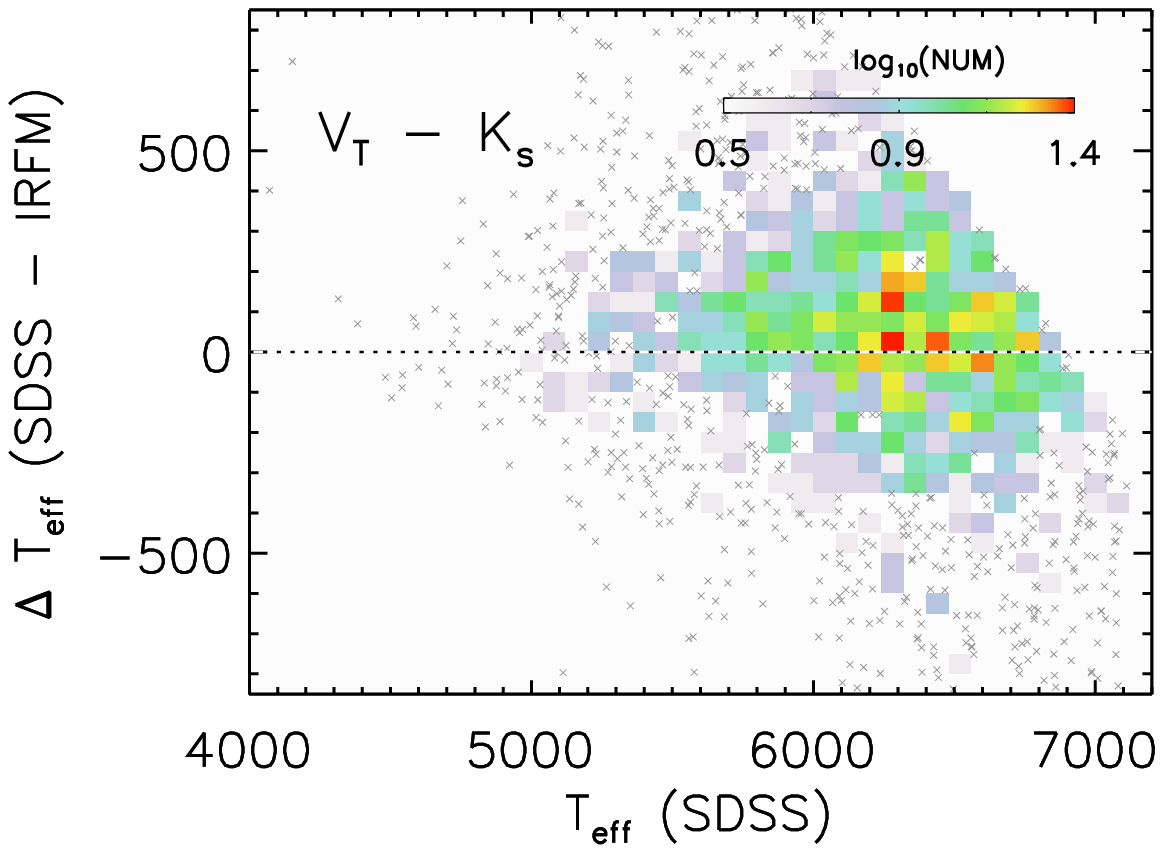}
\caption{Differences in $T_{\rm eff}$ between the IRFM and SDSS scales
as a function of KIC $T_{\rm eff}$:
$T_{\rm eff}(g\, -\, r)$ vs.\ $T_{\rm eff}(J\, -\, K_s)$ (upper left);
$T_{\rm eff}(g\, -\, i)$ vs.\ $T_{\rm eff}(J\, -\, K_s)$ (upper right);
$T_{\rm eff}(g\, -\, z)$ vs.\ $T_{\rm eff}(J\, -\, K_s)$ (lower left);
$T_{\rm eff}(griz)$ vs.\ $T_{\rm eff}(V_T\, -\, K_s)$
(lower right). The color coding defines the logarithmic number density of
points with the indicated temperature and temperature difference (see legend for details).
\label{fig:compteff}}
\end{figure*}

The $T_{\rm eff}$ from the IRFM and the SDSS from individual color indices
($g\, -\, r$, $g\, -\, i$, and $g\, -\, z$) are compared in Figure~\ref{fig:compteff}.
The IRFM scale for the Tycho $V_T$ and 2MASS $K_s$ is used in the bottom right panel.
The central result (that the KIC scale is too cool) is robust, and
can also be seen in comparisons with high-resolution spectroscopic
temperature estimates (see Section~\ref{sec:spec} below). 
In Section~\ref{sec:table} we provide quantitative tabular information on the 
statistical properties of the sample.

The two fundamental scales (IRFM and SDSS) are close, but not
identical, for cooler stars; they deviate from one another and the KIC
above $6000$~K (on the SDSS scale). As discussed in Section~\ref{sec:error} below,
the total internal dispersion in the $griz$ temperature estimates is also 
consistently larger for cool stars than that
expected from random photometric uncertainties alone, and there are modest
but real offsets between the two fundamental scales even for cool stars.
We therefore need to understand the origin of these differences and to
quantify the random and systematic uncertainties in our temperature estimates.

Open clusters provide a good controlled environment for testing the
concordance of the SDSS and IRFM scales. The SDSS scale was developed
to be consistent with Johnson-Cousins-based temperature calibrations in
open clusters, so a comparison of the An et al.\ and IRFM  Johnson-Cousins
systems in clusters will permit us to verify their underlying agreement.
As we show below, the two scales are close for cool stars when $B\, -\, V$,
$V\, -\, I_C$, or $V\, -\, K_s$ indices are employed in the temperature
determinations, but exhibit modest but real systematics for the hotter stars.
The IRFM relation in $J\, -\, K_s$, on the other hand, is found to have
a systematic difference from those of these optical-2MASS indices.
For the reasons discussed in the following section, we therefore adopt
a correction to our SDSS temperatures for hot stars, making the two
photometric systems consistent.

We can also check our methodology against spectroscopic temperature estimates,
and need to consider uncertainties from extinction, binary companions, and
metallicity. We therefore begin by defining an extension of our method to giants,
which can be checked against spectroscopy. We then look at open cluster tests,
spectroscopic tests, binary effects, and the overall error budget.

\subsection {Tests of the temperature scale for giants}\label{sec:giant}

Our YREC $T_{\rm eff}$ estimates are based on calibrated isochrones
(Table~\ref{tab:isochrone}), which do not include evolved stars. 
About $14\%$ of the KIC sample are giants and subgiants 
with $\log{g} \le 3.5$ as estimated in the KIC, so a reliable
method for assigning effective temperatures to such stars is
highly desirable.  Fortunately, this is feasible because 
the color-temperature relations
for the bulk of the long cadence targets are not strong functions
of surface gravity.  For the purposes of the catalog we therefore supplement 
the fundamental dwarf scale with theoretical corrections for the
effect of surface gravity on the colors.

\begin{figure}
\epsscale{1.05}
\plotone{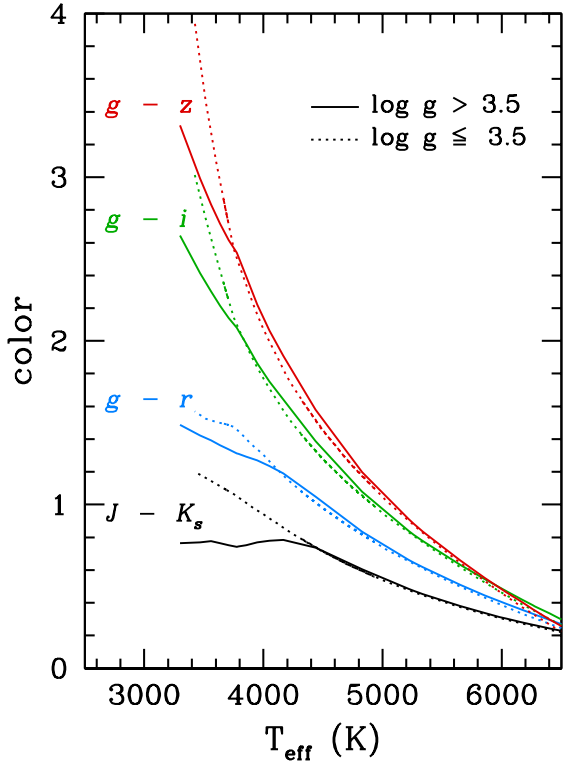}
\caption{Main-sequence (solid; $\log{g} > 3.5$) and post-main-sequence
(dashed; $\log{g} \leq 3.5$) color-temperature relationships for models
along a $1$~Gyr isochrone with solar composition
are shown for illustrative purposes.
Colors illustrated are $g\, -\, z$, $g\, -\, i$, $g\, -\, r$, and
$J\, -\, K_s$ (from top to bottom). 
\label{fig:loggiso}}
\end{figure}

Theoretical model atmospheres can be used to quantify the $\log{g}$
dependence of the color-temperature relations by comparing the spectral
energy distributions of dwarfs and giants. Figure~\ref{fig:loggiso}
shows color-temperature relations along a $1$~Gyr solar abundance
isochrone for $g\, -\, r$, $g\, -\, i$, $g\, -\, z$, and $J\, -\, K_s$
for illustrative purposes.
The model isochrone was taken from the web interface of the Padova isochrone database
\citep{girardi:02,marigo:08}\footnote{\tt http://stev.oapd.inaf.it/cgi-bin/cmd.}.
As seen in Figure~\ref{fig:loggiso} the model color-$T_{\rm eff}$ relations
are moderately dependent on $\log{g}$, and illustrate that our photometric
$T_{\rm eff}$ needs to be adjusted for giants.

We corrected for the difference in $\log{g}$ by taking theoretical
$\log{g}$ sensitivities in $griz$ colors from the ATLAS9 model atmosphere
\citep{castelli:04}. The choice of these models seems internally inconsistent
with our basis model isochrone with MARCS-based colors. Nevertheless, we
adopted the ATLAS9 $\log{g}$-color relations, primarily because our
cluster-based empirical calibration of the color-$T_{\rm eff}$ relations
has not been performed for subgiant and giant branches due to significant
uncertainties in the underlying stellar interior models at these evolved
stages. Therefore, it is just a matter of choice to adopt the ATLAS9 color
tables instead of that of MARCS. Since we generated MARCS color tables in
\citet{an:09a} with a specific set of model parameters for dwarfs
($\log{g} \geq 3.5$), we simply opted to take the ATLAS9 colors, and
estimate a relative sensitivity of theoretical $\log{g}$-color relations.

We convolved synthetic spectra with the SDSS $griz$ filter response
curves\footnote{\tt http://www.sdss3.org/instruments/camera.php.},
and integrated flux with weights given by photon counts \citep{girardi:02}.
Magnitudes were then put onto the AB magnitude system using a flat $3631$~Jy
spectrum \citep{oke:83}. We created a table with synthetic colors from
$\log{g}=0.0$ to $5.0$~dex with a $0.5$~dex increment, and from $4000$~K
to $6000$~K with a $250$~K increment at [M/H]=$-1.0$, $-0.5$, $+0.0$, and
$+0.2$. Because YREC $T_{\rm eff}$ values were estimated at the fiducial
metallicity, [Fe/H]$=-0.2$, we interpolated the color table to obtain
synthetic colors at this metallicity. Note that \citet{castelli:04} adopted
the solar mixture of \citet{grevesse:98}, as in our YREC isochrone models
(A09), so we assumed [M/H] in \citet{castelli:04} is the same as the [Fe/H]
value.

\begin{figure}
\epsscale{1.15}
\plotone{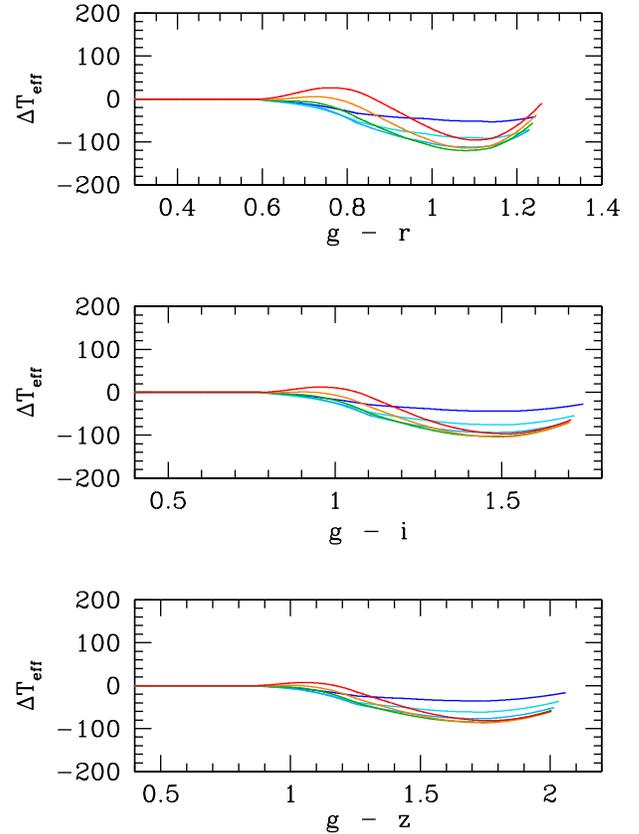}
\caption{Theoretical $T_{\rm eff}$ corrections for various $\Delta \log{g}$ values
with respect to the fiducial isochrones. Corrections from $\Delta \log{g} = 0.5$
to $\Delta \log{g} = 3.0$ with a $0.5$~dex increment are shown. A linear ramp was
used to define smoothly varying $\Delta T_{\rm eff}$ over $4800 < T_{\rm eff} < 5800$~K.
The sense is that giants with lower $\log{g}$ than the base isochrone
tend to have lower $T_{\rm eff}$.
\label{fig:loggcorr}}
\end{figure}

Figure~\ref{fig:loggcorr} shows the correction factors in $T_{\rm eff}$ as
computed from synthetic colors as a function of colors in $g\, -\, r$, $g\, -\, i$,
and $g\, -\, z$. We used our base isochrone to compute $\Delta T_{\rm eff}$ at
$\Delta \log{g} = 0.5$, $1.0$, $1.5$, $2.0$, $2.5$, and $3.0$~dex, where
$\Delta \log{g}$ represents the difference between YREC $\log{g}$ and the
$\log{g}$ in the KIC. The sense of $\Delta T_{\rm eff}$ is that giants with
lower $\log{g}$ than the base model generally tend to have lower $T_{\rm eff}$
than main-sequence dwarfs in the color range considered in this work.

\input{tab4.tex}

In Figure~\ref{fig:loggcorr} we used a linear ramp over
$4800 < T_{\rm eff} < 5800$~K ($0.42 < g\, -\, r < 0.82$), so that the
theoretical $\Delta T_{\rm eff}$ becomes zero at $T_{\rm eff} > 5800$~K.
Otherwise the amplitude of theoretical $T_{\rm eff}$ variations on the blue
side ($g\, -\, r \la 0.6$) would be similar to that of the red colors. Although
this is not strictly true if the $\Delta \log{g}$ is large for blue stars,
those stars are rare because stars on the giant branch  (with the largest
$\Delta \log{g}$) have $g\, -\, r \ga 0.5$ at near solar metallicity.
The correction factors are tabulated in Table~\ref{tab:logg}.
If one wishes to adopt a different $\log{g}$ scale than in our base
isochrone, tabulated $\Delta T_{\rm eff}$ factors can be used to correct
for the $\log{g}$ difference. More importantly, Table~\ref{tab:logg} can
be used to infer $T_{\rm eff}$ for giants, since our base isochrone
(Table~\ref{tab:isochrone}) covers stellar parameters for main-sequence
dwarfs only.

The biggest $\Delta T_{\rm eff}$ in Figure~\ref{fig:loggcorr} is $\sim100$~K.
However, the effects of the $\log{g}$ corrections are moderate in the KIC. If
we take the mean $\Delta T_{\rm eff}$ correction in $g\, -\, r$, $g\, -\, i$,
and $g\, -\, z$, the mean difference in $T_{\rm eff}$ between KIC and YREC
decreases from $190$~K to $166$~K for stars with $\log{g} \le 3.5$.
The $\log{g}$ corrections are insensitive to metallicity. The
$\Delta T_{\rm eff}$ in Figure~\ref{fig:loggcorr} was computed at [Fe/H]$=-0.2$,
but these corrections are within $10$~K away from those computed at [Fe/H]$=-0.5$
($\sim1\sigma$ lower bound for the KIC sample) when $\Delta \log{g}=1$.

The statistical properties of the SDSS giant temperatures are compared
with spectroscopic data in Section~\ref{sec:spec} and with the KIC in
Section~\ref{sec:catalog}.

\subsection{Tests with Open Cluster Data}\label{sec:cluster}

The IRFM technique provides global color-metallicity-$T_{\rm eff}$ correlations
using field samples, while clusters give snapshots at fixed composition, which
define color-$T_{\rm eff}$ trends more precisely. Deviations from color to color
yield the internal systematic within the system, as the color-temperature
relationships defined in \citet{an:07b} are empirical descriptions of actual
cluster data.  The A09 SDSS system, by construction, agrees with the
\citet{an:07b} Johnson-Cousins system; but we can check the concordance between
the two scales within the open cluster system.

We have two basic results from this comparison.  First, $(J\, -\, K_s)$-based 
temperatures from the IRFM are different from other IRFM thermometers.
$J\, -\, K_s$ is also the only IRFM diagnostic available for the 
bulk of the KIC sample.
When accounting for the offset in $J\, -\, K_s$ relative to
other IRFM indicators, the underlying
IRFM system and the SDSS system are in excellent agreement for 
stars below 6000 K.  Second, there is a systematic offset between
the IRFM and SDSS scales above 6000 K.  We therefore correct the
high end temperature estimates for the SDSS to put them on the IRFM
scale, which yields an internally consistent set of photometric
temperature estimates.

\begin{figure}
\epsscale{1.05}
\plotone{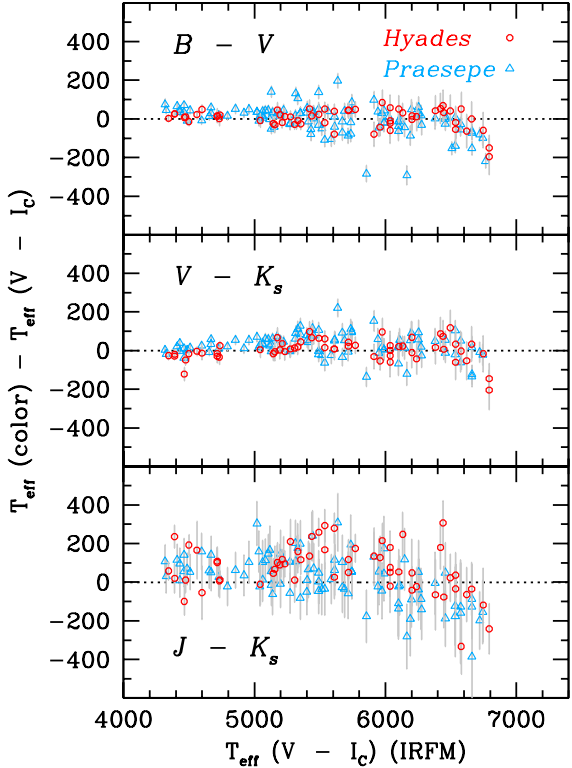}
\caption{Internal consistency of the IRFM $T_{\rm eff}$ estimates for the
Hyades (red circles) and Praesepe stars (blue triangles).
Comparisons are shown for each color index with respect to
the $T_{\rm eff}$ values determined from $V\, -\, I_C$ at [Fe/H]$=0.13$ for
the Hyades and [Fe/H]$=0.14$ for Praesepe.
Error bars represent $\pm1\sigma$ uncertainty propagated from photometric errors.
\label{fig:clusterhyades}}
\end{figure}

\begin{figure}
\epsscale{1.05}
\plotone{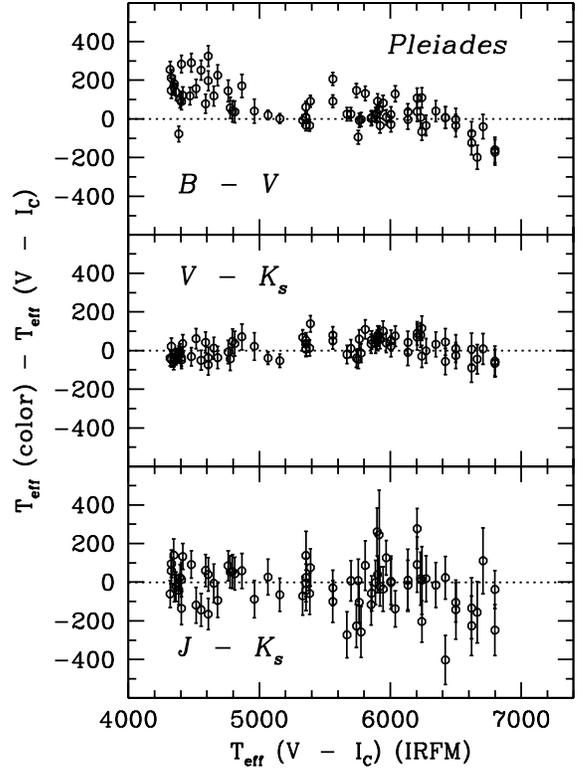}
\caption{Same as in Figure~\ref{fig:clusterhyades}, but for the Pleiades at
[Fe/H]$=0.04$. Note that low-mass Pleiades stars ($T_{\rm eff} \la 5000$~K)
are known to have anomalously blue colors in $B\, -\, V$. These stars could
also have slight near-IR excesses, which may have affected $T_{\rm eff}$
values from $J\, -\, K_s$.
\label{fig:clusterpleiades}}
\end{figure}

\begin{figure}
\epsscale{1.05}
\plotone{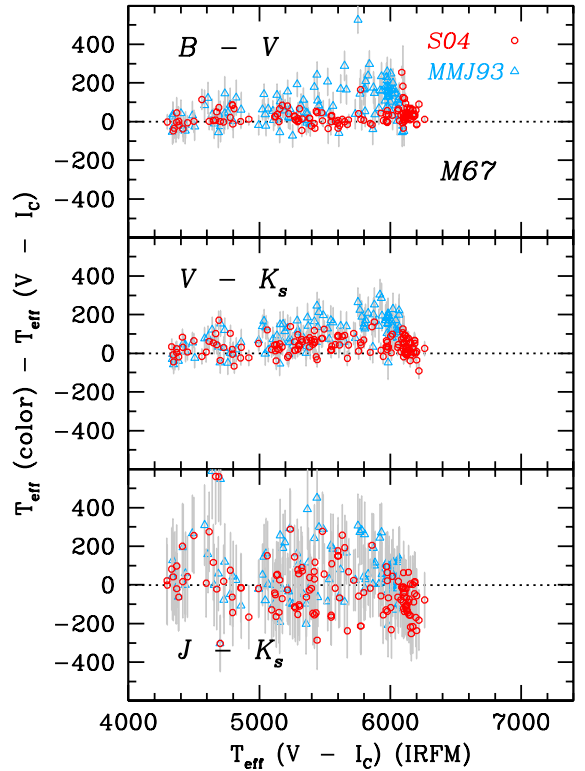}
\caption{Same as in Figure~\ref{fig:clusterhyades}, but for M67
at [Fe/H]$=0.0$. Red circles and blue triangles represent comparisons
based on the \citet[][S04]{sandquist:04} and \citet[][MMJ93]{montgomery:93}
photometry, respectively.
\label{fig:clusterm67}}
\end{figure}

Figures~\ref{fig:clusterhyades}--\ref{fig:clusterm67} show how the IRFM
$T_{\rm eff}$ determinations are internally consistent in the
Jounson-Cousins-2MASS system in $B\, -\, V$, $V\, -\, I_C$, $V\, -\, K_s$,
and $J\, -\, K_s$ using stars in four 
well-studied clusters: The Hyades (red
circles in Figure~\ref{fig:clusterhyades}), 
Praesepe (blue triangles in
Figure~\ref{fig:clusterhyades}), the Pleiades (Figure~\ref{fig:clusterpleiades}),
and M67 (Figure~\ref{fig:clusterm67}). All of the stars shown in these figures
are likely single-star members of each cluster after excluding known (unresolved)
binaries. In Figure~\ref{fig:clusterm67}, we show results based on the
two independent sets of M67 photometry from
\citet[][blue triangles]{montgomery:93} and \citet[][red circles]{sandquist:04}.
The compilation and individual sources of the cluster photometry can
be found in \citet{an:07b}.

To construct Figures~\ref{fig:clusterhyades}--\ref{fig:clusterm67} we corrected
observed magnitudes for extinction using $E(V\, -\, I_C)/E(B\, -\, V) = 1.26$,
$E(V\, -\, K_s)/E(B\, -\, V) = 2.82$, and $E(J\, -\, K_s)/E(B\, -\, V) = 0.53$
\citep{an:07a}. Foreground reddening values of $E(B\, -\, V)=0.000\pm0.002$,
$0.006\pm0.002$, $0.032\pm0.003$, and $0.041\pm0.004$~mag were used for the
Hyades, Praesepe, the Pleiades, and M67 respectively \citep{an:07b}. The
IRFM $T_{\rm eff}$ equations in C10 include metallicity terms, and we adopted
[Fe/H]$=+0.13\pm0.01$, $+0.14\pm0.02$, $+0.04\pm0.02$, and $+0.00\pm0.01$~dex for
the Hyades, Praesepe, the Pleiades, and M67, respectively, based on high-resolution
spectroscopic abundance analysis \citep[see references in][]{an:07b}. Only the
$(B\, -\, V)$-based estimates are significantly impacted by metallicity corrections,
and the relative abundance differences in these well-studied open clusters are
unlikely to be substantial enough to affect our results.

\input{tab5.tex}

The $\pm1\sigma$ error bars in Figures~\ref{fig:clusterhyades}--\ref{fig:clusterm67}
are those propagated from the photometric errors
only. Mean differences in the IRFM $T_{\rm eff}$ and the errors in the
mean are provided in Table~\ref{tab:cluster}. Global differences are shown
for stars at $4000 < T_{\rm eff} \leq 7400$~K, and those cooler and hotter
than $6000$~K are shown in the table. The $\sigma_{\rm sys}$ represents
a total systematic error in this comparison from the reddening and metallicity
errors (summed in quadrature); however, systematic errors are less
important than random errors because of the precise E$(B\, -\, V)$
and [Fe/H] estimates of these well-studied clusters.

The low-mass stars in the Pleiades are known to have anomalously blue
colors related to stellar activity in these heavily spotted, rapidly
rotating, young stars \citep{stauffer:03}. The temperature anomaly for
$B\, -\, V$ at $T_{\rm eff} \la 5000$~K in Figure~\ref{fig:clusterpleiades},
which is $\sim200$~K larger than that for more massive stars, reflects
this known effect and therefore is not a proper test of internal consistency
in old field stars (such as those in the KIC). The M67 data may also be
inappropriate for the test of the IRFM internal consistency, but with
a different reason. Two independent photometry sets lead to a different
conclusion: \citet{montgomery:93} photometry shows internally less
consistent IRFM $T_{\rm eff}$ for M67 stars than \citet{sandquist:04}.
A similar argument was made in \citet{an:07b}, based on the differential
metallicity sensitivities of stellar isochrones in different color indices
(see Figure~$11$ in the above paper);
see also \citet{vandenberg:10} for an independent confirmation of
the systematic zero-point issue with the \citet{montgomery:93} photometry.

Our cluster tests based on the Hyades and Praesepe demonstrate the internal
consistency of the C10 color-$T_{\rm eff}$ relations in $B\, -\, V$,
$V\, -\, I_C$, and $V\, -\, K_s$. The mean differences in $T_{\rm eff}$
among these color indices are typically few tens of degrees for both hot
and cool stars (Table~\ref{tab:cluster}).
Some of these mean differences could be systematic in nature, but they
are generally consistent with the scatter in the C10 IRFM calibrations.
However, the $(J\, -\, K_s)-T_{\rm eff}$ relation tends to produce
hotter $T_{\rm eff}$ than those from other color indices for these
cluster stars (see bottom panel in Figure~\ref{fig:clusterhyades}).
The mean differences between $T_{\rm eff} (V\, -\, I_C)$ and $T_{\rm eff}
(J\, -\, K_s)$ are $95$~K and $44$~K for the Hyades and Praesepe,
respectively. There is also a hint of the downturn in the comparison for
the hot stars in these clusters, where $(J\, -\, K_s)-T_{\rm eff}$ produces
cooler temperatures than $(V\, -\, I_C)-T_{\rm eff}$ relation. The
$\sim100$~K offset between the hot and the cool stars roughly defines
the size of the systematic error in the IRFM technique of C10 in $J\, -\, K_s$.

The Pleiades stars show a weaker systematic $T_{\rm eff}$ trend
for the cool and the hot stars than the Hyades and Praesepe. In spite
of this good agreement, we caution that this could be a lucky coincidence
because the Pleiades low-mass stars probably have slight near-IR
excesses in $K_s$ \citep{stauffer:03}. The main-sequence turn-off of M67
is relatively cool, so the difference is only suggestive.

\begin{figure*}
\centering
\includegraphics[scale=0.62]{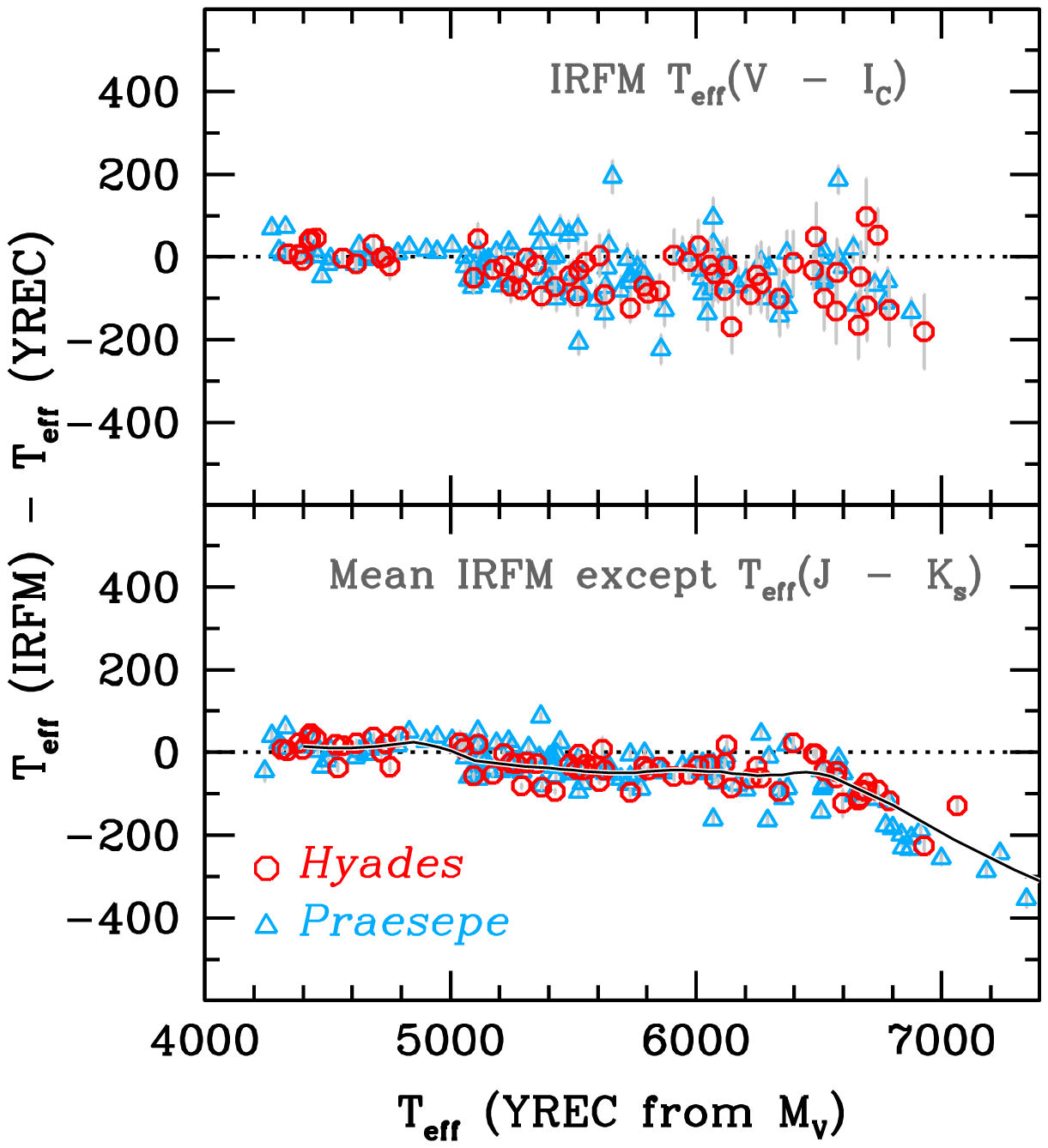}
\includegraphics[scale=0.62]{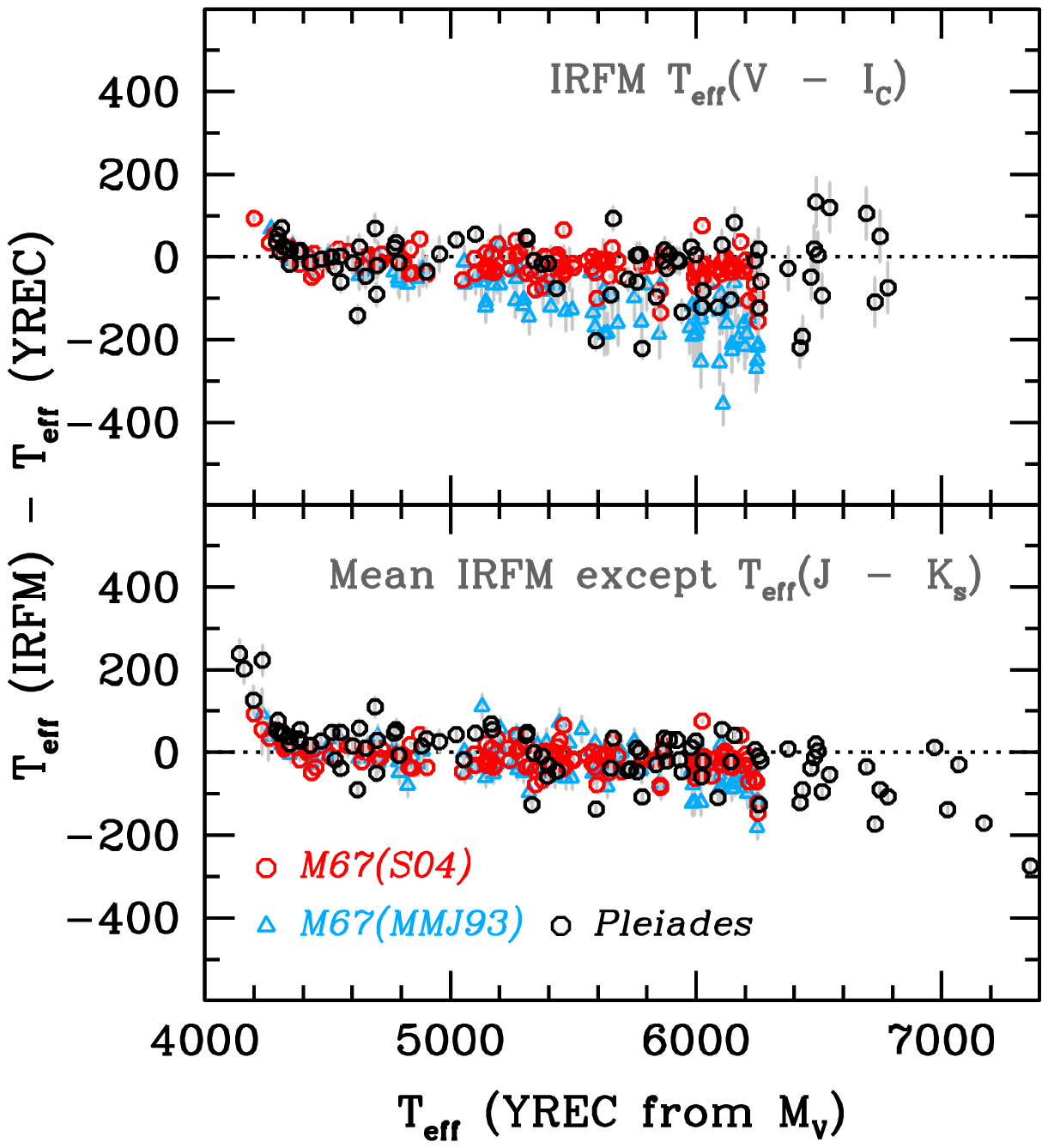}
\caption{
{\it Left}: The $T_{\rm eff}$ comparisons between IRFM and YREC for stars
in the Hyades (red circles) and Praesepe (blue triangles). {\it Right}: Same
as in the left panels, but for the Pleiades stars (black circles), M67 stars
from \citet[][red circles]{sandquist:04} photometry and from that of
\citet[][blue triangles]{montgomery:93}. {\it Top}: IRFM $T_{\rm eff}$ from
$V\, -\, I_C$. {\it Bottom}: Mean IRFM $T_{\rm eff}$ from $B\, -\, V$,
$V\, -\, I_C$, and $V\, -\, K_s$. The YREC $T_{\rm eff}$ was estimated from
the luminosity ($M_V$) of each star. Black line in the bottom left panel shows
a moving averaged trend of the $T_{\rm eff}$ difference.}
\label{fig:teffcluster}
\end{figure*}

Figure~\ref{fig:teffcluster} shows comparisons between the IRFM and
YREC $T_{\rm eff}$ estimates. The left panel shows the comparisons for the
Hyades and Praesepe stars, while the right panel shows those for the
Pleiades and M67 stars. The IRFM $T_{\rm eff}$ on top panels was computed
based on the $(V\, -\, I_C)-T_{\rm eff}$ relation in C10, just as those
used for a principal $T_{\rm eff}$ estimator in the above comparisons
(Figures~\ref{fig:clusterhyades}--\ref{fig:clusterm67}).
The YREC $T_{\rm eff}$ was estimated using \citet{an:07b} isochrones, which
have the same underlying set of interior models as those used in the current
analysis. The model $T_{\rm eff}$ was computed at a constant $M_V$ of
individual stars, assuming $(m\, -\, M)_0 = 3.33\pm0.01$, $6.33\pm0.04$,
$5.63\pm0.02$, and $9.61\pm0.03$~mag for the distance moduli of the
Hyades ($550$~Myr), Praesepe ($550$~Myr), the Pleiades ($100$~Myr),
and M67 ($3.5$~Gyr), respectively \citep[see references in][]{an:07b}.

Table~\ref{tab:cluster} lists weighted mean differences between YREC and
IRFM $T_{\rm eff}$. The mean difference between the $(V\, -\, I_C)$-based
IRFM and the luminosity-based YREC $T_{\rm eff}$ for cool stars
($T_{\rm eff} < 6000$~K) is less than $20$~K, but the differences rise
above $6000$~K to the $50$~K level. The difference between the
$(J\, -\, K_s)$-based IRFM and $M_V$-based YREC $T_{\rm eff}$ shows
different offsets for the cool and hot stars; this trend is consistent
with the above comparison between $(J\, -\, K_s)$-based IRFM and other
IRFM determination.

The bottom panel in Figure~\ref{fig:teffcluster} shows comparisons
between the YREC $T_{\rm eff}$ and the average IRFM $T_{\rm eff}$
from $B\, -\, V$, $V\, -\, I_C$, and $V\, -\, K_s$. Our results using
$J\, -\, K_s$ as a thermometer are consistent with our earlier finding
in Section~\ref{sec:dwarf} that C10 $(J\, -\, K_s)$-based $T_{\rm eff}$
values are systematically cooler than those from the $griz$-based YREC
models for hot stars (above about $6000$~K).  
The $(J\, -\, K_s)$-based
$T_{\rm eff}$ differ both from other IRFM diagnostics
and the values inferred from SDSS colors for cooler stars, 
while the mean values inferred from the IRFM are
close to SDSS for the cooler stars.

We therefore conclude that the cool star temperature scales are
consistent, while there is evidence for a systematic departure
at the hot end. A similar pattern emerges when we compare with spectroscopy,
as discussed in the next section.  Caution is
therefore required in assigning errors for stars with formal temperature
estimates above $6000$~K.

\begin{figure*}
\epsscale{0.75}
\plotone{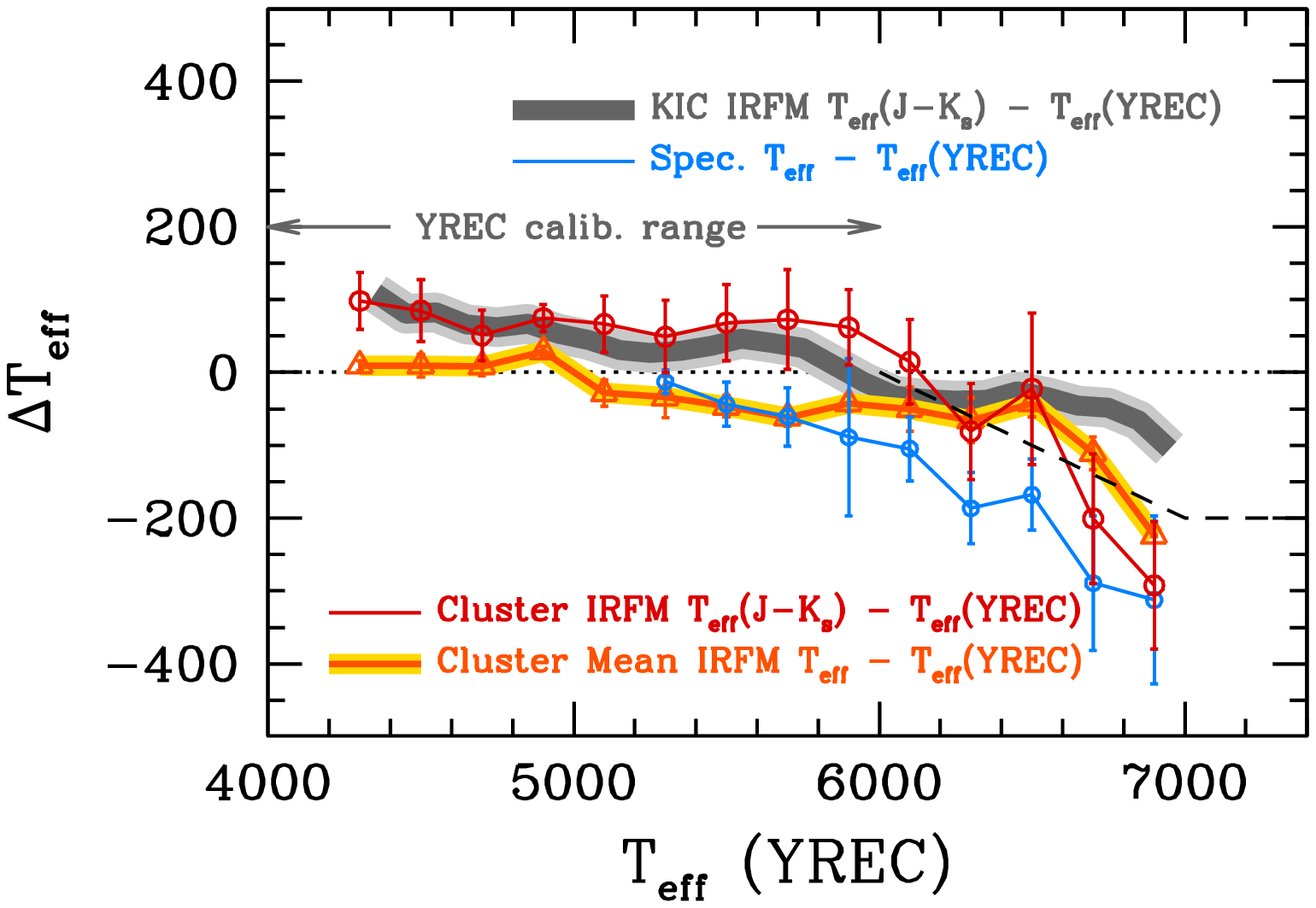}
\caption{
Systematic differences of various $T_{\rm eff}$ estimates with
respect to the YREC scale. Grey line shows the mean trend for the main KIC
sample discussed in this work. The red line represents the difference with
the $(J\, -\, K_s)$-based IRFM $T_{\rm eff}$ for the open cluster sample
(Hyades and Praesepe), while the orange line shows that with respect to
the mean IRFM values from $B\, -\, V$, $V\, -\, I_C$, $V\, -\, K_s$.
The blue line shows the trend for the B11 spectroscopic sample.
Error bars in all cases represent $\pm1\sigma$ error in the mean difference.
Our adopted hot-$T_{\rm eff}$ corrections are shown with a black dashed line.
Note that the empirical color-$T_{\rm eff}$ corrections in YREC are defined at
$4000 \leq T_{\rm eff} \leq 6000$~K in SDSS colors.
\label{fig:stat}}
\end{figure*}

Systematic $T_{\rm eff}$ differences are shown in Figure~\ref{fig:stat}.
The red line represents the difference with
the $(J\, -\, K_s)$-based IRFM $T_{\rm eff}$ for the open cluster sample
(Hyades and Praesepe), while the orange line shows that with respect to
the mean IRFM values from $B\, -\, V$, $V\, -\, I_C$, $V\, -\, K_s$.
Error bars indicate $\pm1\sigma$ error in the mean difference.
The difference between the average IRFM scale and the SDSS scale in the clusters is
less than $25$~K on average from $4000$--$6000$~K, which we take as a conservative
systematic temperature uncertainty in that domain.  The differences are
moderately larger for the IRFM $J\, -\, K_s$ temperature alone, but that
diagnostic is also different from other IRFM thermometers for cool stars.

The differences in the hot cluster stars reflect actual differences in
the calibrations, not issues peculiar to the photometry, extinction, or blending. 
We therefore attribute the comparable differences seen in the KIC stars
(gray band) as caused by calibration issues in $J\, -\, K_s$ rather than
as a reflection of systematics between the IRFM and SDSS systems.
Furthermore, the SDSS calibration was based on M67 data, where the hotter 
turnoff stars ($T_{\rm eff} > 6000$~K) were saturated.
As a result, we believe that an adjustment 
closer to the IRFM scale is better justified.

A simple correction term, of the form below
\begin{eqnarray}
T_{\rm eff,SDSS} < 6000~K             &:& T_{\rm eff,corr}
 = T_{\rm eff}{\rm (SDSS)},\\
6000~K \leq T_{\rm eff,SDSS} < 7000~K &:& \\ T_{\rm eff,corr}
 = 0.8\ (T_{\rm eff}{\rm (SDSS)} &-& 6000~K) + 6000~K,\\
T_{\rm eff,SDSS} \geq 7000~K          : T_{\rm eff,corr}
 &=& T_{\rm eff}{\rm (SDSS)} - 200~K
\label{eq:phot}
\end{eqnarray}
brings the two scales into close agreement across their mutual range of validity.  
This empirical correction is indicated by the black dashed line in
Figure~\ref{fig:stat}.
Below we find offsets similar in magnitude and opposite in sign
between the IRFM and spectroscopic temperatures for hotter stars.
Although this does not necessarily indicate problems with the fundamental scales,
it does imply that systematic temperature scale differences are important for
these stars.

\subsection{Comparison with Spectroscopy}\label{sec:spec}

Spectroscopy provides a powerful external check on the precision of photometric
temperature estimates.  Spectroscopic temperatures are independent of
extinction, and can be less sensitive to unresolved binary companions
and crowding. In this
section we therefore compare the photometric and spectroscopic temperature
estimates for two well-studied samples in the {\it Kepler} fields.
\citet[][hereafter B11]{bruntt:11} reported results for $93$ stars with
asteroseismic data, including 83 stars in our sample.
\citet[][hereafter MZ11]{molenda:11} reported results for $78$ stars,
including $45$ targets in common with our sample.  
The MZ11 data for cool stars are
mostly subgiants and giants, while the bulk of the dwarf sample is
hotter than $6000$~K.  The B11 sample is similarly
distributed, with the transition from the cool evolved to the hot
unevolved sample occurring at $5500$~K.

All comparisons below are for the corrected photometric scale, adjusted
for concordance with the IRFM at the hot end. We compare spectroscopic
methods both with the fixed-metallicity ([Fe/H]$=-0.2$) temperatures in the catalog and
the refined temperature estimates made possible with the addition
of metallicity information and theoretical metallicity corrections.
We excluded outliers in the following statistical comparisons using
a $3\sigma$ outlier rejection.

As demonstrated below, we find that the two 
spectroscopic samples have different zero-points with 
respect to both the SDSS and KIC samples, indicating the 
importance of systematic errors in such comparisons.  
The photometric scale for the cool dwarfs and giants are in good agreement
with the B11 scale, while both are offset relative
to MZ11.  The situation is different for hot dwarfs.  The 
IRFM scale was cooler than the uncorrected SDSS scale.  The
spectroscopic samples are cooler than both.  We interpret
this as evidence of additional systematic uncertainties for
the F stars, and discuss possible causes.

The stellar parameters for the MZ11 sample were derived
using the \citet{molenda:07} template approach.  The spectra were
compared with a library of reference stars.  The surface gravity,
effective temperature, and metallicity were derived from a weighted
average of the five closest spectral matches in the catalog. B11
used asteroseismic surface gravities and
derived effective temperatures from traditional Boltzmann-Saha
consistency arguments.

\begin{figure*}
\epsscale{0.95}
\plotone{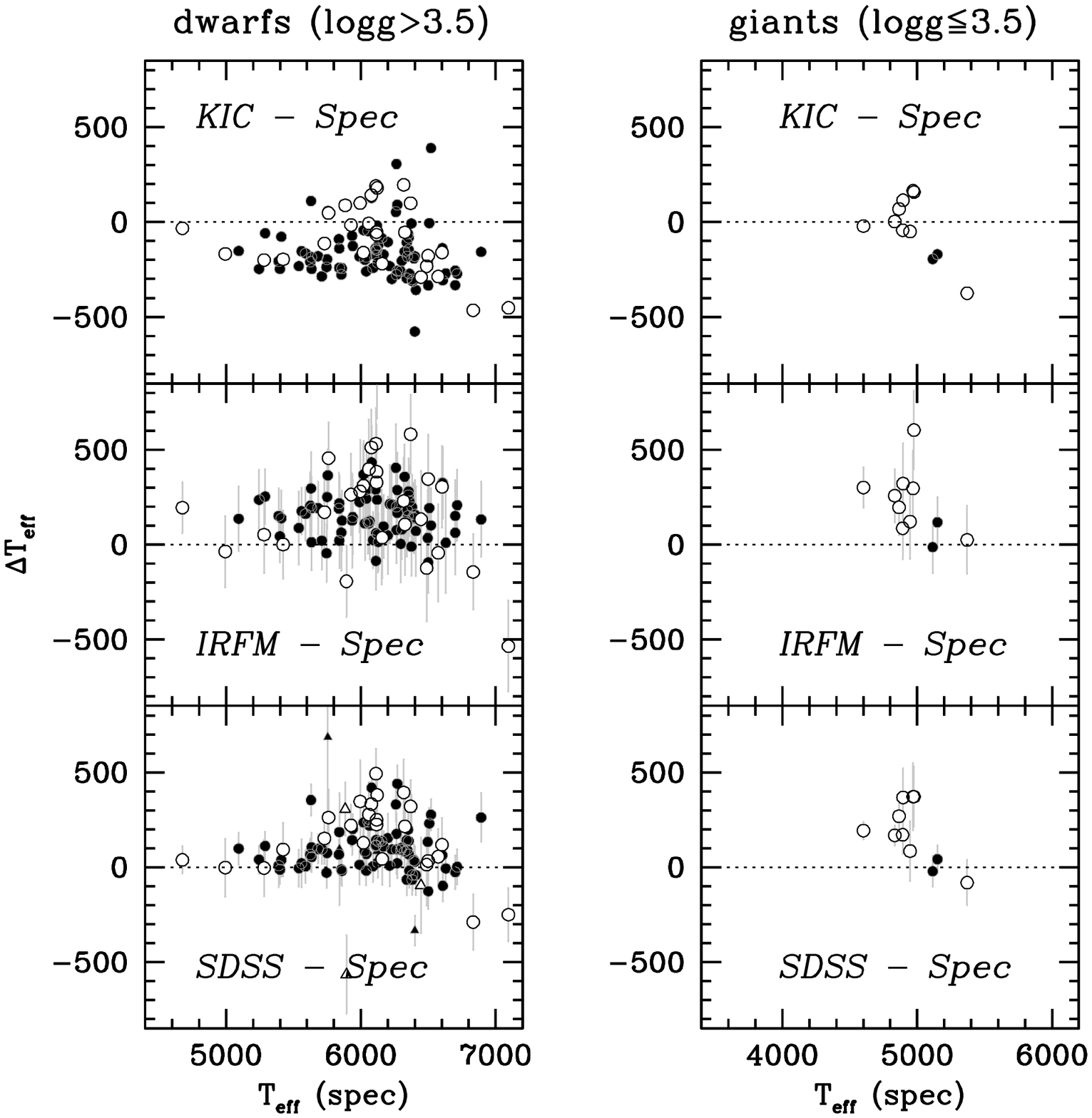}
\caption{Comparisons of spectroscopic $T_{\rm eff}$ with KIC (top),
IRFM from $J\, -\, K_s$ (middle), and SDSS estimates from $griz$ (bottom).
Filled and open points are from \citet{bruntt:11} and \citet{molenda:11},
respectively. Left panels show dwarf comparisons (KIC $\log{g} > 3.5$),
while the right panels show giant comparisons (KIC $\log{g} \leq 3.5$).
Triangles in the bottom two panels represent stars flagged as having
internally inconsistent effective temperature estimates (see text).
\label{fig:spec}}
\end{figure*}

We compare the spectroscopic and photometric temperature estimates 
in Figure~\ref{fig:spec}. The top, middle, and bottom panels compare
spectroscopic temperatures to those of the  KIC, IRFM ($J\, -\, K_s$),
and SDSS, respectively. Left panels show comparisons for dwarfs (KIC
$\log{g} > 3.5$), while the right panels show those for giants (KIC
$\log{g} \leq 3.5$). Filled circles are the B11 data,
while open circles are the MZ11 data. In total,
$83$ out of $93$ sample stars in B11 were used in this
comparison; the remaining $10$ stars do not have $griz$ photometry
in all passbands, so were not included in our KIC subsample.
For the same reason, we initially included 45 spectroscopic targets from MZ11,
but later excluded 8 more stars with $g\, -\, r < 0.1$ or $g\, -\, r > 1.0$.
Triangles in the bottom two panels represent stars flagged as having internally
inconsistent effective temperature estimates (Section~\ref{sec:table4}).
Error bars show the
expected random errors, with a $70$~K error adopted in the temperature for the
individual B11 sample stars.

In the above comparisons, we corrected the IRFM temperature estimates
for the spectroscopic metallicity measurement of each sample, although
the $T_{\rm eff}$ corrections in C10 were negligible ($\Delta T_{\rm eff}\approx18$~K)
in $J\, -\, K_s$. We also used individual stellar isochrones at each
spectroscopic metallicity to estimate SDSS $T_{\rm eff}$ from $griz$,
assuming a constant age of $1$~Gyr at all metallicity bins.
However, the net effect of these corrections was small ($\Delta T_{\rm eff}\approx25$~K),
because $griz$-$T_{\rm eff}$ relations are relatively insensitive to metallicity
and the mean metallicities of the spectroscopic samples are close to our
fiducial value ($\langle {\rm [Fe/H]} \rangle=-0.07\pm0.02$ and $-0.11\pm0.03$
for the B11 and MZ11 samples, respectively). The SDSS $T_{\rm eff}$ values
for giants were corrected for the $\log{g}$ difference from the dwarf
temperature scale as described in Section~\ref{sec:giant}.

Both spectroscopic samples for dwarfs are systematically hotter than the KIC
(top left panel in Figure~\ref{fig:spec}). The weighted average
difference between the B11
sample and the KIC, in the sense of the KIC minus spectroscopic values, is
$-170$~K with a dispersion of $116$~K, after a $3\sigma$ outlier rejection.
The MZ11 sample is closer to the KIC, with a $-82$~K mean
difference and a dispersion of $172$~K. This difference of $88$~K is
a reflection of the systematic errors in the spectroscopic temperature scales.
In the above comparisons, we did not include stars with inconsistent SDSS
temperature measurements (triangles in Figure~\ref{fig:spec}).

The weighted average difference between the B11 sample and the SDSS (in the sense SDSS
$-$ Spec) for dwarfs is $85$~K with a $95$~K dispersion,
after excluding those flagged as having discrepant $T_{\rm eff}$(YREC) values.
If the metallicity corrections to the
SDSS values were not taken into account (i.e., based on models at [Fe/H]$=-0.2$),
 the mean difference becomes $73$~K, but the dispersion increases
to $111$~K.

However, there is
a strong temperature dependence in the offset. Below $6000$~K the mean difference
is $50$~K with a dispersion of $47$~K. For the hotter stars the mean difference
is $101$~K with a dispersion of $118$~K. 
The blue line in Figure~\ref{fig:stat} shows a moving averaged
difference between the B11 spectroscopic values and SDSS $T_{\rm eff}$
without the hot-end $T_{\rm eff}$ corrections (equations~6--8).

Although the size of the dwarf sample in MZ11 is small, it is
found that the effective temperatures are systematically cooler than the SDSS
values, with a weighted mean offset of $152$~K (SDSS $-$ Spec) and a dispersion of $175$~K.
The difference is temperature dependent, being $53$~K for the stars below $6000$~K
and $178$~K above it. These differences are $3$~K and $77$~K larger,
respectively, than the results from the B11 sample. The temperature
differences between photometry and spectroscopy are therefore smaller than the
differences between the spectroscopic measurements and the KIC, while there
is a real difference at the hot end even when systematic differences
between the two spectroscopic samples are accounted for.

The B11 sample includes only two giants (KIC $\log{g}\leq3.5$),
but their spectroscopic temperatures are consistent with both IRFM and SDSS
temperatures (see middle and bottom right panels in Figure~\ref{fig:spec}).
On the other hand, the MZ11 sample shows a large
offset from IRFM ($\Delta T_{\rm eff}=245$~K) and SDSS
($\Delta T_{\rm eff}=206$~K), while
the KIC and the MZ11 values agree with each other ($\Delta T_{\rm eff}=9$~K).

The cool MZ11 stars are mostly subgiants and giants, while
the B11 cool sample includes a large dwarf population between
$5000$~K and $6000$~K. The difference between the two
cool end results - good agreement with B11 for cool dwarfs,
but not with MZ11 - is real.  This could reflect systematic
differences between the dwarf and corrected giant results for the SDSS or the
templates adopted by MZ11 for the evolved and unevolved
stars.  The scatter between the MZ11 results and the
photometric ones is substantially larger than that between B11
and photometric temperature estimates. It would be worth
investigating the zero-point of the templates used in the former
method, as well as the random errors, in light of the results reported here.

In the section above we have focused on differences between the scales;
it is fair to ask how both might compare to the true temperatures.
The photometric scale is at heart simply an empirical relationship
between color and the definition of the effective temperature itself
($L = 4 \pi R^2 \sigma T_{\rm eff}^4$), and therefore the scale itself
should be sound where the photometric relations are well-defined.
However, the photometric methods can fail if there is more than one
contributor to the photometry, or if the reddening is incorrectly
measured. Spectroscopic temperatures measure physical conditions in
the atmosphere, and are only indirectly tied to the fundamental flux
per unit area, which defines the effective temperature. There are also
systematic uncertainties between different methods for inferring
effective temperatures, for example, fitting the wings of strong lines,
or the use of Boltzmann-Saha solutions based on ionization and excitation
balance. Finally, both photometric and spectroscopic estimates are only
as good as their assumptions; stars with large surface temperature
differences will be poorly modeled by both methods.

Our primary conclusion is therefore that the various dwarf temperature
methods, spectroscopic and photometric, are in good agreement for the
cooler stars.  Systematic effects are at or below the 50 K level.
The hotter stars in the sample have real systematic differences between spectroscopic
and photometric temperatures, and similar discrepancies are also present between the
photometric methods themselves. This is further evidence that work is needed
to tie down more precisely the temperature scale above $6000$~K, and that
larger systematic errors should be assigned in this domain until such an
analysis is performed.  We have less data for the giants, but there does appear to 
be a real difference between the photometric  results and the temperatures inferred
for the MZ11 sample.

\subsection{Effects of Binaries on Colors}\label{sec:binary}

Unresolved binaries in the sample could bias a color-based $T_{\rm eff}$ estimate.
Unless the mass ratio of the primary and secondary components in the binary system
is close to either unity (twins) or zero (negligible contributions from the secondary),
composite colors of the system are redder than those from the primaries alone, leading towards
systematically lower $T_{\rm eff}$.  It is difficult 
to directly flag potential binaries given the filters
available to us, and as a result we do not include
star by star corrections in the table.  However, such
a systematic bias will be important when evaluating the
bulk properties of the KIC sample.  
In this section, we therefore estimate the size of the bias
due to unresolved binaries in the KIC, and provide statistical 
corrections for the effect of unresolved binary companions on 
average effective temperature estimates.

Binary contamination effects on the color-$T_{\rm eff}$ relations were derived
by performing artificial star tests. We used a $1$~Gyr old Padova models at
solar abundance \citep{girardi:04}. These models include stellar masses down to
$0.15\ M_{\odot}$, allowing us to include low-mass systems outside the formal
range of the SDSS color calibration. The absolute color-$T_{\rm eff}$
relations in these models are not exactly the same as in our base calibration,
and the adoption of a solar metallicity isochrone is not strictly
self-consistent with our application of the base model at [Fe/H]$=-0.2$.
However, our main purpose is to evaluate the {\it relative} temperature errors
induced by companions, and the effects of these offsets are presumably small.

We assumed a $50\%$ binary fraction with $10,000$ single stars and $10,000$ binary
systems. Primary masses were randomly drawn from a Salpeter mass function, while
we explored three different choices for the relative masses of the secondaries:
Salpeter, flat, and one drawn from the open cluster M35 \citep{barrado:01}.
A flat mass function is expected for short-period binaries, which will be
a minority of the sample; this is thus a limiting case.
In the artificial star simulations, we derived empirical color-color sequences
in $g\, -\, i$, $g\, -\, z$, and $J\, -\, K_s$ with $g\, -\, r$ as the principal color index.
We simulated photometric errors by injecting dispersions of $0.01$~mag in $gri$,
$0.03$~mag in $z$, $0.024$~mag in $J$, and $0.028$~mag in $K_s$. These 2MASS errors
are median values of the actual photometric errors in the KIC sample.

\begin{figure}
\epsscale{1.05}
\plotone{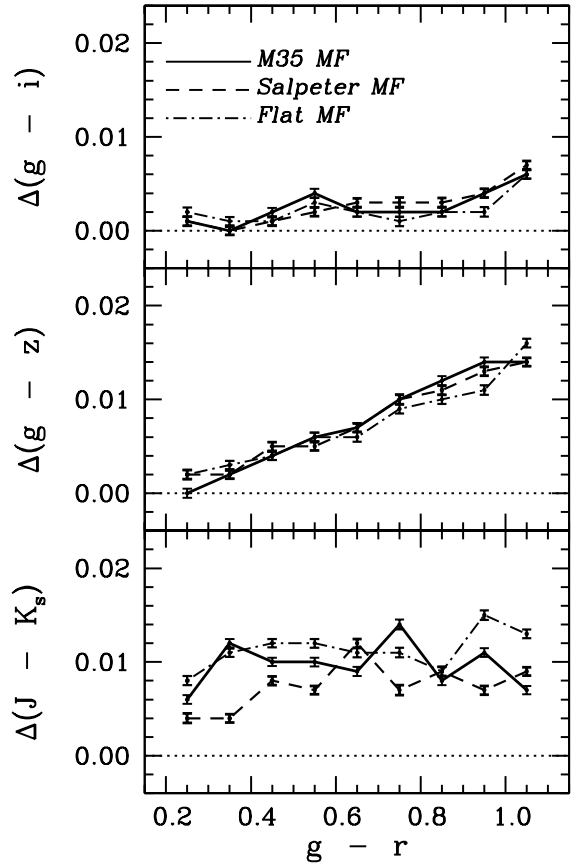}
\caption{Average color bias in $g\, -\, i$, $g\, -\, z$, and $J\, -\, K_s$ at fixed
$g\, -\, r$ due to unresolved binaries for three different assumptions about
the secondary mass function. Points and error bars are the centroid and
the error in the mean distribution from the simulations. A $50\%$ binary
fraction is assumed.
\label{fig:binarycolor}}
\end{figure}

The result of these binary simulations is presented in Figure~\ref{fig:binarycolor},
which shows the mean deviations in $g\, -\, i$, $g\, -\, z$, and $J\, -\, K_s$ from those with
primaries alone. For Figure~\ref{fig:binarycolor} we fitted a Gaussian for each
$g\, -\, r$ bin to estimate the mean color offset and the uncertainty as shown by
circles and error bars. The three curves indicate
results from three different relative mass functions for secondaries.

\input{tab6.tex}

The sizes of these color shifts are shown in Table~\ref{tab:binary}. The systematic
color shift due to unresolved binaries is less strongly dependent on the choice
of secondary mass functions. Typical sizes of these color shifts are $\sim0.003$~mag,
$0.008$~mag, and $0.010$~mag in $g\, -\, i$, and $g\, -\, z$, and $J\, -\, K_s$, respectively.
To correct for the unresolved binaries in the KIC, the above color shifts should
be subtracted before estimating $T_{\rm eff}$. The last four columns in Table~\ref{tab:binary}
list the average $T_{\rm eff}$ difference between 
a population with a $50\%$ unresolved binary fraction and that of primaries alone.
The sense is that unresolved binary stars have lower temperatures than
expected from primaries alone. Different SDSS color indices have similar 
binary sensitivities, and temperatures based on these filters are less
affected by unrecognized companions than those derived using $J\, -\, K_s$.
These color shifts are small for any given star, but significant when applied
to the entire catalog.  We therefore recommend including them when using large
samples of photometric effective temperature estimates, and include this effect
in our global error budget below.

\subsection{Other Sources of Uncertainties and Error Budget}\label{sec:error}

We can assess our overall errors by comparing the real to the observed
dispersions in the color-color plane.  Photometric errors, unresolved binaries,
and metallicity all induce scatter; so would extinction uncertainties.
Significant mismatches between the two reflect unrecognized or overestimated error sources.

\begin{figure}
\centering 
\includegraphics[scale=0.65]{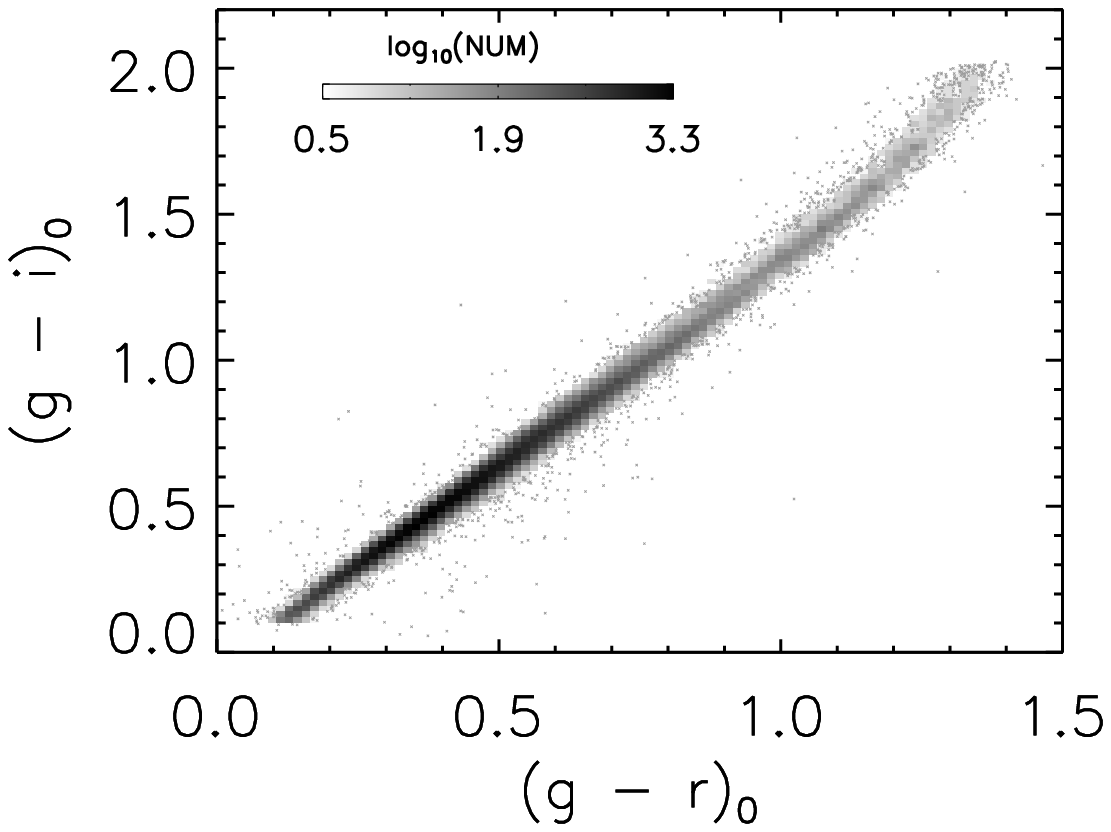}
\includegraphics[scale=0.65]{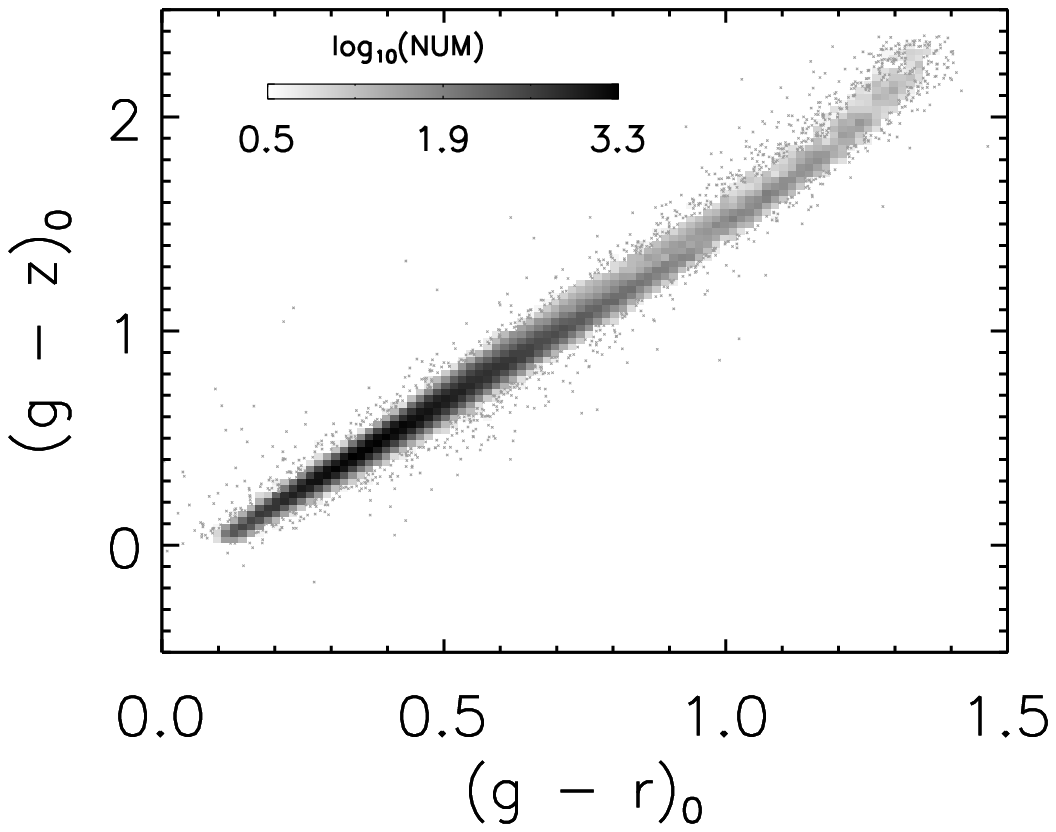}
\includegraphics[scale=0.65]{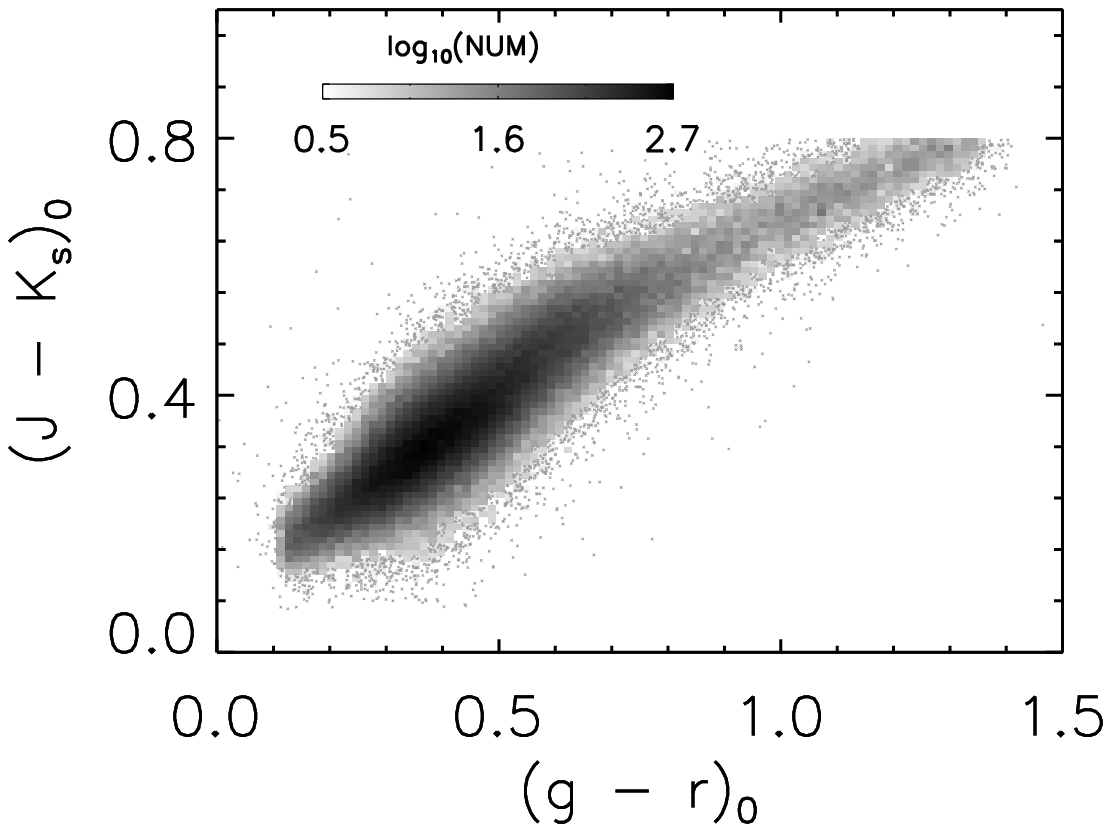}
\caption{Extinction corrected color-color relations in the KIC, after the zero-point
corrections as described in Section~\ref{sec:phot}. Only those with $\log{g} > 3.5$
are shown.
\label{fig:binsim}}
\end{figure}

\begin{figure}
\epsscale{1.05}
\plotone{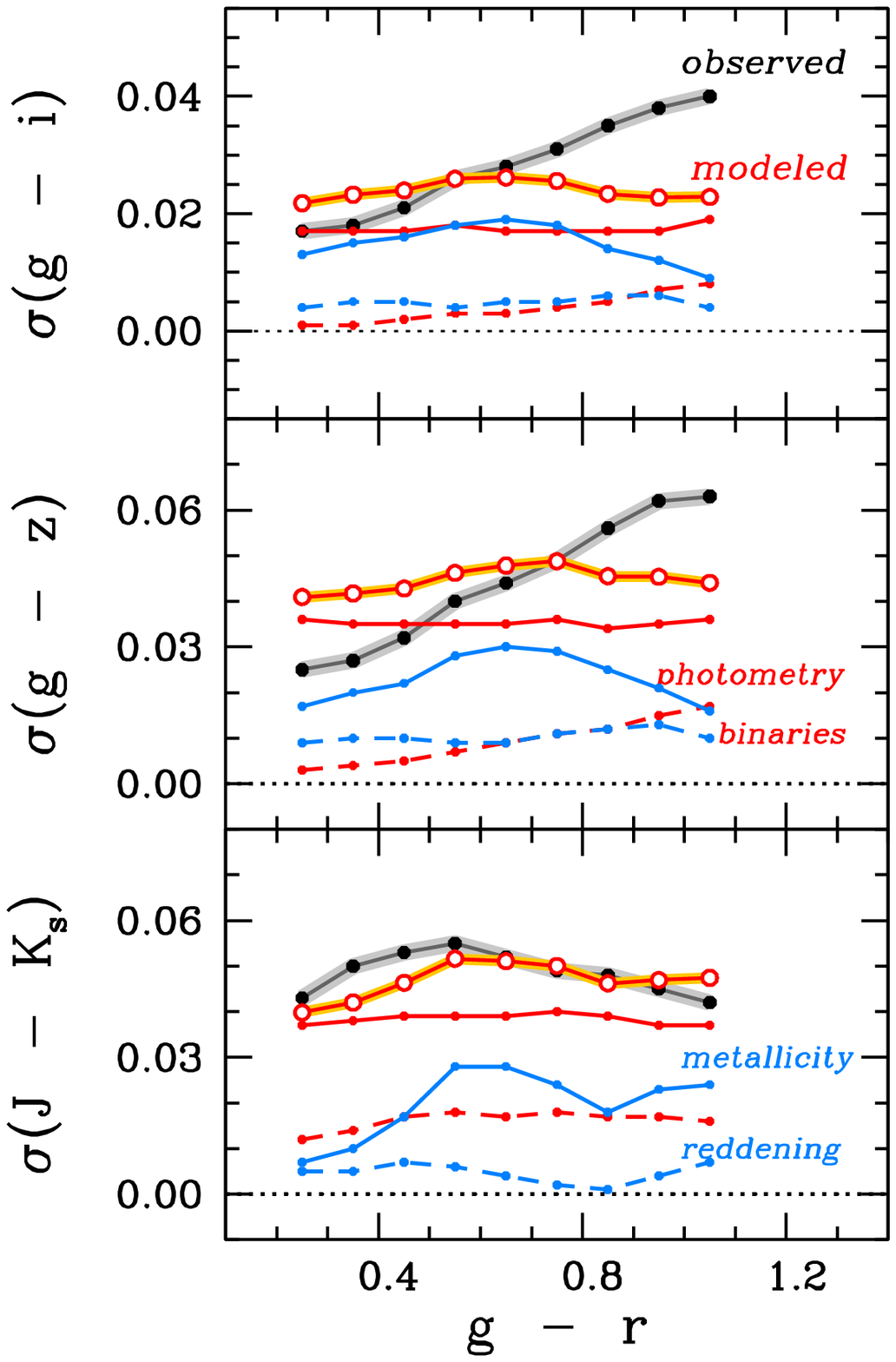}
\caption{Comparison between observed (thick black line with closed circles)
and modeled (thick red line with open circles) dispersions of the color-color
sequence as a function of $g\, -\, r$. The modeled dispersion is
a quadrature sum of individual error contributions: photometric errors
(red solid), unresolved binaries (red dashed), metallicity (blue solid),
and reddening (blue dashed).
\label{fig:binaryerror}}
\end{figure}

Figure~\ref{fig:binsim} shows the observed color-color diagrams in the KIC,
after the extinction corrections and the zero-point adjustment as described
in Section~\ref{sec:phot}. From Figure~\ref{fig:binsim},
we estimated the standard deviation of the color dispersion from a fiducial
line (fit using a $5^{th}$ order polynomial) in $g\, -\, i$, $g\, -\, z$,
and $J\, -\, K_s$ at each $g\, -\, r$ bin.
These observed dispersions with good $T_{\rm eff}$ estimates are shown
as solid black curves with closed circles in Figure~\ref{fig:binaryerror}.
Here the criteria for the good $T_{\rm eff}$ are
that the standard deviation of individual $T_{\rm eff}$ from three color
indices ($g\, -\, r$, $g\, -\, i$, and $g\, -\, z$) is less than $130$~K
or that the difference between SDSS and IRFM measurements is no larger
than three times the random errors of these measurements
(see also Section~\ref{sec:table4}).
There is a strong overlap between the two criteria.  Since the formal
random SDSS errors are of order $40$~K, and the systematics between the
colors are typically at that level as well, differences of $130$~K
represent clear evidence of a breakdown in the color-temperature
relationships, likely from unresolved blends.  Excluding extreme outliers
is essential because they would otherwise dominate the dispersion measure,
and we are interested in testing the properties of the majority of the sample.

Other lines in Figure~\ref{fig:binaryerror} represent contributions from
random photometric errors, unresolved binaries or photometric blends,
metallicity, and dust extinction as described below. Red lines with open
circles are the quadrature sum of all of these error sources.

We assumed $0.01$~mag errors in $gri$, $0.03$~mag errors in $z$, $0.024$~mag
in $J$, and $0.028$~mag in $K_s$ to estimate color dispersions from photometric
errors alone (red curve in Figure~\ref{fig:binaryerror}). To perform this
simulation in $grizJK_s$, we combined our base model (Table~\ref{tab:isochrone})
with our earlier set of isochrones in the 2MASS system \citep{an:07b}\footnote{Available
at http://www.astronomy.ohio-state.edu/iso/pl.html.} at the same metallicity
([Fe/H]$=-0.2$) and age ($1$~Gyr) as those for the base isochrone.
As with the binary simulations described in the previous section, we employed
a $1$~Gyr old, solar metallicity Padova model \citep{girardi:04} to generate
color-color sequences with a $50\%$ binary fraction based on the M35 mass
function for secondaries.
Again, running this isochrone in the simulation is not strictly consistent
with the usage of our base model, but the relative effects induced by unresolved
companions would be rather insensitive to the small metallicity difference.
The dispersion induced by unresolved binaries is shown in a red dashed curve
in Figure~\ref{fig:binaryerror}.

The KIC sample has a mean [Fe/H]$=-0.2$ with a standard deviation
of $0.28$~dex. If the KIC [Fe/H] values are accurate enough for these stars,
this metallicity spread would induce a significant spread in $T_{\rm eff}$.
The color dispersion due to metal abundances was estimated by taking the
color difference between our base model ([Fe/H]$=-0.2$) and the models at
[Fe/H]$=+0.1$ and $-0.5$ as an effective $\pm1\sigma$ uncertainty. The metallicity error
contribution is shown in blue solid curves.
The KIC sample has a wide range of reddening values ($0 \la E(B\, -\, V) \la 0.2$).
We took $0.02$~mag error as an approximate $\pm1\sigma$ error in
E$(B\, -\, V)$, roughly equivalent to a 15 \% fractional 
uncertainty for a typical star.
Stars on the simulated color-color sequence were randomly displaced from their
original positions assuming this E$(B\, -\, V)$ dispersion. The resulting color
dispersion is shown with the blue dashed curves in Figure~\ref{fig:binaryerror}.

In Figure~\ref{fig:binaryerror} there is a color-dependent trend in the error budget,
where observed color dispersion increases for cooler 
stars in $g\, -\, i$ and $g\, -\, z$. On the other hand, the simulated
dispersions (open circles connected with solid red curves) are essentially flat.  Our results are consistent with
expectations in $J\, -\, K$; if anything, the random errors appear to
be overstated.  This is probably caused by correlated errors in $J$ and $K_s$,
which were treated as uncorrelated in the temperature error
estimates.

Based on this exercise, we conclude that our error model is reasonable
for the hot stars in the sample, especially when the stars most
impacted by blends are removed.  There is excess color scatter for red
stars, which correspond to effective temperatures below $\sim5000$~K
in our sample.
About $16\%$ of the sample are found in this temperature domain.
This could reflect contamination of the dwarf sample by
giants, which have different color-color relationships; or a breakdown
in the photometric error model for red stars.  It would be useful to
revisit this question when we have a solid estimate of the giant
contamination fraction for the cool dwarfs in the sample.

\section {The Revised $T_{\rm eff}$ Catalog}\label{sec:catalog}

\subsection{A Recipe for Estimating $T_{\rm eff}$}\label{sec:recipe}

We present results for the long-cadence sample with the overall
properties of the catalog and systematic error estimates in this section.
We have not provided corrected values for the entire KIC, because the
additional quality control is outside the scope of our effort.  However,
our method could be applied in general to the KIC, employing the following steps.

\input{tab7.tex}

\begin{itemize}

\item Correct the KIC $griz$ photometry onto the SDSS DR8 system
using equations 1--4.

\item Apply the KIC extinctions and the extinction coefficients
in Section~\ref{sec:method} to obtain dereddened colors.

\item Use our $griz$-$T_{\rm eff}$ polynomials (Table~\ref{tab:poly})
or the original isochrone (Table~\ref{tab:isochrone}) to obtain temperature
estimates. If complementary IRFM estimates are desired,
use the C10 polynomials (for $V_TJK_s$).

\item Adjust hot-end temperatures above $6000$~K using equations 5--7.
The polynomials in Table~\ref{tab:poly} are for the original SDSS temperature
calibration (Table~\ref{tab:isochrone}) without the hot-end adjustment
described in Section~\ref{sec:cluster}.

\item In Table~\ref{tab:main} we adopted a metallicity [Fe/H]$=-0.2$ and a
dispersion of $0.3$~dex for error purposes.  We also adopted a
fractional error of $15\%$ in the extinction.

\item The SDSS temperatures are inferred from the weighted
average of the independent color estimates using the photometric errors
discussed in Section~\ref{sec:method}, and the
random uncertainties are the maximum of the formal random errors and the
dispersion in those inferred from different $griz$ colors.

\item If the metallicity is known independent of the KIC, the SDSS temperatures
can be corrected using the values in Table~\ref{tab:feh} and if desired
the IRFM temperatures can be corrected for metallicity by adopting
star-by-star metallicities in the C10 formulae.

\item Apply gravity corrections in Table~\ref{tab:logg} for giants
with $\log{g} {\rm (KIC)} \leq 3.5$.

\item Outside the temperature range of the SDSS calibration,
zero-point shifts of $223$~K at the hot end and $150$~K at the
cool end should be applied to the KIC $T_{\rm eff}$
to avoid artificial discontinuities in the
temperature scale at the edges of validity of the method.

\item In our revised $T_{\rm eff}$ table, we did not apply statistical corrections for
binaries, but the current Table~\ref{tab:binary} could be employed to do so, and this
should be included in population studies.

\item We expect about $4 \% $ of the sample to have photometry impacted by blends.
Such stars could be identified as those having an excess dispersion from individual
SDSS colors on the order of $130$~K or more, and/or as
those showing more than a $3\sigma$ deviation from the mean
difference between the IRFM and SDSS temperatures.

\end{itemize}

\subsection{Main $T_{\rm eff}$ Catalog}\label{sec:table}

Our main result, the revised $T_{\rm eff}$ for $161,977$ stars in the
long-cadence KIC, is presented in Table~\ref{tab:main}.\footnote{
Only a portion of this table is shown here to demonstrate its form and
content. A machine-readable version of the full table is available online.}
All of our revised $T_{\rm eff}$ estimates in the catalog are based
on the recalibrated $griz$ magnitudes in the KIC (Section~\ref{sec:phot}).
In addition to the $griz$-based SDSS $T_{\rm eff}$, Table~\ref{tab:main}
contains $(J\, -\, K_s)$-based IRFM $T_{\rm eff}$ using the original
C10 relation, and KIC values along with $\log{g}$ and [Fe/H] in the KIC.
The null values in the SDSS $T_{\rm eff}$ column are those outside of the
color range in the model ($4043$~K $< T_{\rm eff} < 7509$~K). Similarly,
the C10 IRFM $T_{\rm eff}$ are defined at $0.07 \leq (J\, -\, K_s)_0 \leq 0.80$.

\input{tab8.tex}

\begin{figure*}
\centering
\includegraphics[scale=0.6]{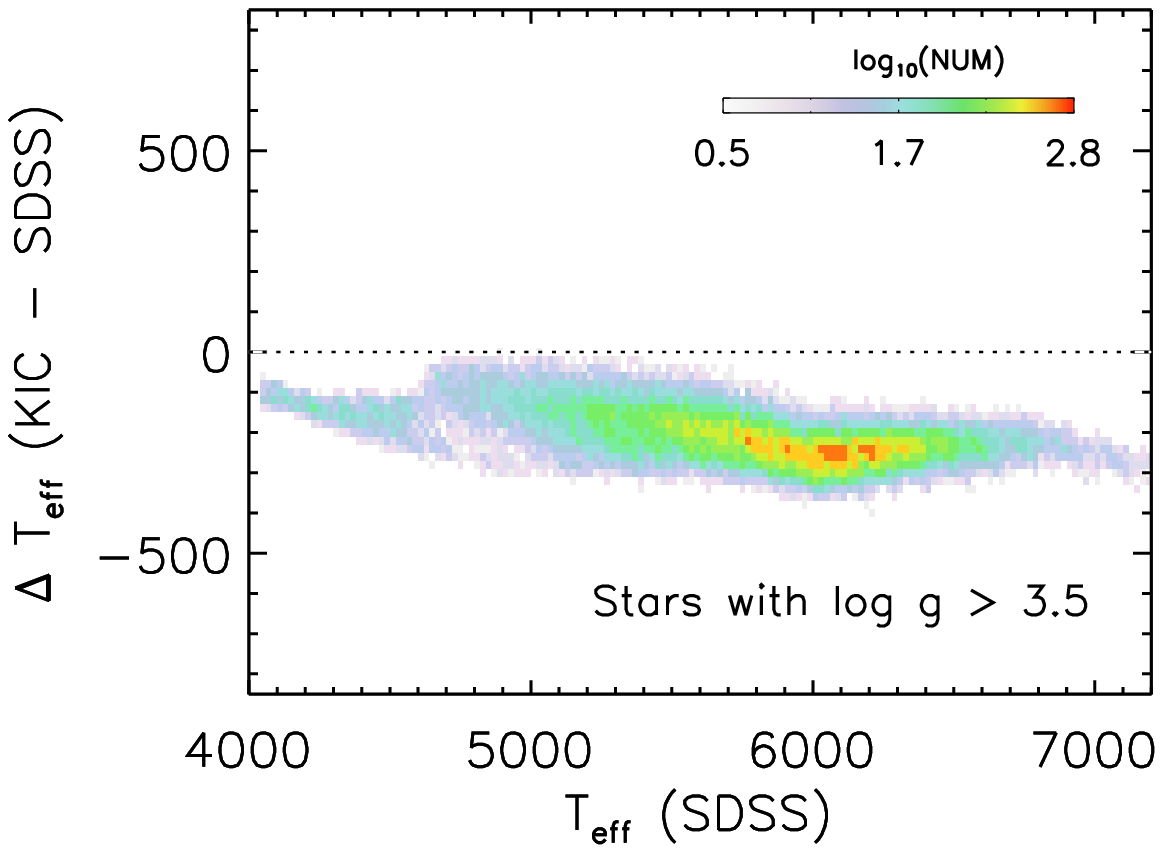}
\includegraphics[scale=0.6]{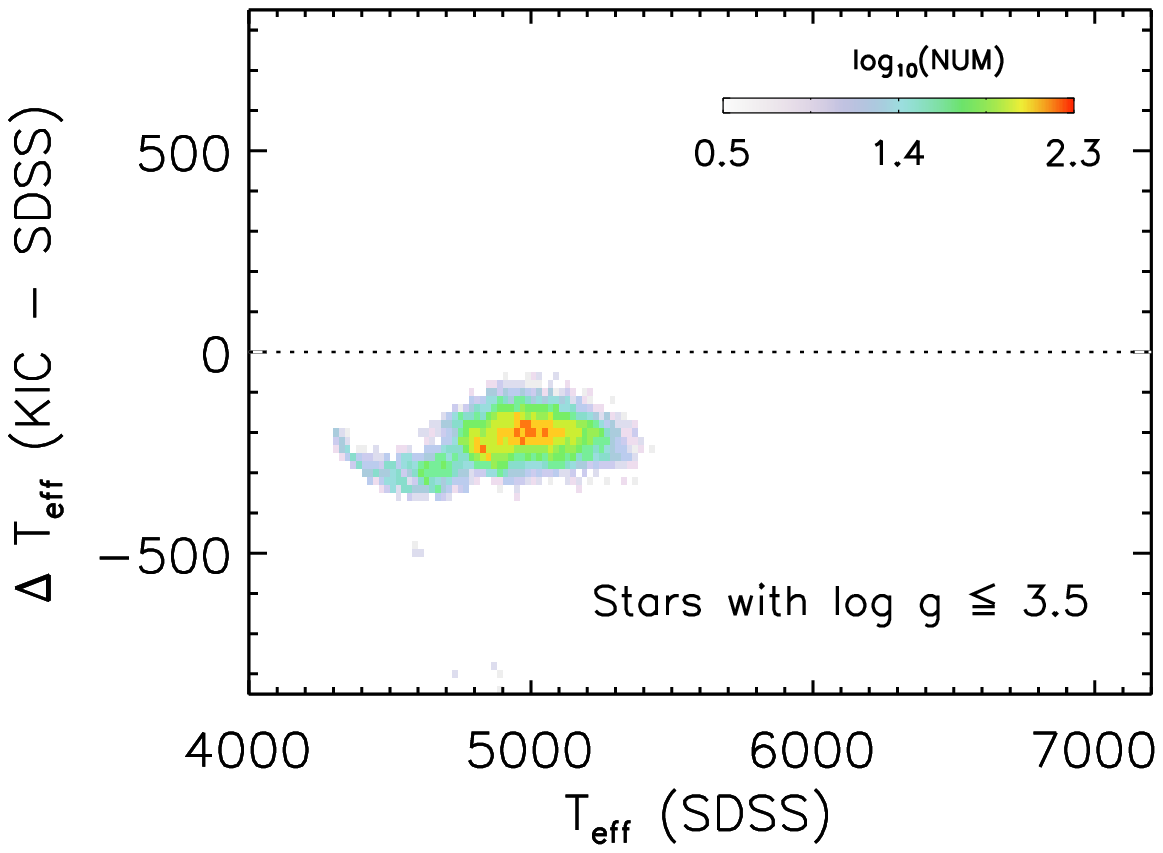}
\includegraphics[scale=0.6]{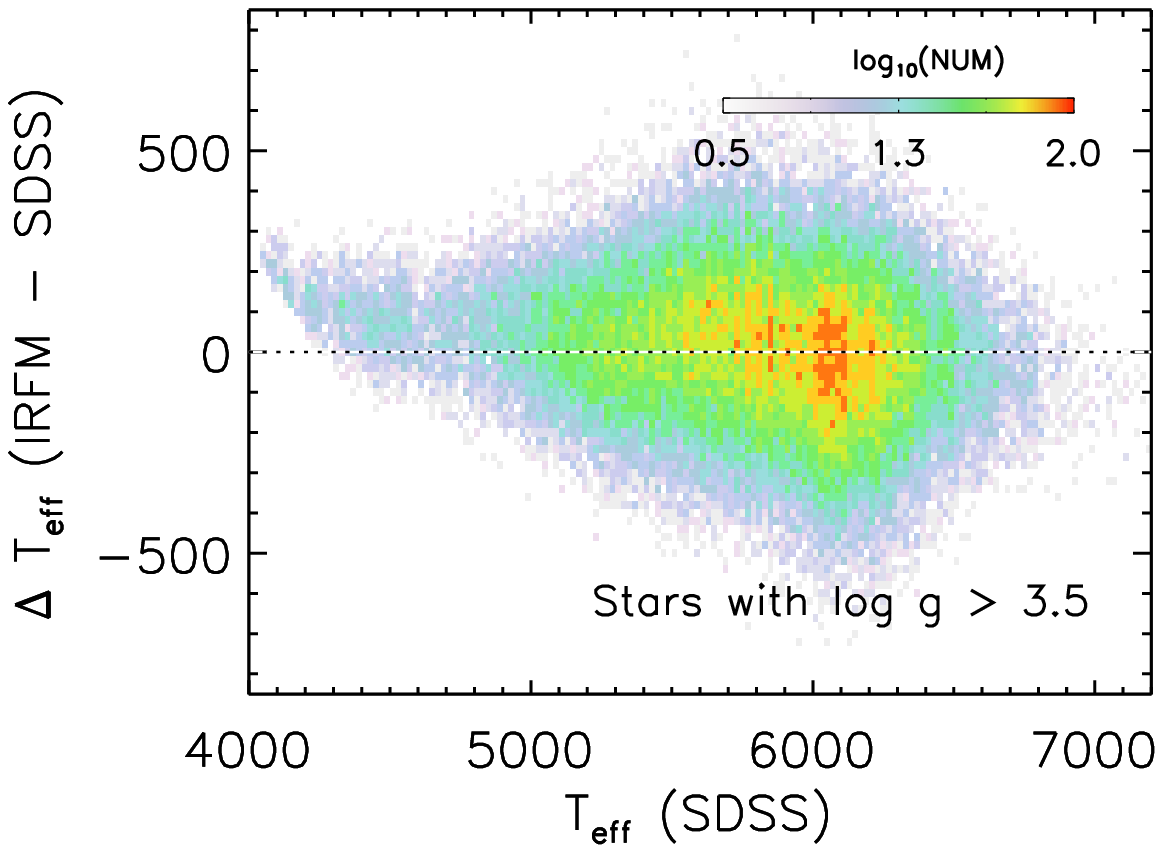}
\includegraphics[scale=0.6]{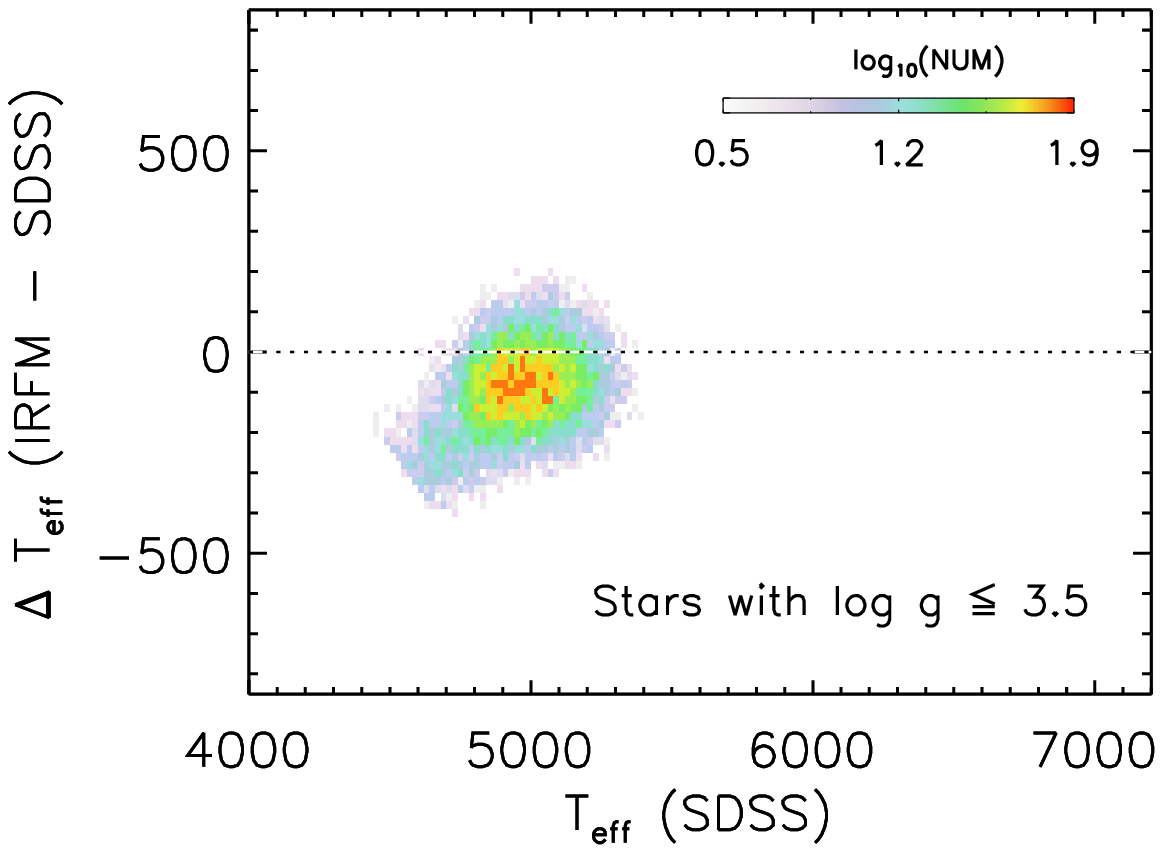}
\caption{Comparisons of $T_{\rm eff}$ using the final SDSS $T_{\rm eff}$ estimates.
Comparisons are shown for the original KIC $T_{\rm eff}$ for dwarfs (top left)
and giants (top right), and for the $(J\, -\, K_s)$-based IRFM estimates for
dwarfs (bottom left) and giants (bottom right).
\label{fig:final}}
\end{figure*}

Statistical properties of our final temperature estimates are listed
in Table~\ref{tab:stat} for dwarfs and for giants, separately.
The relative KIC, IRFM, and SDSS temperatures for dwarfs and giants
in the final catalog are compared in  Figure~\ref{fig:final}.  These
comparisons include the adjustment to the hot end published SDSS scale 
described in Section~\ref{sec:cluster}.  We did not correct the IRFM
temperature estimates for gravity effects in the giants.  The discrepancy
between the two scales for the cool giants is consistent with being
caused by this effect, as can be seen from the gravity sensitivity of
$(J\, -\, K_s)$ in Figure~\ref{fig:loggiso}.

Below we describe each column of Table~\ref{tab:main}, and provide
a summary on how to correct $T_{\rm eff}$ for different $\log{g}$,
binarity (blending), and metallicity.

\subsubsection{Error Estimates in $T_{\rm eff}$}\label{sec:table1}

For the SDSS and IRFM, we estimated total ($\sigma_{\rm tot}$) and random
($\sigma_{\rm ran}$) errors for individual stars as follows. The random
errors for the SDSS were taken from two approaches, tabulating whichever yields the 
larger value: a propagated error from the photometric precision and the one from
measurements of $T_{\rm eff}$ from individual color indices ($g\, -\, r$,
$g\, -\, i$, and $g\, -\, z$). For the former, we repeated our procedures
of solving for $T_{\rm eff}$ with $0.01$~mag photometric errors in $gri$
and $0.03$~mag errors in $z$: we added corresponding $T_{\rm eff}$ errors
from individual determinations. The random errors for the IRFM were
estimated from the 2MASS-reported photometric errors in $J$ and $K_s$
(combined in quadrature).

In Table~\ref{tab:main} we included systematic errors from $\pm15\%$ error
in the foreground dust extinction and $\pm0.3$~dex error in [Fe/H] from
our fiducial case ([Fe/H]$=-0.2$) for both SDSS and IRFM measurements.
The total error ($\sigma_{\rm tot}$) is a quadrature sum of both random
and systematic error components.  The total errors are dominated by 
the extinction uncertainties, which relate to both galactic position and 
distance.  The quoted values yield dispersions in temperature between YREC, 
IRFM, and spectroscopy consistent with the data.
We present effective temperatures defined at a fixed [Fe/H]$=-0.2$. If
it is desired to correct for metallicities different from this fiducial
[Fe/H], $T_{\rm eff}$ corrections in Table~\ref{tab:feh} can be used.

\subsubsection{Corrections for different $\log{g}$}\label{sec:table2}

Our application of the isochrone assumes that all of the stars are
main-sequence dwarfs. To correct for differences between the KIC and
the model $\log{g}$ values, we used $\log{g}$ sensitivities of the
$griz$ colors using \citet{castelli:04} ATLAS9 models, as described
in Section~\ref{sec:giant}. Table~\ref{tab:logg} lists the correction
factors in $T_{\rm eff}$ as a function of each color index over
$\Delta \log{g} = 0.5$--$3.0$ in a $0.5$ dex increment. For a given
color in each of these color indices, a difference between the KIC
and the model $\log{g}$ can be estimated ($\Delta \log{g} = 
\log{g}_{\rm KIC}\, -\, \log{g}_{\rm YREC}$), and the corresponding
$\Delta T_{\rm eff}$ values in Table~\ref{tab:logg} can be found in
$g\, -\, r$, $g\, -\, i$, and $g\, -\, z$, respectively. The mean
$\Delta T_{\rm eff}$ correction was then added to the dwarf-based
$T_{\rm eff}$ estimates. Our catalog (Table~\ref{tab:main}) lists
SDSS $T_{\rm eff}$ estimates already corrected using these $\log{g}$
corrections for those with $\log{g}{\rm (KIC)} \le 3.5$ at $T_{\rm eff}{\rm (SDSS)} < 5300$~K.
If it is desired to recover the dwarf-based solution, correction terms
($\Delta T_{\rm eff}$) in Table~\ref{tab:main} should be subtracted
from the listed $T_{\rm eff}{\rm (SDSS)}$.

\subsubsection{Corrections for Binaries}\label{sec:table3}

As described in Section~\ref{sec:binary}, unresolved binaries and blending
can have an impact on the overall distribution of photometric $T_{\rm eff}$.
If the population effect is of greater importance than individual $T_{\rm eff}$,
correction factors in Table~\ref{tab:binary} should be added to the SDSS and
IRFM $T_{\rm eff}$ (making them hotter) in Table~\ref{tab:main}. With
$1\%-3\%$ errors in $griz$ photometry, it is difficult to distinguish
between single stars with unresolved binaries and/or blended sources in
the catalog.

\subsubsection{Quality Control Flag}\label{sec:table4}

The last column in Table~\ref{tab:main} shows a quality control flag.
If the flag is set (${\tt flag}=1$), the SDSS $T_{\rm eff}$ values should
be taken with care. The flag was set
\begin{itemize}

\item if the standard deviation of individual $T_{\rm eff}$ from three
color indices ($g\, -\, r$, $g\, -\, i$, and $g\, -\, z$) exceeds $130$~K ($N=1,402$)

\item or if the difference between SDSS and IRFM measurements is greater than $3\sigma$
random errors (summed in quadrature) with respect to the mean trend ($N=4,388$).
Only those at $4700 < T_{\rm eff} < 7000$~K for dwarfs and $4700 < T_{\rm eff} < 5400$~K
for giants were flagged this way, to avoid a biased $\Delta T_{\rm eff}$ distribution at the
cool and hot temperature range (see Figure~\ref{fig:final}).

\item or if any of the $griz$ measurements are not reported in the KIC ($N=257$).

\end{itemize}
In total, $5,798$ stars (about $4\%$ of $154,931$ stars with a valid SDSS $T_{\rm eff}$)
were flagged this way.

\subsection{IRFM $T_{\rm eff}$ from Tycho-2MASS System}\label{sec:tycho}

\input{tab9.tex}

In addition to our main catalog in Table~\ref{tab:main}, we also present
in Table~\ref{tab:irfm} the IRFM $T_{\rm eff}$ in {\it Tycho} $V_T$ and
2MASS $JHK_s$ colors for $7,912$ stars. These stars are a subset of
the long-cadence KIC sample, which are bright enough to have $V_T$ magnitudes,
and can be used as an independent check on our $T_{\rm eff}$ scale
(see lower left panel in Figure~\ref{fig:compteff}).
The IRFM $T_{\rm eff}$ values are presented using $V_T\, -\, J$, $V_T\, -\, H$,
$V_T\, -\, K_s$, and $J\, -\, K_s$, with both random $(\sigma_{\rm ran})$
and total $(\sigma_{\rm tot})$ errors. As in Table~\ref{tab:main}, random
errors are propagated from photometric uncertainties, and total errors
are a quadrature sum of random and systematic errors ($15\%$ error in reddening
and $0.3$~dex error in [Fe/H]).

\section{Summary and Future Directions}\label{sec:summary}

The {\it Kepler} mission has a rich variety of applications, all of which
are aided by better knowledge of the fundamental stellar properties.
We have focused on the effective temperature scale, which is a
well-posed problem with the existing photometry.   However, in addition to the
revised KIC temperature there are two significant independent results
from our investigation.  We have identified a modest color-dependent offset between the KIC
and SDSS DR8 photometry, whose origin should be investigated.
Applying the relevant corrections to the KIC photometry significantly
improves the internal consistency of temperature estimates.  We have
also verified that the independent temperature scales (Johnson-Cousins
and SDSS) of An et al. and those from recent IRFM studies (Casagrande
et al.) are in good agreement, permitting a cross-calibration of the
latter to the SDSS filter system.  
Below we summarize our main results for the KIC, then turn to the major
limitations of our main catalog, a brief discussion of the implications,
and prospects for future improvements.

\subsection{Summary}

Our main result
is a shift to higher effective temperatures than those included in the
existing KIC.  We have employed
multiple diagnostic tools, including two distinct photometric scales
and some high-resolution spectroscopy.  In the case of cool (below
$6000$~K) dwarfs, the various methods for assigning effective
temperature have an encouraging degree of consistency.
The Johnson-Cousins measurements of \citet{an:07a}
are in good agreement with the independent IRFM temperatures
from C10 in star clusters.  In Table~\ref{tab:cluster}, for example,
the $V\, -\, I_c$ results agree within 15 K for all clusters
if we adopt the \citet{sandquist:04} dataset for M67.  The SDSS-based A09 system
is constructed to be on the same absolute scale as the \citet{an:07a}
system, so a similar level of agreement is expected between
the IRFM and the temperatures that we derive from 
the SDSS filters. A comparison of the IRFM and SDSS temperatures in
the KIC confirms this pattern, with agreement to better than 100 K
for the cool stars.  Even this level of disagreement overestimates
the underlying accord in the systems, because the IRFM ($J\, -\, K_s$)
diagnostic that was available to us in the KIC has systematic
offsets relative to other IRFM thermometers even in the open
clusters.  When we correct for these offsets, the agreement for cool stars 
between the SDSS-based method of A09 and the IRFM ($J\, -\, K_s$)
temperatures is  very good, with average differences below $25$~K and maximum differences
below the $50$~K level. Our cool dwarf temperatures are also within
$50$~K on average when compared to the spectroscopic results from B11.
The spectroscopic sample of MZ11 is cooler at the $88$~K level, which
we take as a measure of systematic uncertainties in the spectroscopic scale
\citep[See][for a further comparison of the spectroscopic and fundamental temperature scales]{bruntt:10}.

For hotter dwarfs the revised temperature estimates are higher than in
the KIC, but the magnitude of the offset is not consistent between the
two photometric scales and the spectroscopic data.  Motivated by
this offset, we adjusted the SDSS-based system of A09 to be cooler on average
by 100 K between 6000 K and 7000 K on the IRFM system.  The
consistency between photometric and spectroscopic scales degrades for
stars in this range.
This could reflect defects in the
fundamental temperature scale for hotter stars; the existing
fundamental data for the IRFM include relatively few solar-abundance
dwarfs above $6000$~K.  There could also be errors in photometric or
spectroscopic temperature estimates from the
onset of rapid rotation above 6300 K, or color anomalies from chemically
peculiar hot stars. On the
spectroscopic side, it would be valuable to compare the atmospheric
temperatures inferred from Boltzman and Saha constraints to
fundamental ones; as discussed in C10, there can be
significant systematic offsets between these scales for some systems.
This issue deserves future scrutiny and additional fundamental data would be very helpful.

In the case of evolved stars we also found a hotter temperature scale
than in the KIC.  We had to employ theoretical estimates of gravity
sensitivity, however, to temperature diagnostics derived for dwarfs.
An extension of the fundamental work to giants has been performed for
other colors in the past, and it would be beneficial to test the
theoretical predictions against actual radius data.

\subsection{Cautions and Caveats in Usage of the Catalog}

There are some significant drawbacks of the existing catalog, and care
is required in its proper application.  Binary companions will modify
the colors and temperatures of stars; we have provided tables for
statistical corrections, but have not included this in the tabulated
effective temperatures.  Blending can also impact colors, and there is
clear evidence of some blended objects in our comparison of the KIC to
SDSS DR8 data with superior resolution.  The major error source for the temperature estimates is the
uncertainty in the extinction.  We have adopted a global percentage value based
on typical errors in extinction maps, but there could be larger local
variations.  The color combinations available to us have limited
diagnostic power for star-by-star extinction and binary corrections.
For population studies, the stars in the long-cadence KIC sample were selected for a
planet transit survey, and do not represent an unbiased set of the
underlying population.

The KIC abundance estimates have significant errors, largely because
the filters with the greatest metallicity sensitivity were not
available.  As a result, we have adopted metallicity insensitive
temperature diagnostics, but the temperatures should be corrected for
individual metallicities if available.  These effects are at the 100 K
per dex level, and will therefore be smaller than the extinction
uncertainties for most stars in the sample.  The $\log{g}$ values for hot
stars are not well-constrained in the KIC, but we have adopted KIC
gravities for cool stars.  Our results would be affected at a modest
level by changes in the derived gravities, and the appropriate
corrections should be made if precise values are available.

There are two open areas for further discussion as well: the
appropriate temperature scale for the hot dwarfs and errors in the
photometry.  In the former case, we recommend adjustments above 6000 K to
the SDSS scale.  For the entire domain we also note inconsistencies
between the $J\, -\, K_s$ calibration and the other color-temperature
relationships in the IRFM.  Even after putting the fundamental photometric 
temperature scale on a common system, however, there is a
difference between it and the spectroscopic scale
for stars above $6000$~K.  Until it is resolved we recommend
inclusion of systematic temperature errors in this domain.  The impact of
the $\log{g}$ determinations on the extinction estimates for the hot stars
should be investigated as well.  The gravity diagnostics for the hot
stars are not well measured, and asteroseismic gravities confirm this expected
lack of precision.  The KIC catalog included this as an ingredient in 
the distance estimates, but it is difficult to reconstruct the weights 
and importance of this uncertainty after the fact.  Star-by-star 
extinctions would be useful for this purpose.

The origin
of the differences between the SDSS (DR8) and KIC photometry should
also be tracked down, and there may be spatially dependent or magnitude
dependent terms.  We also noted some cases with severe internal
inconsistency in the photometric temperature diagnostics, and flagged
those which we identified.
We believe that unresolved blends are a promising candidate,
but further work
on this front is warranted.  In a small fraction of cases these
photometric issues can cause severe errors in the temperatures.
Effective temperatures for stars where different colors return very
different estimates should be treated with caution.

Despite these reservations, we believe that the addition of
temperatures more closely tied to the fundamental scale will
significantly improve the reliability of inferences about the
underlying stellar populations.

\subsection{Implications and Future Directions}

A shift to higher effective temperatures will have consequences for
both planetary and stellar science.  On the main sequence, hotter stars will
be on average more massive and larger.  This would imply larger planet radii
on average for such objects.  The radii of evolved stars require more information
(especially from surface gravity effects), and the consequences of the 
temperature scale shift for them are more difficult to predict from first 
principles.  Stars of known asteroseismic
radius will be on average more luminous, which could partially explain
discrepancies in the mass-radius relationship for evolved stars
\citep{chap:11}.  Asteroseismic parameters defined with scaling
relationships will also be impacted. A more precise absolute effective
temperature scale will also permit more stringent constraints on
asteroseismic properties from detailed modeling of the frequency spectrum
\citep[see][]{metcalfe:10}.

However, the full potential will be realized as complementary
information becomes available on the {\it Kepler} sample.  Blue data (such
as Johnson $UB$ or SDSS $u$) could be employed to infer more reliable
photometric metallicities; Johnson-Cousins $UBV(RI)_C$ data would enable more
reliable extinction estimates, binary discrimination, and broader
application of the IRFM directly to stars in the sample.  Photometric
systems naturally designed for F-type stars, such as Str\"omgren, would be
useful for addressing the temperature and surface gravity scales in
that regime.  

A more robust set of input data
would provide an important control sample for the measured planet
population; it will be challenging to obtain spectroscopic
temperatures of both the planet candidates and the background stellar
population.  A better calibration of the fundamental temperature scale
is possible once asteroseismic radii are combined with parallaxes in the {\it Kepler} field,
either via {\it Kepler} data or through the Gaia mission.  The time domain
data from the satellite are exquisite; a proper application of
complementary tools from stellar astrophysics is now essential to
fully realize their considerable scientific promise.

\acknowledgements

We thank Timothy Brown, Luca Casagrande, and Constance Rockosi for useful
discussions. We also thank the anonymous referee for careful and detailed comments.
M.P.\ acknowledges support from NASA ATP grant NNX11AE04G.
D.A.\ acknowledges support from the Ewha Womans University Research
Grant of 2010, as well as support by the National Research Foundation
of Korea to the Center for Galaxy Evolution Research.
JM-\.Z acknowledges the Polish Ministry grant no N\,N203\,405139.
WJC acknowledges financial support from
the UK Science and Technology Facilities Council.
T.S.M.\ acknowledges support from NASA grant NNX09AE59G.

\clearpage

\appendix

\section{Erratum: ``A Revised Effective Temperature Scale for the {\it Kepler}
Input Catalog'' (2012, \apjs, 199, 30)}

\subsection{Sign Errors in the Gravity Corrections}

We derived surface gravity corrections to the color-temperature relationships
for red giants.  This step was required because we adopted a dwarf-based color
calibration, and the theoretical correction is given in Figure~8.  However, the
sign of the correction was reversed in the note at the bottom of Table 4. In
Table 7 the sense in which the corrections should have been applied is noted
correctly, but the actual corrections used were reversed. In addition, null
values ($-999$) in the gravity correction table were accidentally included in
the interpolation by the sign flip error, which resulted in unreasonably large
correction terms at the very cool end ($T_{\rm eff} \la 4200$~K).

We have revised the data in Tables 4 and 7 accordingly, and updated the figures
(Figures 8 and 18) and statistical properties (Table 8) that were impacted.
Gravity terms were applied for stars with low gravities ($\log{g} \leq 3.5$), to
correct for differences between the the {\it Kepler} Input Catalog (KIC) and the
model $\log{g}$ values.  The $T_{\rm eff}$ estimates for dwarfs with $\log{g} >
3.5$ remain valid in the original version of Table~$7$.

A revised version of the gravity corrections is found in Table~\ref{taba:logg},
with the same form and content as in the original table, except its signs. Here,
$\log{g}{\rm (YREC)}$ is the $\log{g}$ in our YREC model ($3^{\rm rd}$ column in
Table~\ref{taba:logg}).  Negative values in the table mean that giants have lower $T_{\rm
eff}$ than dwarfs at fixed colors.  Therefore, $\Delta T_{\rm eff}$ values in
Table~\ref{taba:logg} should be added to dwarf-based $T_{\rm eff}$ estimates, if one wishes to
infer $T_{\rm eff}$ for giants or subgiants.

We also revised our scheme to handle the gravity corrections in the very cool
end ($T_{\rm eff} \la 4200$~K).  A revised plot of the gravity corrections is
presented in Figure~\ref{figa:loggcorr}. In this revision, a simple quadratic
relation was used to extend theoretical $T_{\rm eff}$ corrections beyond
theoretical computations, in a way that the correction terms become zero at
$T_{\rm eff} \sim 4000$~K (see Figure~$7$, where dwarf color-$T_{\rm eff}$
relations cross giant relations at $T_{\rm eff} \sim 4000$~K). However, only a
minor fraction of stars in our sample ($\sim 400$ giants) are affected by this
change.

A revised main catalog (Table~\ref{taba:main}) shows SDSS $T_{\rm eff}$
estimates, after correcting for the difference in gravity, at $\log{g}{\rm
(KIC)} \le 3.5$. For giants, the mean and median changes from the original
version are $59$~K and $36$~K, respectively, at $T_{\rm eff} > 4200$~K.  The
dwarf-based solution can be obtained by subtracting the tabulated correction
terms ($\Delta T_{\rm eff}$) in Table~\ref{taba:main}.

Figure~\ref{figa:final} is a replacement for the original figure, and shows
comparisons of our revised $T_{\rm eff}$ with those of the KIC ({\it top}
panels) and Infrared Flux Method (IRFM; bottom panels). Comparisons for dwarfs
in the left panels are unaffected by the sign flips in the gravity corrections.
However, an improved agreement is seen between the IRFM and SDSS (YREC) scales
for red giants (lower right panel). The discrepancy with the original KIC
remains, but the differences for cool giants are significantly smaller.

The corrected statistical properties of the temperature differences between
SDSS, IRFM, and KIC estimates are shown in Table~\ref{taba:stat}, a replacement
for the original table.  For hotter giants the differences from prior results
are zero, while the magnitude of the error rises to $210$~K at the cool end of
the calibration range. The weighted mean differences for giants are $-165$~K (a
median difference of $-161$~K) for the KIC minus SDSS temperatures and $-47$~K (a
median difference of $-42$~K) for the IRFM minus SDSS temperatures. The
differences were $-252$~K (median $-215$~K) and $-109$~K (median $-92$~K),
respectively, from prior results. A significant offset between the KIC and SDSS
values remains, which is one of our main results in the paper.

The main focus of the original paper concerned the $T_{\rm eff}$ scale for
dwarfs, which is not impacted.  With the revised $\log{g}$ corrections, a
smaller number of stars ($5,347$ versus $5,798$ among $154,931$ stars with a
valid SDSS $T_{\rm eff}$) are now flagged as having internally inconsistent
$T_{\rm eff}$ estimates, according to our quality criteria (Section~4.2.4).
Comparisons of the revised $T_{\rm eff}$ for giants with spectroscopic samples
in Figure~14 are unaffected, as the gravity corrections have correctly been
applied in this case.

\subsection{Detailed Steps to Compute Mean $T_{\rm eff}$ and its Errors}

Regardless of the sign flip errors in the gravity corrections, we provide below
detailed descriptions on how to obtain mean $T_{\rm eff}$ and its random and
systematic errors using YREC isochrones in $griz$ passbands.

The mean $T_{\rm eff}$ can be determined in the following way. This compliments
a description in the third step of ``A Recipe for Estimating $T_{\rm eff}$'' in
Section~4.1 of the original paper. For a given set of $griz$ magnitudes, which
were corrected for the photometry zero-point errors and interstellar extinctions
in the previous steps, one can determine a mean distance modulus for each star
using model magnitudes in $griz$:
\begin{equation}
(m\, -\, M)_0 = \frac {\Sigma_i [(m_{{\rm obs},i} - m_{{\rm model},i})/\sigma^2_i]}
                      {\Sigma_i (1/\sigma^2_i)},
\end{equation}
where the subscript $i$ indicates each of the $griz$ passbands. The $m_{{\rm
obs},i}$ and $\sigma_i$ are observed magnitude and its error ($0.01$~mag in
$gri$ and $0.03$~mag in $z$) in each passband. The $m_{{\rm model},i}$ is the
model magnitude in each passband from our base isochrone in Table~$1$.

From this, one can compute a $\chi^2$ value of the model fit for each star as follows:
\begin{equation}
\chi^2 = \Sigma_i \frac{[(m_{{\rm obs},i} - m_{{\rm model},i}) - (m\, -\, M)_0]^2}
                       {\sigma^2_i}.
\end{equation}
We searched for a minimum $\chi^2$ of the model fit in the entire mass grid of
our base isochrone, and determined a mean $T_{\rm eff}$ as the one that gives
the most consistent fit overall to the data in $griz$ passbands. The best-fitting
$T_{\rm eff}$ is shown in the second column in Table~\ref{taba:main}.

Random errors in $T_{\rm eff}$ were obtained in the following way. We applied
photometric errors to each passband ($\pm0.01$~mag in $gri$ and $\pm0.03$~mag in $z$),
and computed $T_{\rm eff}$. Taking the mean difference from the original $T_{\rm
eff}$ as an effective $1\sigma$ error, we added in quadrature $T_{\rm eff}$
errors from all photometric passbands in $griz$. We also computed a random error
by taking a standard deviation of individual $T_{\rm eff}$ estimates from $g\,
-\, r$, $g\, -\, i$, and $g\, -\, z$, respectively. We then took a larger one
from the above two approaches as the size of a representative random source of
error.  This error is shown in the fourth column ($\sigma_{\rm ran}$) in
Table~\ref{taba:main}.

Systematic errors in $T_{\rm eff}$ were estimated as follows. For the error in
the reddening, we repeated computing $T_{\rm eff}$ with $15\%$ lower and higher
$E(B\, -\, V)$ values than in the KIC, and took the mean difference from the
original $T_{\rm eff}$ as an effective $\pm1\sigma$ error.  For the error from
the metallicity, we computed $T_{\rm eff}$ from models at [Fe/H]$=-0.5$ and
[Fe/H]$=+0.1$ (one can use our tabulated metallicity corrections in Table~$3$),
and took the mean difference from an original $T_{\rm eff}$ as an effective
$\pm1\sigma$ error in $T_{\rm eff}$.  From both errors in reddening and
metallicity, we obtained a total systematic error in $T_{\rm eff}$ by adding
individual errors in quadrature.  Based on both random and systematic errors,
the total error was computed as a quadrature sum of these errors, and is
tabulated in the $3^{\rm rd}$ column ($\sigma_{\rm tot}$) in Table~\ref{taba:main}.

\setcounter{table}{3}
\input{taba4.tex}

\setcounter{table}{6}
\input{taba7.tex}

\setcounter{table}{7}
\input{taba8.tex}

\setcounter{figure}{7}
\begin{figure}
\epsscale{0.65}
\plotone{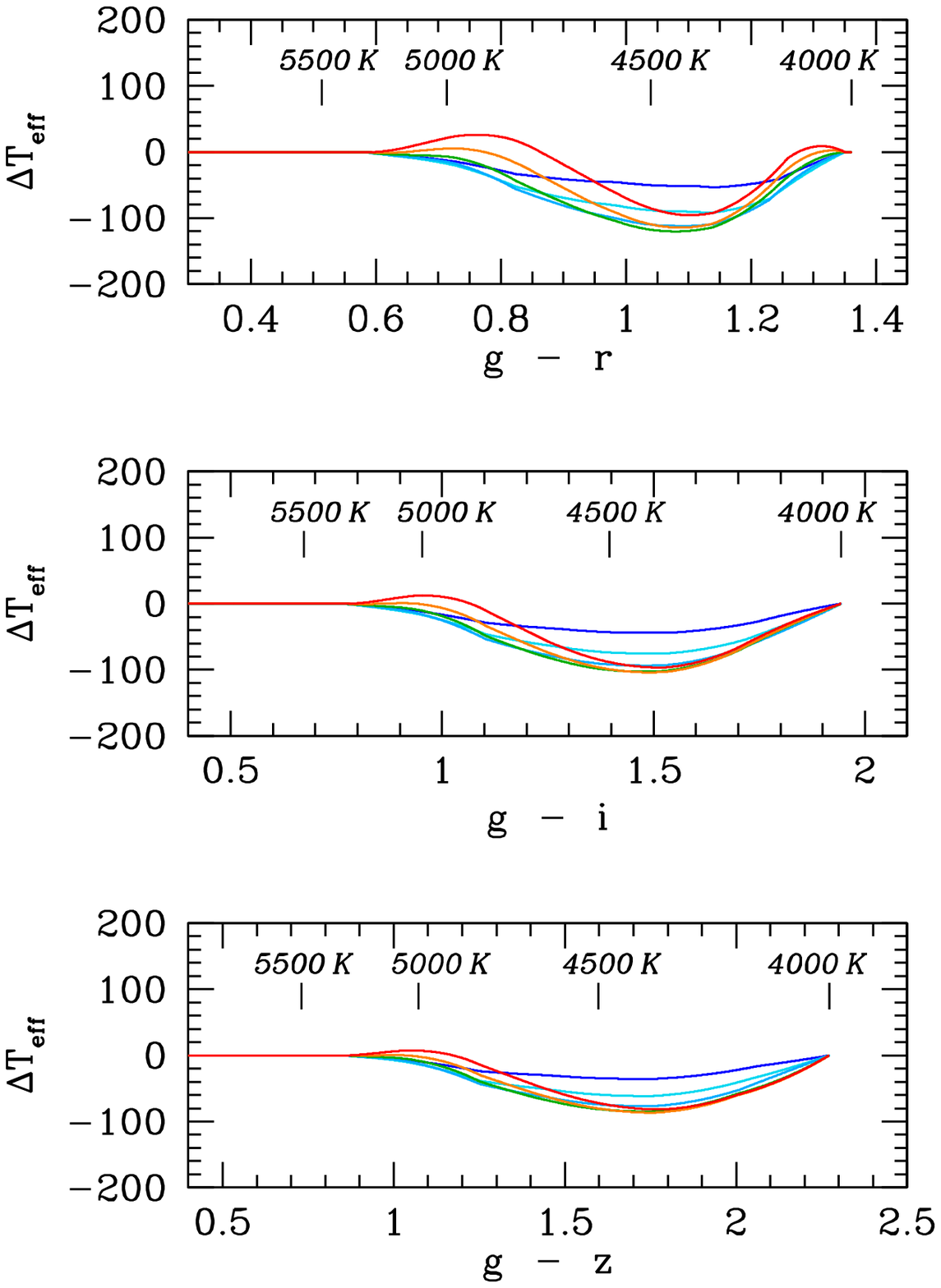}
\caption{Theoretical $T_{\rm eff}$ corrections for various $\Delta \log{g}$
values with respect to the fiducial isochrones. Corrections from $\Delta \log{g}
= 0.5$ (blue) to $\Delta \log{g} = 3.0$ (red) with a $0.5$~dex increment are
shown.  The sense is that giants tend to have lower $T_{\rm eff}$ than dwarfs at
fixed colors. A linear ramp was used to define smoothly varying $\Delta T_{\rm
eff}$ over $4800$~K $< T_{\rm eff} < 5800$~K.  In this revision, a simple
quadratic relation is used to extend theoretical $T_{\rm eff}$ corrections to
the very cool end ($g\, -\, r \ga 1.25$), where correction factors become zero
at $T_{\rm eff} \sim 4000$~K. Only a minor fraction of giants in the sample
($\sim400$ stars) are affected by this change.
\label{figa:loggcorr}}
\end{figure}

\setcounter{figure}{17}
\begin{figure}
\centering
\includegraphics[scale=0.6]{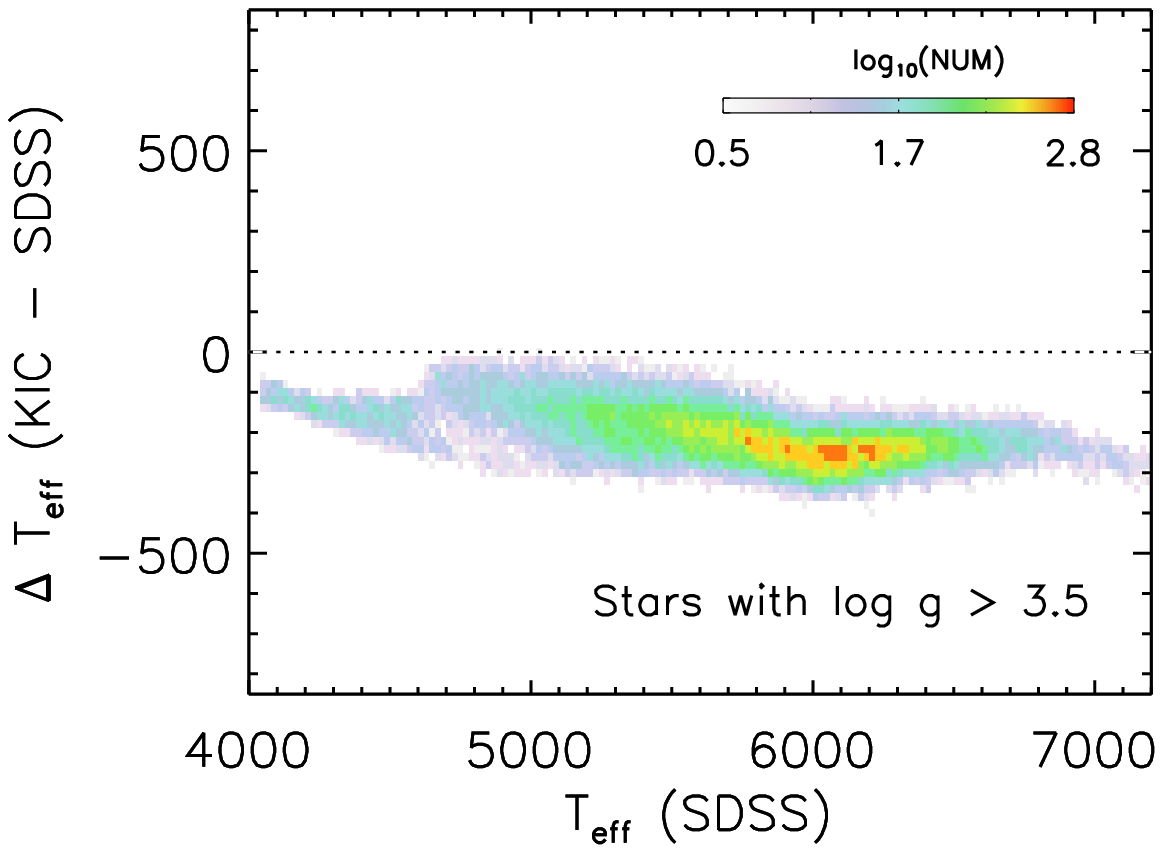}
\includegraphics[scale=0.6]{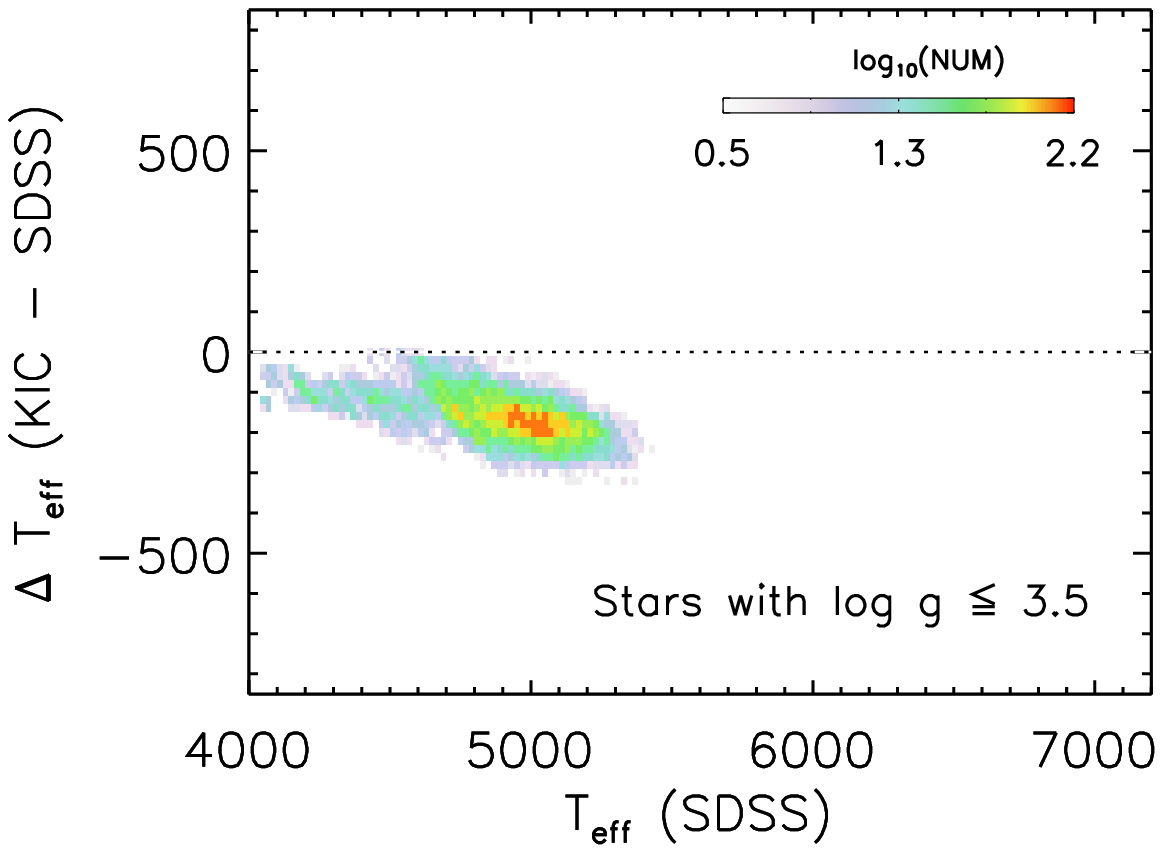}
\includegraphics[scale=0.6]{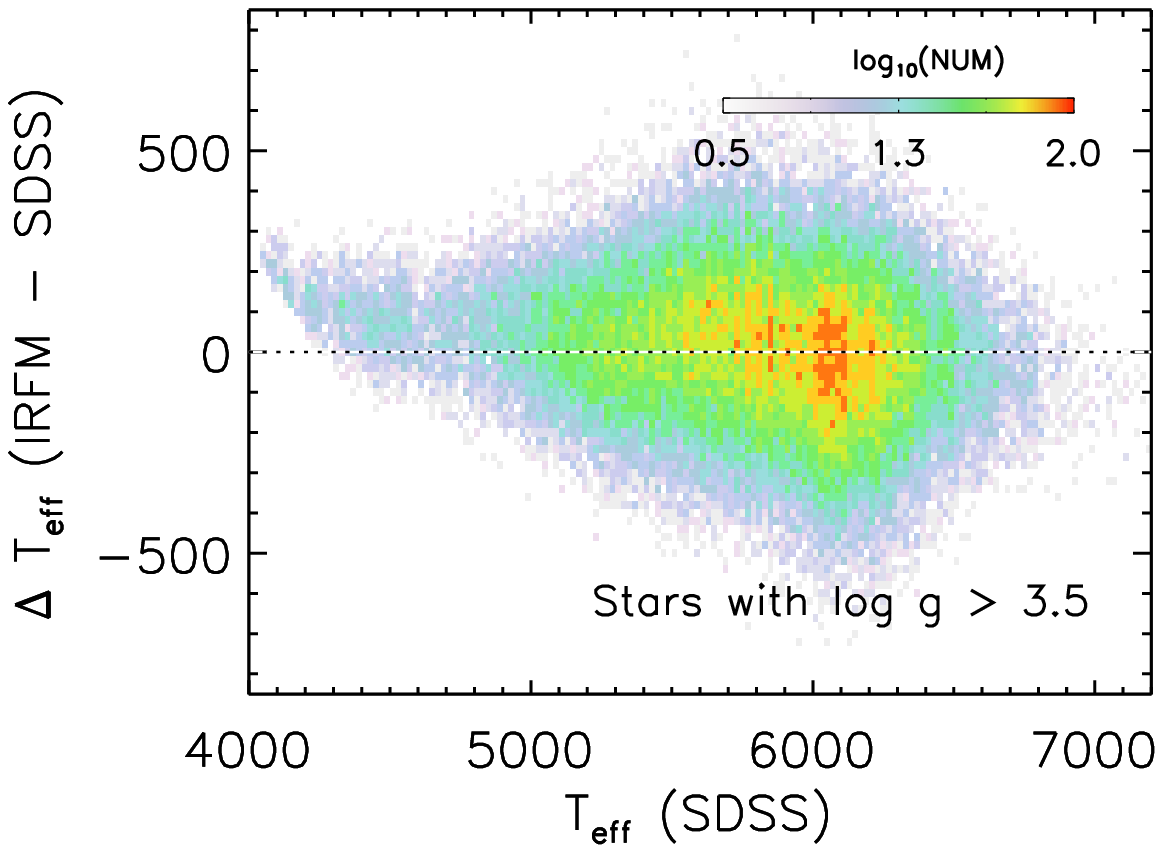}
\includegraphics[scale=0.6]{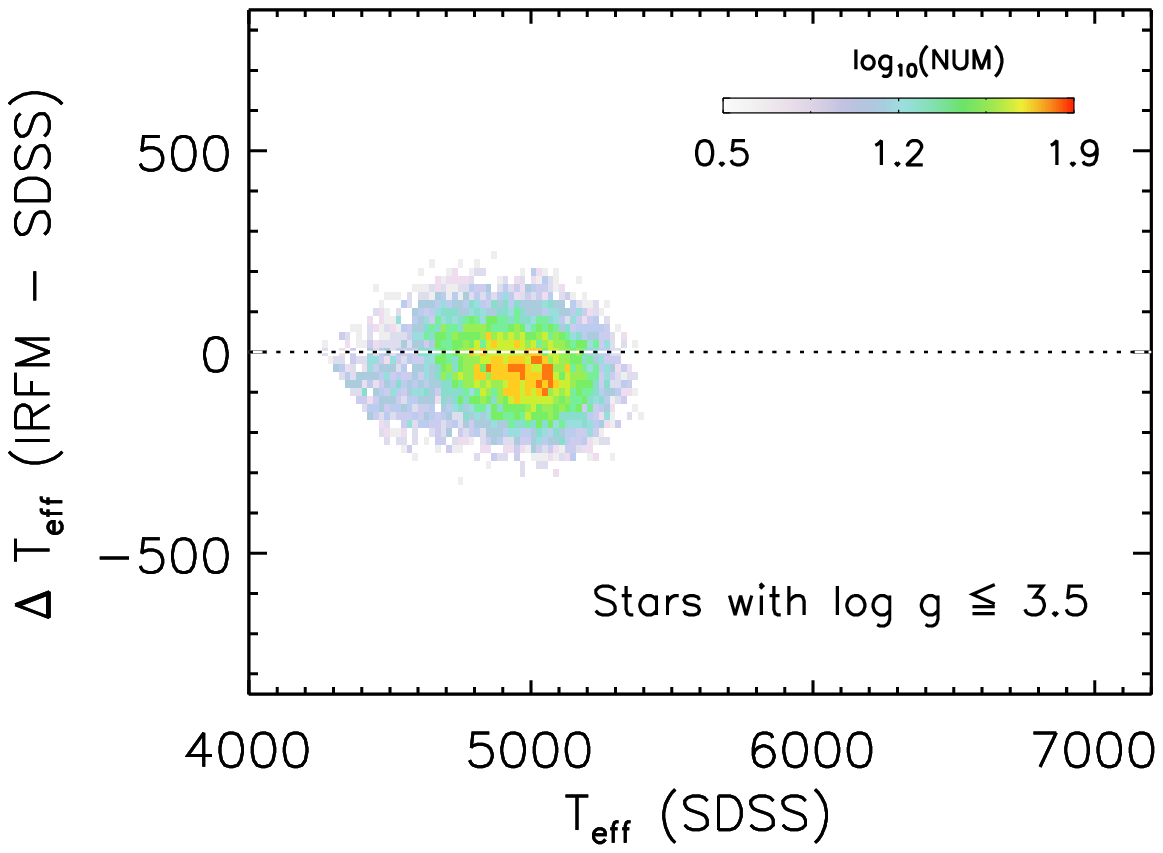}
\caption{Comparisons of $T_{\rm eff}$ using the final SDSS $T_{\rm eff}$
estimates.  Comparisons are shown for the original KIC $T_{\rm eff}$ for
dwarfs (top left) and giants (top right), and for the $(J\, -\, K_s)$-based
IRFM estimates for dwarfs (bottom left) and giants (bottom right). Comparisons
for giants shown in the right panels are affected by the sign flip errors in
the gravity corrections, but comparisons for dwarfs in the left panels are
unaffected.
\label{figa:final}}
\end{figure}

\end{document}

%% file: tab1.tex
\begin{deluxetable*}{cccccccc}
\tablewidth{0pt}
\tabletypesize{\scriptsize}
\tablecaption{Base Isochrone at [Fe/H]$=-0.2$\label{tab:isochrone}}
\tablehead{
  \colhead{Mass/$M_\odot$} &
  \colhead{$T_{\rm eff}$} &
  \colhead{$\log{L/L_\odot}$} &
  \colhead{$\log{g}$} &
  \colhead{$M_r$} &
  \colhead{$g\, -\, r$} &
  \colhead{$g\, -\, i$} &
  \colhead{$g\, -\, z$}
}
\startdata
$1.50$ & $7506.2$ & $ 0.85$ & $4.22$ & $2.60$ & $0.032$ & $-0.022$ & $-0.111$ \nl
$1.46$ & $7409.2$ & $ 0.81$ & $4.23$ & $2.70$ & $0.051$ & $ 0.007$ & $-0.076$ \nl
$1.43$ & $7306.3$ & $ 0.76$ & $4.24$ & $2.80$ & $0.070$ & $ 0.038$ & $-0.039$ \nl
$1.40$ & $7200.2$ & $ 0.72$ & $4.25$ & $2.90$ & $0.091$ & $ 0.069$ & $-0.000$ \nl
$1.37$ & $7091.8$ & $ 0.68$ & $4.25$ & $3.00$ & $0.113$ & $ 0.102$ & $ 0.039$ \nl
$1.35$ & $6992.9$ & $ 0.64$ & $4.26$ & $3.10$ & $0.133$ & $ 0.132$ & $ 0.076$ \nl
$1.32$ & $6902.5$ & $ 0.59$ & $4.27$ & $3.20$ & $0.152$ & $ 0.160$ & $ 0.111$ \nl
$1.30$ & $6817.8$ & $ 0.55$ & $4.29$ & $3.30$ & $0.169$ & $ 0.187$ & $ 0.143$ \nl
$1.27$ & $6737.8$ & $ 0.51$ & $4.30$ & $3.40$ & $0.186$ & $ 0.212$ & $ 0.175$ \nl
$1.25$ & $6662.9$ & $ 0.47$ & $4.31$ & $3.50$ & $0.203$ & $ 0.236$ & $ 0.204$ \nl
$1.22$ & $6592.6$ & $ 0.43$ & $4.33$ & $3.60$ & $0.218$ & $ 0.259$ & $ 0.232$ \nl
$1.20$ & $6524.8$ & $ 0.39$ & $4.34$ & $3.70$ & $0.234$ & $ 0.281$ & $ 0.259$ \nl
$1.18$ & $6458.9$ & $ 0.34$ & $4.36$ & $3.80$ & $0.249$ & $ 0.304$ & $ 0.286$ \nl
$1.16$ & $6394.6$ & $ 0.30$ & $4.37$ & $3.90$ & $0.265$ & $ 0.326$ & $ 0.313$ \nl
$1.14$ & $6332.4$ & $ 0.26$ & $4.39$ & $4.00$ & $0.280$ & $ 0.347$ & $ 0.339$ \nl
$1.12$ & $6271.2$ & $ 0.22$ & $4.41$ & $4.10$ & $0.295$ & $ 0.369$ & $ 0.366$ \nl
$1.10$ & $6210.7$ & $ 0.18$ & $4.42$ & $4.20$ & $0.310$ & $ 0.391$ & $ 0.392$ \nl
$1.08$ & $6151.0$ & $ 0.14$ & $4.44$ & $4.30$ & $0.326$ & $ 0.413$ & $ 0.418$ \nl
$1.06$ & $6092.1$ & $ 0.10$ & $4.45$ & $4.40$ & $0.341$ & $ 0.434$ & $ 0.445$ \nl
$1.04$ & $6033.6$ & $ 0.06$ & $4.47$ & $4.50$ & $0.357$ & $ 0.457$ & $ 0.472$ \nl
$1.02$ & $5975.0$ & $ 0.02$ & $4.49$ & $4.60$ & $0.373$ & $ 0.479$ & $ 0.498$ \nl
$1.00$ & $5915.8$ & $-0.02$ & $4.50$ & $4.70$ & $0.389$ & $ 0.501$ & $ 0.524$ \nl
$0.98$ & $5855.5$ & $-0.06$ & $4.51$ & $4.80$ & $0.406$ & $ 0.524$ & $ 0.550$ \nl
$0.96$ & $5793.9$ & $-0.10$ & $4.53$ & $4.90$ & $0.423$ & $ 0.548$ & $ 0.579$ \nl
$0.95$ & $5731.4$ & $-0.14$ & $4.54$ & $5.00$ & $0.441$ & $ 0.573$ & $ 0.609$ \nl
$0.93$ & $5669.8$ & $-0.18$ & $4.55$ & $5.10$ & $0.460$ & $ 0.598$ & $ 0.640$ \nl
$0.91$ & $5606.9$ & $-0.22$ & $4.56$ & $5.20$ & $0.479$ & $ 0.625$ & $ 0.672$ \nl
$0.90$ & $5538.8$ & $-0.26$ & $4.57$ & $5.30$ & $0.500$ & $ 0.655$ & $ 0.709$ \nl
$0.88$ & $5472.0$ & $-0.29$ & $4.58$ & $5.40$ & $0.522$ & $ 0.686$ & $ 0.747$ \nl
$0.86$ & $5406.0$ & $-0.33$ & $4.59$ & $5.50$ & $0.545$ & $ 0.717$ & $ 0.785$ \nl
$0.85$ & $5340.0$ & $-0.37$ & $4.60$ & $5.60$ & $0.568$ & $ 0.750$ & $ 0.826$ \nl
$0.83$ & $5273.8$ & $-0.41$ & $4.61$ & $5.70$ & $0.593$ & $ 0.785$ & $ 0.869$ \nl
$0.82$ & $5207.7$ & $-0.44$ & $4.61$ & $5.80$ & $0.619$ & $ 0.822$ & $ 0.914$ \nl
$0.81$ & $5142.1$ & $-0.48$ & $4.62$ & $5.90$ & $0.647$ & $ 0.861$ & $ 0.961$ \nl
$0.79$ & $5077.2$ & $-0.51$ & $4.63$ & $6.00$ & $0.676$ & $ 0.901$ & $ 1.010$ \nl
$0.78$ & $5013.1$ & $-0.55$ & $4.63$ & $6.10$ & $0.707$ & $ 0.943$ & $ 1.061$ \nl
$0.77$ & $4949.9$ & $-0.58$ & $4.64$ & $6.20$ & $0.739$ & $ 0.987$ & $ 1.115$ \nl
$0.75$ & $4887.6$ & $-0.62$ & $4.64$ & $6.30$ & $0.773$ & $ 1.034$ & $ 1.170$ \nl
$0.74$ & $4826.3$ & $-0.65$ & $4.65$ & $6.40$ & $0.808$ & $ 1.082$ & $ 1.228$ \nl
$0.73$ & $4766.1$ & $-0.68$ & $4.65$ & $6.50$ & $0.846$ & $ 1.132$ & $ 1.288$ \nl
$0.72$ & $4707.4$ & $-0.72$ & $4.66$ & $6.60$ & $0.885$ & $ 1.185$ & $ 1.350$ \nl
$0.71$ & $4650.3$ & $-0.75$ & $4.66$ & $6.70$ & $0.926$ & $ 1.239$ & $ 1.414$ \nl
$0.70$ & $4595.1$ & $-0.78$ & $4.66$ & $6.80$ & $0.967$ & $ 1.295$ & $ 1.479$ \nl
$0.69$ & $4541.8$ & $-0.81$ & $4.67$ & $6.90$ & $1.007$ & $ 1.350$ & $ 1.544$ \nl
$0.68$ & $4490.6$ & $-0.84$ & $4.67$ & $7.00$ & $1.047$ & $ 1.406$ & $ 1.610$ \nl
$0.67$ & $4441.5$ & $-0.87$ & $4.68$ & $7.10$ & $1.084$ & $ 1.461$ & $ 1.675$ \nl
$0.66$ & $4394.5$ & $-0.90$ & $4.68$ & $7.20$ & $1.121$ & $ 1.515$ & $ 1.740$ \nl
$0.65$ & $4349.3$ & $-0.93$ & $4.69$ & $7.30$ & $1.155$ & $ 1.568$ & $ 1.804$ \nl
$0.64$ & $4306.0$ & $-0.96$ & $4.69$ & $7.40$ & $1.188$ & $ 1.619$ & $ 1.867$ \nl
$0.63$ & $4264.4$ & $-0.99$ & $4.70$ & $7.50$ & $1.218$ & $ 1.670$ & $ 1.928$ \nl
$0.62$ & $4224.3$ & $-1.02$ & $4.70$ & $7.60$ & $1.247$ & $ 1.719$ & $ 1.989$ \nl
$0.61$ & $4185.9$ & $-1.04$ & $4.71$ & $7.70$ & $1.273$ & $ 1.766$ & $ 2.048$ \nl
$0.60$ & $4148.9$ & $-1.07$ & $4.71$ & $7.80$ & $1.298$ & $ 1.812$ & $ 2.105$ \nl
$0.59$ & $4113.7$ & $-1.10$ & $4.72$ & $7.90$ & $1.320$ & $ 1.856$ & $ 2.160$ \nl
$0.58$ & $4079.8$ & $-1.13$ & $4.73$ & $8.00$ & $1.340$ & $ 1.898$ & $ 2.213$ \nl
$0.58$ & $4047.1$ & $-1.16$ & $4.74$ & $8.10$ & $1.358$ & $ 1.939$ & $ 2.265$ \nl
\enddata
\end{deluxetable*}

%% file: tab2.tex
\begin{deluxetable}{crrr}
\tablewidth{0pt}
\tabletypesize{\scriptsize}
\tablecaption{Coefficients for Polynomial Color-$T_{\rm eff}$ Relations\label{tab:poly}}
\tablehead{
  \colhead{Coeff.} &
  \colhead{$g\, -\, r$} &
  \colhead{$g\, -\, i$} &
  \colhead{$g\, -\, z$}
}
\startdata
$a_0$ & $ 0.6676$ &  $ 0.6888$ & $ 0.7053$ \nl
$a_1$ & $ 0.3434$ &  $ 0.2012$ & $ 0.2022$ \nl
$a_2$ & $ 0.5851$ &  $ 0.4518$ & $ 0.2733$ \nl
$a_3$ & $-0.6919$ &  $-0.4871$ & $-0.2844$ \nl
$a_4$ & $ 0.1445$ &  $ 0.1926$ & $ 0.1079$ \nl
$a_5$ & $ 0.0594$ &  $-0.0256$ & $-0.0144$
\enddata
\tablecomments{Coefficients in equation~\ref{eq:teff}.  These cofficients
are valid at $4080 \leq T_{\rm eff} < 7000$~K, or 
$0.13 < (g\, -\, r)_0 < 1.34$,
$0.13 < (g\, -\, i)_0 < 1.90$, and
$0.07 < (g\, -\, z)_0 < 2.21$, respectively.}
\end{deluxetable}

%% file: tab3.tex
\begin{deluxetable*}{lrrrrrrrrr}
\tabletypesize{\scriptsize}
\tablewidth{0pt}
\tablecaption{$T_{\rm eff}$ Corrections for Different [Fe/H]\label{tab:feh}}
\tablehead{
  \colhead{} &
  \multicolumn{9}{c}{[Fe/H]} \nl
  \cline{2-10}
  \colhead{color} &
  \colhead{$-2.0$} &
  \colhead{$-1.5$} &
  \colhead{$-1.0$} &
  \colhead{$-0.6$} &
  \colhead{$-0.4$} &
  \colhead{$-0.2$\tablenotemark{a}} &
  \colhead{$+0.0$} &
  \colhead{$+0.2$} &
  \colhead{$+0.4$}
}
\startdata
\multicolumn{10}{c}{$g\, -\, r$} \nl
\hline
$0.1$&$   44$&$   36$&$   20$&$   19$&$    7$&$0$&$    0$&\nodata&\nodata\\
$0.2$&$  -32$&$  -32$&$  -27$&$   -6$&$   -4$&$0$&$   20$&$   61$&$  100$\\
$0.3$&$ -120$&$ -103$&$  -74$&$  -33$&$  -20$&$0$&$   33$&$   84$&$  145$\\
$0.4$&$ -162$&$ -144$&$ -110$&$  -60$&$  -32$&$0$&$   44$&$  104$&$  176$\\
$0.5$&$ -168$&$ -165$&$ -130$&$  -75$&$  -42$&$0$&$   44$&$  116$&$  194$\\
$0.6$&$ -186$&$ -185$&$ -144$&$  -80$&$  -42$&$0$&$   46$&$  113$&$  192$\\
$0.7$&$ -214$&$ -195$&$ -149$&$  -84$&$  -44$&$0$&$   48$&$  104$&$  178$\\
$0.8$&$ -238$&$ -198$&$ -149$&$  -85$&$  -42$&$0$&$   45$&$   97$&$  157$\\
$0.9$&$ -262$&$ -204$&$ -153$&$  -86$&$  -43$&$0$&$   44$&$   83$&$  131$\\
$1.0$&$ -282$&$ -212$&$ -160$&$  -94$&$  -46$&$0$&$   45$&$   77$&$  113$\\
$1.1$&$ -289$&$ -223$&$ -174$&$ -107$&$  -51$&$0$&$   45$&$   69$&$   93$\\
$1.2$&\nodata&$ -233$&$ -207$&$ -136$&$  -60$&$0$&$   42$&$   45$&$   28$\\
$1.3$&\nodata&\nodata&\nodata&\nodata&$  -83$&$0$&$   37$&$  -33$&\nodata\\
\hline
\multicolumn{10}{c}{$g\, -\, i$} \nl
\hline
$0.1$&$   86$&$   67$&$   33$&$   20$&$    9$&$0$&$   -5$&$  -23$&\nodata\\
$0.3$&$    5$&$   -3$&$  -13$&$   -5$&$   -3$&$0$&$   16$&$   35$&$   61$\\
$0.5$&$  -69$&$  -70$&$  -61$&$  -37$&$  -21$&$0$&$   27$&$   63$&$  108$\\
$0.7$&$  -84$&$  -99$&$  -88$&$  -52$&$  -27$&$0$&$   33$&$   75$&$  127$\\
$0.9$&$ -126$&$ -142$&$ -118$&$  -65$&$  -34$&$0$&$   35$&$   75$&$  125$\\
$1.1$&$ -168$&$ -156$&$ -128$&$  -72$&$  -36$&$0$&$   39$&$   71$&$  112$\\
$1.3$&$ -204$&$ -166$&$ -134$&$  -78$&$  -40$&$0$&$   41$&$   68$&$   99$\\
$1.5$&$ -233$&$ -180$&$ -143$&$  -83$&$  -42$&$0$&$   44$&$   68$&$   95$\\
$1.7$&\nodata&$ -206$&$ -172$&$  -95$&$  -47$&$0$&$   50$&$   70$&$   93$\\
$1.9$&\nodata&\nodata&\nodata&\nodata&$  -56$&$0$&$   61$&$   80$&$  101$\\
\hline
\multicolumn{10}{c}{$g\, -\, z$} \nl
\hline
$0.1$&$   94$&$   68$&$   33$&$   15$&$    6$&$0$&$   -4$&$  -16$&\nodata\\
$0.3$&$   45$&$   25$&$    2$&$   -1$&$    0$&$0$&$   10$&$   19$&$   33$\\
$0.5$&$  -11$&$  -23$&$  -35$&$  -24$&$  -13$&$0$&$   20$&$   41$&$   72$\\
$0.7$&$  -27$&$  -43$&$  -47$&$  -29$&$  -17$&$0$&$   20$&$   50$&$   87$\\
$0.9$&$  -63$&$  -84$&$  -78$&$  -46$&$  -23$&$0$&$   26$&$   54$&$   92$\\
$1.1$&$ -109$&$ -119$&$ -102$&$  -59$&$  -29$&$0$&$   31$&$   60$&$   93$\\
$1.3$&$ -145$&$ -137$&$ -116$&$  -66$&$  -34$&$0$&$   35$&$   60$&$   90$\\
$1.5$&$ -177$&$ -147$&$ -121$&$  -70$&$  -36$&$0$&$   39$&$   61$&$   87$\\
$1.7$&$ -205$&$ -160$&$ -128$&$  -73$&$  -36$&$0$&$   44$&$   66$&$   92$\\
$1.9$&$ -225$&$ -182$&$ -144$&$  -77$&$  -39$&$0$&$   49$&$   72$&$   97$\\
$2.1$&\nodata&\nodata&\nodata&$  -87$&$  -44$&$0$&$   58$&$   82$&$  109$\\
\enddata
\tablecomments{The sense of the difference is the model $T_{\rm eff}$ at
a given [Fe/H] minus that of the fiducial metallicity, [Fe/H]$=-0.2$.
The $T_{\rm eff}$ at a fixed color generally becomes cooler at a lower [Fe/H].
In other words, the above correction factor should be added to the SDSS
$T_{\rm eff}$, if the metallicity effects should be taken into account.}
\tablenotetext{a}{Fiducial metallicity.}
\end{deluxetable*}

%% file: tab4.tex
\begin{deluxetable*}{cccrrrrrr}
\tabletypesize{\scriptsize}
\tablewidth{0pt}
\tablecaption{$\log{g}$ Corrections\label{tab:logg}}
\tablehead{
  \colhead{} &
  \colhead{} &
  \colhead{Ref.} &
  \multicolumn{6}{c}{$\Delta \log{g}$} \nl
  \cline{4-9}
  \colhead{$g - r$} &
  \colhead{$g - i$/$g - z$} &
  \colhead{$\log{g}$\tablenotemark{a}} &
  \colhead{$0.5$} &
  \colhead{$1.0$} &
  \colhead{$1.5$} &
  \colhead{$2.0$} &
  \colhead{$2.5$} &
  \colhead{$3.0$}
}
\startdata
\multicolumn{9}{c}{$\Delta T_{\rm eff}$ from $g - r$} \nl
\hline
$0.50$ & \nodata & $4.57$ & $ 0.0$ & $ 0.0$ & $  0.0$ & $  0.0$ & $  0.0$ & $  0.0$ \nl
$0.55$ & \nodata & $4.59$ & $ 0.0$ & $ 0.0$ & $  0.0$ & $  0.0$ & $  0.0$ & $  0.0$ \nl
$0.60$ & \nodata & $4.61$ & $ 1.5$ & $ 2.3$ & $  2.4$ & $  2.0$ & $  1.2$ & $ -0.2$ \nl
$0.65$ & \nodata & $4.62$ & $ 5.8$ & $ 8.1$ & $  7.3$ & $  4.5$ & $ -0.2$ & $ -7.3$ \nl
$0.70$ & \nodata & $4.63$ & $11.2$ & $15.3$ & $ 12.9$ & $  6.1$ & $ -4.4$ & $-18.5$ \nl
$0.75$ & \nodata & $4.64$ & $18.5$ & $26.7$ & $ 24.6$ & $ 14.0$ & $ -3.8$ & $-25.8$ \nl
$0.80$ & \nodata & $4.65$ & $28.2$ & $43.2$ & $ 44.3$ & $ 31.6$ & $  7.3$ & $-22.8$ \nl
$0.85$ & \nodata & $4.65$ & $36.2$ & $58.6$ & $ 65.1$ & $ 55.1$ & $ 28.9$ & $ -6.7$ \nl
$0.90$ & \nodata & $4.66$ & $41.2$ & $69.3$ & $ 81.1$ & $ 76.4$ & $ 54.4$ & $ 19.1$ \nl
$0.95$ & \nodata & $4.66$ & $44.6$ & $77.2$ & $ 93.7$ & $ 94.8$ & $ 78.5$ & $ 45.3$ \nl
$1.00$ & \nodata & $4.67$ & $47.1$ & $83.4$ & $103.8$ & $109.3$ & $ 97.9$ & $ 69.8$ \nl
$1.05$ & \nodata & $4.67$ & $50.6$ & $88.9$ & $110.8$ & $118.8$ & $111.4$ & $ 88.2$ \nl
$1.10$ & \nodata & $4.68$ & $51.5$ & $90.4$ & $111.8$ & $119.4$ & $114.0$ & $ 95.5$ \nl
$1.15$ & \nodata & $4.69$ & $53.1$ & $90.2$ & $106.4$ & $109.7$ & $104.0$ & $ 88.4$ \nl
$1.20$ & \nodata & $4.69$ & $48.4$ & $80.1$ & $ 88.0$ & $ 84.0$ & $ 76.0$ & $ 62.0$ \nl
\hline
\multicolumn{9}{c}{$\Delta T_{\rm eff}$ from $g - i$} \nl
\hline
$0.50$ & $0.655$ & $4.57$ & $ 0.0$ & $ 0.0$ & $ 0.0$ & $  0.0$ & $  0.0$ & $  0.0$ \nl
$0.55$ & $0.725$ & $4.59$ & $ 0.0$ & $ 0.0$ & $ 0.0$ & $  0.0$ & $  0.0$ & $  0.0$ \nl
$0.60$ & $0.795$ & $4.61$ & $ 1.2$ & $ 1.9$ & $ 1.8$ & $  1.4$ & $  0.6$ & $ -0.5$ \nl
$0.65$ & $0.865$ & $4.62$ & $ 5.2$ & $ 7.5$ & $ 6.6$ & $  4.0$ & $ -0.2$ & $ -6.0$ \nl
$0.70$ & $0.934$ & $4.63$ & $10.3$ & $14.5$ & $13.2$ & $  8.2$ & $ -0.4$ & $-12.0$ \nl
$0.75$ & $1.003$ & $4.64$ & $16.9$ & $25.2$ & $25.8$ & $ 19.4$ & $  7.0$ & $-10.3$ \nl
$0.80$ & $1.071$ & $4.65$ & $24.7$ & $38.7$ & $43.4$ & $ 37.3$ & $ 22.4$ & $  0.8$ \nl
$0.85$ & $1.138$ & $4.65$ & $31.1$ & $50.8$ & $60.1$ & $ 57.1$ & $ 43.4$ & $ 21.5$ \nl
$0.90$ & $1.205$ & $4.66$ & $34.9$ & $58.6$ & $71.4$ & $ 72.3$ & $ 61.9$ & $ 42.8$ \nl
$0.95$ & $1.272$ & $4.66$ & $37.9$ & $65.1$ & $80.7$ & $ 85.2$ & $ 78.6$ & $ 63.0$ \nl
$1.00$ & $1.341$ & $4.67$ & $41.3$ & $70.9$ & $88.3$ & $ 95.4$ & $ 92.1$ & $ 79.3$ \nl
$1.05$ & $1.411$ & $4.67$ & $43.4$ & $74.3$ & $92.5$ & $101.1$ & $100.6$ & $ 90.4$ \nl
$1.10$ & $1.483$ & $4.68$ & $43.8$ & $75.3$ & $93.6$ & $102.7$ & $103.9$ & $ 95.9$ \nl
$1.15$ & $1.559$ & $4.69$ & $43.1$ & $73.7$ & $90.8$ & $ 98.7$ & $100.4$ & $ 94.2$ \nl
$1.20$ & $1.639$ & $4.69$ & $38.3$ & $66.2$ & $80.6$ & $ 85.9$ & $ 86.7$ & $ 82.0$ \nl
\hline
\multicolumn{9}{c}{$\Delta T_{\rm eff}$ from $g - z$} \nl
\hline 
$0.50$ & $0.708$ & $4.57$ & $ 0.0$ & $ 0.0$ & $ 0.0$ & $ 0.0$ & $ 0.0$ & $ 0.0$ \nl
$0.55$ & $0.795$ & $4.59$ & $ 0.0$ & $ 0.0$ & $ 0.0$ & $ 0.0$ & $ 0.0$ & $ 0.0$ \nl  
$0.60$ & $0.881$ & $4.61$ & $ 0.8$ & $ 1.2$ & $ 1.1$ & $ 0.9$ & $ 0.4$ & $-0.4$ \nl  
$0.65$ & $0.966$ & $4.62$ & $ 3.6$ & $ 5.1$ & $ 4.4$ & $ 2.5$ & $-0.3$ & $-4.4$ \nl  
$0.70$ & $1.050$ & $4.63$ & $ 7.7$ & $11.2$ & $10.1$ & $ 6.2$ & $ 0.5$ & $-7.5$ \nl  
$0.75$ & $1.133$ & $4.64$ & $13.5$ & $20.6$ & $20.7$ & $15.6$ & $ 7.5$ & $-4.2$ \nl  
$0.80$ & $1.215$ & $4.65$ & $20.4$ & $32.5$ & $35.5$ & $30.6$ & $21.0$ & $ 6.4$ \nl  
$0.85$ & $1.294$ & $4.65$ & $25.5$ & $41.7$ & $48.3$ & $46.5$ & $37.4$ & $22.5$ \nl  
$0.90$ & $1.373$ & $4.66$ & $28.1$ & $46.9$ & $56.5$ & $58.3$ & $50.9$ & $37.4$ \nl  
$0.95$ & $1.452$ & $4.66$ & $30.2$ & $51.5$ & $63.8$ & $68.4$ & $63.3$ & $51.8$ \nl  
$1.00$ & $1.533$ & $4.67$ & $33.0$ & $56.5$ & $70.5$ & $76.7$ & $74.0$ & $64.3$ \nl  
$1.05$ & $1.616$ & $4.67$ & $34.8$ & $59.9$ & $74.9$ & $82.2$ & $81.7$ & $73.9$ \nl  
$1.10$ & $1.703$ & $4.68$ & $35.4$ & $61.3$ & $76.7$ & $84.6$ & $85.9$ & $80.0$ \nl  
$1.15$ & $1.794$ & $4.69$ & $34.6$ & $60.0$ & $75.2$ & $83.1$ & $85.4$ & $81.2$ \nl  
$1.20$ & $1.891$ & $4.69$ & $30.3$ & $53.7$ & $67.8$ & $74.9$ & $77.6$ & $75.2$ \nl  
\enddata
\tablecomments{The sense of the difference is that a positive $\Delta T_{\rm eff}$
means a higher $T_{\rm eff}$ at a lower $\log{g}$.}
\tablenotetext{a}{The $\log{g}$ values in the YREC model.}
\end{deluxetable*}

%% file: tab5.tex
\begin{deluxetable*}{lrrrr}
\tabletypesize{\scriptsize}
\tablewidth{0pt}
\tablecaption{Statistical Properties of Clusters Comparisons\label{tab:cluster}}
\tablehead{
  \colhead{} &
  \multicolumn{4}{c}{$\Delta T_{\rm eff}$ (K)} \nl
  \cline{2-5}
  \colhead{Cluster Data} &
  \colhead{$4000 < T_{\rm eff} \leq 7400$} &
  \colhead{$4000 < T_{\rm eff} \leq 6000$} &
  \colhead{$6000 < T_{\rm eff} \leq 7400$} &
  \colhead{$\sigma_{\rm sys}$\tablenotemark{a}}
}
\startdata
\multicolumn{4}{c}{T$_{\rm eff} (B\, -\, V, {\rm IRFM}) -$ T$_{\rm eff} (V\, -\, I_C, {\rm IRFM})$} \nl
\hline
 Hyades                      & $ 10.3\pm 6.7$ & $ 11.5\pm 6.0$ & $ -2.9\pm15.3$ & $4.0$ \\
 Praesepe                    & $ 11.3\pm 7.4$ & $ 23.0\pm 7.4$ & $-59.4\pm15.3$ & $6.7$ \\
 Pleiades                    & $ 54.3\pm12.1$ & $ 67.5\pm13.3$ & $ -5.5\pm19.0$ & $6.5$ \\
 M67 (MMJ93)\tablenotemark{b}& $ 74.1\pm 7.9$ & $ 71.5\pm 8.5$ & $ 93.2\pm21.0$ & $4.3$ \\
 M67 (S04)\tablenotemark{b}  & $ 28.8\pm 4.0$ & $ 16.7\pm 4.2$ & $ 36.1\pm 7.7$ & $5.3$ \\
\hline
\multicolumn{4}{c}{T$_{\rm eff} (V\, -\, K_s, {\rm IRFM}) -$ T$_{\rm eff} (V\, -\, I_C, {\rm IRFM})$} \nl
\hline
 Hyades                      & $ -3.8\pm 7.4$ & $ -4.0\pm 7.6$ & $ -1.9\pm16.7$ & $0.9$ \\
 Praesepe                    & $ 23.2\pm 6.2$ & $ 25.0\pm 6.7$ & $  1.6\pm13.7$ & $1.0$ \\
 Pleiades                    & $ 10.8\pm 5.9$ & $  9.0\pm 6.9$ & $ 19.2\pm13.9$ & $1.1$ \\
 M67 (MMJ93)\tablenotemark{b}& $ 87.3\pm 7.7$ & $ 81.8\pm 8.7$ & $133.8\pm19.6$ & $0.5$ \\
 M67 (S04)\tablenotemark{b}  & $ 34.1\pm 3.9$ & $ 36.7\pm 4.9$ & $ 29.9\pm 6.2$ & $1.2$ \\
\hline
\multicolumn{4}{c}{T$_{\rm eff} (J\, -\, K_s, {\rm IRFM}) -$ T$_{\rm eff} (V\, -\, I_C, {\rm IRFM})$} \nl
\hline
 Hyades                      & $ 94.7\pm17.0$ & $108.3\pm15.7$ & $ 36.0\pm32.9$ & $3.9$ \\
 Praesepe                    & $ 44.2\pm13.1$ & $ 66.2\pm13.7$ & $-97.8\pm34.8$ & $4.0$ \\
 Pleiades                    & $-10.1\pm13.9$ & $ -0.6\pm14.8$ & $-55.3\pm30.7$ & $5.4$ \\
 M67 (MMJ93)\tablenotemark{b}& $ 51.8\pm16.6$ & $ 61.4\pm18.3$ & $  9.1\pm38.6$ & $6.3$ \\
 M67 (S04)\tablenotemark{b}  & $-47.3\pm14.1$ & $-18.7\pm18.5$ & $-86.7\pm21.7$ & $7.3$ \\
\hline
\multicolumn{4}{c}{T$_{\rm eff} (V\, -\, I_C, {\rm IRFM}) -$ T$_{\rm eff} (M_V, {\rm YREC})$} \nl
\hline
 Hyades                      & $-12.1\pm8.7$ & $ -8.3\pm8.3$ & $ -56.4\pm15.4$ & $ 6.4$ \\
 Praesepe                    & $-14.7\pm6.6$ & $-12.2\pm7.4$ & $ -52.9\pm11.9$ & $21.3$ \\
 Pleiades                    & $-14.6\pm8.4$ & $ -9.9\pm8.4$ & $ -50.5\pm20.3$ & $13.9$ \\
 M67 (MMJ93)\tablenotemark{b}& $-52.9\pm8.6$ & $-34.4\pm7.5$ & $-178.4\pm13.7$ & $22.0$ \\
 M67 (S04)\tablenotemark{b}  & $-20.4\pm3.4$ & $-13.5\pm3.5$ & $ -37.1\pm 8.2$ & $24.1$ \\
\hline
\multicolumn{4}{c}{T$_{\rm eff} (J\, -\, K_s, {\rm IRFM}) -$ T$_{\rm eff} (M_V, {\rm YREC})$} \nl
\hline
 Hyades                      & $ 62.2\pm16.6$ & $ 90.9\pm12.5$ & $ -35.1\pm32.5$ & $ 6.9$ \\
 Praesepe                    & $  7.9\pm14.9$ & $ 43.9\pm13.0$ & $-152.5\pm27.5$ & $17.5$ \\
 Pleiades                    & $  1.2\pm15.9$ & $ 22.6\pm17.6$ & $ -92.8\pm26.9$ & $11.4$ \\
 M67 (MMJ93)\tablenotemark{b}& $-50.1\pm15.9$ & $-16.5\pm19.2$ & $-123.8\pm28.4$ & $35.3$ \\
 M67 (S04)\tablenotemark{b}  & $-60.1\pm15.8$ & $-31.9\pm18.8$ & $-127.7\pm29.1$ & $30.4$ \\
\enddata
\tablenotetext{a}{Systematic errors from reddening and metallicity, summed in
quadrature. In the comparisons between IRFM and YREC, we also include effects
of the cluster age and distance modulus errors.}
\tablenotetext{b}{MMJ=\citet{montgomery:93}; S04=\citet{sandquist:04}.}
\end{deluxetable*}

%% file: tab6.tex
\begin{deluxetable*}{cccccccc}
\tabletypesize{\scriptsize}
\tablewidth{0pt}
\tablecaption{Binary Corrections\label{tab:binary}}
\tablehead{
  \colhead{} &
  \colhead{} &
  \colhead{} &
  \colhead{} &
  \multicolumn{4}{c}{$\langle \Delta T_{\rm eff} \rangle$\tablenotemark{a}} \nl
  \cline{5-8}
  \colhead{$g\, -\, r$} &
  \colhead{$\Delta (g\, -\, i)$} &
  \colhead{$\Delta (g\, -\, z)$} &
  \colhead{$\Delta (J\, -\, K_s)$} &
  \colhead{$g\, -\, r$} &
  \colhead{$g\, -\, i$} &
  \colhead{$g\, -\, z$} &
  \colhead{$J\, -\, K_s$} \nl
  \colhead{(mag)} &
  \colhead{(mag)} &
  \colhead{(mag)} &
  \colhead{(mag)} &
  \colhead{(K)} &
  \colhead{(K)} &
  \colhead{(K)} &
  \colhead{(K)}
}
\startdata
\multicolumn{8}{c}{M35 Mass Function\tablenotemark{b}} \nl
\hline
$ 0.25$&$ 0.001$&$ 0.000$&$ 0.006$&$  18$&$  21$&$  23$&$  83$\nl
$ 0.35$&$ 0.000$&$ 0.002$&$ 0.012$&$  25$&$  28$&$  32$&$  89$\nl
$ 0.45$&$ 0.002$&$ 0.004$&$ 0.010$&$  27$&$  32$&$  35$&$  70$\nl
$ 0.55$&$ 0.004$&$ 0.006$&$ 0.010$&$  31$&$  38$&$  43$&$  77$\nl
$ 0.65$&$ 0.002$&$ 0.007$&$ 0.009$&$  32$&$  36$&$  42$&$  93$\nl
$ 0.75$&$ 0.002$&$ 0.010$&$ 0.014$&$  30$&$  34$&$  40$&$ 121$\nl
$ 0.85$&$ 0.002$&$ 0.012$&$ 0.008$&$  25$&$  31$&$  37$&$  97$\nl
$ 0.95$&$ 0.004$&$ 0.014$&$ 0.011$&$  22$&$  29$&$  35$&$  71$\nl
$ 1.05$&$ 0.006$&$ 0.014$&$ 0.007$&$  18$&$  26$&$  31$&$  35$\nl
\hline
\multicolumn{8}{c}{Flat Mass Function\tablenotemark{b}} \nl
\hline
$ 0.25$&$ 0.002$&$ 0.002$&$ 0.008$&$  34$&$  38$&$  42$&$  98$\nl
$ 0.35$&$ 0.001$&$ 0.003$&$ 0.011$&$  40$&$  43$&$  48$&$ 106$\nl
$ 0.45$&$ 0.001$&$ 0.004$&$ 0.012$&$  42$&$  45$&$  51$&$ 104$\nl
$ 0.55$&$ 0.003$&$ 0.006$&$ 0.012$&$  45$&$  51$&$  56$&$  93$\nl
$ 0.65$&$ 0.002$&$ 0.006$&$ 0.011$&$  44$&$  48$&$  53$&$ 115$\nl
$ 0.75$&$ 0.001$&$ 0.009$&$ 0.011$&$  39$&$  42$&$  50$&$ 129$\nl
$ 0.85$&$ 0.002$&$ 0.010$&$ 0.009$&$  34$&$  39$&$  45$&$ 104$\nl
$ 0.95$&$ 0.002$&$ 0.011$&$ 0.015$&$  28$&$  33$&$  39$&$  91$\nl
$ 1.05$&$ 0.006$&$ 0.016$&$ 0.013$&$  21$&$  30$&$  35$&$  54$\nl
\hline
\multicolumn{8}{c}{Salpeter Mass Function\tablenotemark{b}}\nl
\hline
$ 0.25$&$ 0.001$&$ 0.002$&$ 0.004$&$   7$&$   9$&$  10$&$  50$\nl
$ 0.35$&$ 0.000$&$ 0.002$&$ 0.004$&$  11$&$  12$&$  15$&$  49$\nl
$ 0.45$&$ 0.001$&$ 0.005$&$ 0.008$&$  12$&$  15$&$  18$&$  43$\nl
$ 0.55$&$ 0.002$&$ 0.005$&$ 0.007$&$  12$&$  16$&$  19$&$  43$\nl
$ 0.65$&$ 0.003$&$ 0.007$&$ 0.012$&$  13$&$  18$&$  22$&$  78$\nl
$ 0.75$&$ 0.003$&$ 0.010$&$ 0.007$&$  13$&$  18$&$  24$&$  73$\nl
$ 0.85$&$ 0.003$&$ 0.011$&$ 0.009$&$  11$&$  17$&$  23$&$  74$\nl
$ 0.95$&$ 0.004$&$ 0.013$&$ 0.007$&$  10$&$  17$&$  23$&$  51$\nl
$ 1.05$&$ 0.007$&$ 0.014$&$ 0.009$&$  10$&$  18$&$  23$&$  21$\nl
\enddata
\tablecomments{The sense of the bias is that populations mixed with unresolved
binaries look redder (cooler) at a given $g\, -\, r$ in the above color indices.}
\tablenotetext{a}{Mean difference in $T_{\rm eff}$ between
a population with a $50\%$ unresolved binary fraction and that of primaries alone.
The sense is that unresolved binary stars have lower temperatures than
expected from primaries alone.}
\tablenotetext{b}{Mass function for secondary components in the binary system.
All simulation results are based on a $50\%$ unresolved binary fraction.}
\end{deluxetable*}

%% file: tab7.tex
\begin{deluxetable*}{rccccccccccccc}
\tablewidth{0pt}
\tabletypesize{\scriptsize}
\tablecaption{Catalog with Revised $T_{\rm eff}$\label{tab:main}}
\tablehead{
  \colhead{} &
  \multicolumn{3}{c}{SDSS} &
  \colhead{} &
  \multicolumn{3}{c}{IRFM\tablenotemark{a}} &
  \colhead{} &
  \multicolumn{3}{c}{KIC} &
  \colhead{} &
  \colhead{} \nl
  \cline{2-4}
  \cline{6-8}
  \cline{10-12}
  \colhead{\tt KIC ID} & 
  \colhead{\tt $T_{\rm eff}$} &
  \colhead{\tt $\sigma_{\rm tot}$} &
  \colhead{\tt $\sigma_{\rm ran}$} &
  \colhead{} &
  \colhead{\tt $T_{\rm eff}$} &
  \colhead{\tt $\sigma_{\rm tot}$} &
  \colhead{\tt $\sigma_{\rm ran}$} &
  \colhead{} &
  \colhead{\tt $T_{\rm eff}$} &
  \colhead{\tt $\log{g}$} &
  \colhead{\tt [Fe/H]} &
  \colhead{\tt $\Delta T_{\rm eff}$\tablenotemark{b}} &
  \colhead{\tt flag\tablenotemark{c}} \nl
  \colhead{} &
  \colhead{(K)} &
  \colhead{(K)} &
  \colhead{(K)} &
  \colhead{} &
  \colhead{(K)} &
  \colhead{(K)} &
  \colhead{(K)} &
  \colhead{} &
  \colhead{(K)} &
  \colhead{(dex)} &
  \colhead{(dex)} &
  \colhead{(K)} &
  \colhead{}
}
\startdata
757076 & $5137$ & $ 85$ & $55$ && $5150$ & $ 98$ & $ 94$ && $5174$ & $3.60$ & $-0.08$ & $ 0$ & $0$ \nl
757099 & $5523$ & $ 97$ & $34$ && $5270$ & $110$ & $101$ && $5589$ & $3.82$ & $-0.21$ & $ 0$ & $0$ \nl
757137 & $4822$ & $ 74$ & $42$ && $4536$ & $101$ & $ 99$ && $4879$ & $2.58$ & $-0.08$ & $49$ & $0$ \nl
757218 & $4728$ & $ 79$ & $17$ && $4489$ & $ 90$ & $ 75$ && $4555$ & $2.28$ & $-0.12$ & $67$ & $0$ \nl
757231 & $4909$ & $116$ & $64$ && $4974$ & $111$ & $ 89$ && $4825$ & $2.60$ & $-0.08$ & $24$ & $0$ \nl
\enddata
\tablecomments{Only a portion of this table is shown here to demonstrate
its form and content. A machine-readable version of the full table is available.}
\tablecomments{Effective temperatures presented here were computed at a fixed [Fe/H]$=-0.2$.}
\tablenotetext{a}{$T_{\rm eff}$ estimates based on $J\, -\, K_s$ using the original
formula in C10.}
\tablenotetext{b}{$T_{\rm eff}$ correction for giants. The sense is that
this correction factor has been subtracted from the SDSS $T_{\rm eff}$ estimate
in the above table.}
\tablenotetext{c}{Quality flag indicating stars with unusually discrepant
SDSS $T_{\rm eff}$ estimates (see text).}
\end{deluxetable*}

%% file: tab8.tex
\begin{deluxetable*}{ccrrrrrrrrrrrrrrrrrrr}
\tablewidth{0pt}
\tabletypesize{\scriptsize}
\tablecaption{Statistical Properties of $T_{\rm eff}$\label{tab:stat}}
\tablehead{
\colhead{$\langle T_{\rm eff} \rangle$} &
\colhead{} &
\colhead{} &
\multicolumn{3}{c}{IRFM $-$ KIC\tablenotemark{a}} &
\colhead{} &
\multicolumn{3}{c}{SDSS $-$ KIC\tablenotemark{a}} &
\colhead{} &
\multicolumn{3}{c}{SDSS $-$ IRFM\tablenotemark{a}} &
\colhead{} &
\multicolumn{3}{c}{$T_{\rm eff}$(color) $-$ $T_{\rm eff}$($griz$)} &
\colhead{} &
\multicolumn{2}{c}{SDSS} \nl
\cline{4-6} \cline{8-10} \cline{12-14} \cline{16-18} \cline{20-21}
\colhead{(KIC)} &
\colhead{$\langle (g\, -\, r)_0 \rangle$} &
\colhead{${\rm N_{stars}}$} &
\colhead{$\Delta T_{\rm eff}$} &
\colhead{$\sigma$} &
\colhead{$\sigma_{\rm prop}$} &
\colhead{} &
\colhead{$\Delta T_{\rm eff}$} &
\colhead{$\sigma$} &
\colhead{$\sigma_{\rm prop}$} &
\colhead{} &
\colhead{$\Delta T_{\rm eff}$} &
\colhead{$\sigma$} &
\colhead{$\sigma_{\rm prop}$} &
\colhead{} &
\colhead{$g\, -\, r$} &
\colhead{$g\, -\, i$} &
\colhead{$g\, -\, z$} &
\colhead{} &
\colhead{$\sigma_{griz}$\tablenotemark{b}} &
\colhead{$\sigma_{\rm prop}$\tablenotemark{c}}
} 
\startdata
\multicolumn{21}{c}{dwarfs (KIC $\log{g} > 3.5$)} \nl
\hline
$6597$ & $0.13$ & $ 1032$ & $165$ & $182$ & $184$ && $223$ & $41$ & $46$ && $  54$ & $172$ & $191$ && $ -9$ & $-2$ & $14$ && $33$ & $43$ \nl
$6501$ & $0.15$ & $ 1480$ & $167$ & $172$ & $180$ && $224$ & $36$ & $46$ && $  50$ & $170$ & $188$ && $-11$ & $-1$ & $13$ && $33$ & $43$ \nl
$6393$ & $0.18$ & $ 2156$ & $180$ & $182$ & $184$ && $228$ & $35$ & $46$ && $  40$ & $176$ & $191$ && $-15$ & $-2$ & $19$ && $34$ & $42$ \nl
$6296$ & $0.21$ & $ 3029$ & $190$ & $185$ & $188$ && $231$ & $34$ & $46$ && $  34$ & $181$ & $195$ && $-18$ & $-2$ & $23$ && $36$ & $41$ \nl
$6201$ & $0.24$ & $ 4239$ & $211$ & $183$ & $186$ && $237$ & $32$ & $45$ && $  23$ & $181$ & $194$ && $-20$ & $-1$ & $24$ && $37$ & $40$ \nl
$6095$ & $0.27$ & $ 6551$ & $209$ & $193$ & $195$ && $242$ & $30$ & $45$ && $  32$ & $195$ & $202$ && $-24$ & $-1$ & $28$ && $39$ & $39$ \nl
$5995$ & $0.30$ & $ 8154$ & $203$ & $199$ & $197$ && $250$ & $30$ & $45$ && $  42$ & $201$ & $205$ && $-28$ & $-1$ & $29$ && $41$ & $37$ \nl
$5899$ & $0.33$ & $ 9685$ & $220$ & $192$ & $194$ && $258$ & $31$ & $45$ && $  34$ & $197$ & $202$ && $-31$ & $ 0$ & $30$ && $43$ & $36$ \nl
$5802$ & $0.36$ & $11632$ & $225$ & $194$ & $193$ && $266$ & $32$ & $44$ && $  35$ & $199$ & $202$ && $-36$ & $ 1$ & $31$ && $44$ & $36$ \nl
$5697$ & $0.39$ & $12398$ & $235$ & $191$ & $192$ && $265$ & $42$ & $45$ && $  20$ & $201$ & $201$ && $-41$ & $ 2$ & $37$ && $44$ & $35$ \nl
$5596$ & $0.43$ & $11492$ & $240$ & $188$ & $187$ && $244$ & $44$ & $43$ && $  -6$ & $200$ & $196$ && $-41$ & $ 3$ & $33$ && $42$ & $34$ \nl
$5502$ & $0.46$ & $ 9946$ & $234$ & $180$ & $181$ && $227$ & $45$ & $41$ && $ -20$ & $194$ & $190$ && $-40$ & $ 3$ & $31$ && $42$ & $32$ \nl
$5400$ & $0.49$ & $ 8914$ & $230$ & $176$ & $176$ && $216$ & $44$ & $39$ && $ -26$ & $189$ & $184$ && $-35$ & $ 2$ & $27$ && $38$ & $30$ \nl
$5302$ & $0.53$ & $ 7370$ & $214$ & $170$ & $168$ && $206$ & $47$ & $37$ && $ -23$ & $183$ & $176$ && $-34$ & $ 2$ & $27$ && $38$ & $29$ \nl
$5201$ & $0.57$ & $ 6119$ & $195$ & $157$ & $158$ && $203$ & $49$ & $35$ && $  -9$ & $173$ & $166$ && $-30$ & $ 1$ & $25$ && $37$ & $27$ \nl
$5099$ & $0.60$ & $ 6112$ & $184$ & $149$ & $149$ && $201$ & $50$ & $34$ && $   1$ & $166$ & $156$ && $-26$ & $ 1$ & $25$ && $36$ & $25$ \nl
$5002$ & $0.65$ & $ 4619$ & $177$ & $134$ & $140$ && $192$ & $53$ & $32$ && $  -3$ & $156$ & $147$ && $-29$ & $ 1$ & $26$ && $36$ & $22$ \nl
$4901$ & $0.70$ & $ 3587$ & $180$ & $129$ & $135$ && $187$ & $54$ & $30$ && $ -11$ & $149$ & $142$ && $-24$ & $ 1$ & $26$ && $35$ & $20$ \nl
$4804$ & $0.75$ & $ 2829$ & $178$ & $122$ & $131$ && $177$ & $59$ & $29$ && $ -23$ & $145$ & $138$ && $-22$ & $ 0$ & $26$ && $34$ & $18$ \nl
$4703$ & $0.81$ & $ 1887$ & $179$ & $120$ & $128$ && $162$ & $57$ & $26$ && $ -39$ & $137$ & $134$ && $-16$ & $ 0$ & $25$ && $31$ & $16$ \nl
$4605$ & $0.86$ & $ 1384$ & $182$ & $123$ & $126$ && $161$ & $61$ & $23$ && $ -36$ & $140$ & $130$ && $ -7$ & $-2$ & $19$ && $27$ & $15$ \nl
$4498$ & $0.91$ & $  809$ & $201$ & $118$ & $123$ && $174$ & $52$ & $20$ && $ -26$ & $127$ & $127$ && $  5$ & $-4$ & $11$ && $23$ & $14$ \nl
$4396$ & $0.98$ & $ 1258$ & $223$ & $115$ & $118$ && $179$ & $41$ & $18$ && $ -45$ & $118$ & $121$ && $  7$ & $-4$ & $ 9$ && $20$ & $13$ \nl
$4302$ & $1.06$ & $ 1421$ & $237$ & $109$ & $110$ && $176$ & $34$ & $17$ && $ -69$ & $102$ & $113$ && $  7$ & $-3$ & $11$ && $18$ & $12$ \nl
$4200$ & $1.14$ & $ 1157$ & $257$ & $ 97$ & $104$ && $169$ & $28$ & $16$ && $ -95$ & $ 94$ & $107$ && $ 11$ & $-3$ & $10$ && $16$ & $11$ \nl
$4099$ & $1.21$ & $ 1022$ & $279$ & $ 74$ & $ 97$ && $150$ & $19$ & $17$ && $-134$ & $ 73$ & $101$ && $ 26$ & $-3$ & $10$ && $19$ & $11$ \nl

\hline
\multicolumn{21}{c}{giants (KIC $\log{g} \leq 3.5$)} \nl
\hline
$5292$ & $0.51$ & $  35$ & $246$ & $204$ & $122$ && $216$ & $ 38$ & $35$ && $-62$ & $198$ & $132$ && $ -5$ & $-4$ & $29$ && $34$ & $29$ \nl
$5184$ & $0.56$ & $ 175$ & $167$ & $111$ & $112$ && $214$ & $ 41$ & $36$ && $ 30$ & $132$ & $121$ && $-30$ & $-2$ & $37$ && $39$ & $27$ \nl
$5086$ & $0.60$ & $ 676$ & $159$ & $100$ & $108$ && $216$ & $ 43$ & $34$ && $ 38$ & $102$ & $117$ && $-30$ & $-1$ & $33$ && $37$ & $25$ \nl
$4995$ & $0.65$ & $2098$ & $135$ & $ 99$ & $105$ && $215$ & $ 42$ & $33$ && $ 58$ & $105$ & $114$ && $-34$ & $ 1$ & $32$ && $38$ & $23$ \nl
$4897$ & $0.69$ & $3376$ & $129$ & $ 96$ & $101$ && $220$ & $ 41$ & $32$ && $ 68$ & $101$ & $110$ && $-33$ & $ 2$ & $29$ && $37$ & $21$ \nl
$4800$ & $0.74$ & $4316$ & $124$ & $ 91$ & $ 98$ && $225$ & $ 40$ & $30$ && $ 76$ & $ 96$ & $105$ && $-30$ & $ 3$ & $26$ && $36$ & $19$ \nl
$4702$ & $0.79$ & $3435$ & $118$ & $ 91$ & $ 94$ && $236$ & $ 39$ & $28$ && $ 93$ & $ 95$ & $101$ && $-26$ & $ 4$ & $22$ && $34$ & $17$ \nl
$4599$ & $0.85$ & $3002$ & $110$ & $ 95$ & $ 91$ && $254$ & $ 35$ & $23$ && $124$ & $100$ & $ 96$ && $-18$ & $ 5$ & $16$ && $28$ & $16$ \nl
$4509$ & $0.91$ & $1148$ & $ 71$ & $106$ & $ 87$ && $261$ & $ 32$ & $20$ && $174$ & $112$ & $ 91$ && $-12$ & $ 5$ & $11$ && $23$ & $14$ \nl
$4401$ & $0.97$ & $ 930$ & $ 58$ & $ 97$ & $ 84$ && $294$ & $ 27$ & $17$ && $227$ & $103$ & $ 87$ && $ -3$ & $ 4$ & $ 3$ && $18$ & $13$ \nl
$4307$ & $1.03$ & $ 861$ & $ 64$ & $ 80$ & $ 81$ && $313$ & $ 27$ & $16$ && $239$ & $ 86$ & $ 84$ && $  3$ & $ 4$ & $-2$ && $16$ & $12$ \nl
$4202$ & $1.10$ & $ 665$ & $ 97$ & $ 53$ & $ 79$ && $313$ & $ 30$ & $14$ && $208$ & $ 58$ & $ 81$ && $ 12$ & $ 3$ & $-5$ && $13$ & $12$ \nl
$4105$ & $1.20$ & $ 631$ & $169$ & $ 29$ & $ 80$ && $321$ & $148$ & $14$ && $129$ & $ 87$ & $ 83$ && $-64$ & $98$ & $-5$ && $15$ & $11$ \nl
\enddata
\tablecomments{Statistical properties derived from the full long-cadence sample, after applying
the hot $T_{\rm eff}$ corrections. No metallicity and binary corrections were applied.}
\tablenotetext{a}{Weighted mean difference ($T_{\rm eff}$), weighted standard deviation ($\sigma$), and
the expected dispersion propagated from random errors ($\sigma_{\rm prop}$).}
\tablenotetext{b}{Median standard deviation of $griz$-based temperature estimates from
$g\, -\, r$, $g\, -\, i$, and $g\, -\, z$.}
\tablenotetext{c}{Median dispersion expected from photometric errors in $griz$.}
\end{deluxetable*}

%% file: tab9.tex
\begin{deluxetable*}{rrrrrrrrrrrrrrrr}
\tablewidth{0pt}
\tabletypesize{\scriptsize}
\tablecaption{Tycho-2MASS-based IRFM $T_{\rm eff}$\label{tab:irfm}}
\tablehead{
  \colhead{} &
  \multicolumn{3}{c}{$T_{\rm eff} (V_T - J)$} &
  \colhead{} &
  \multicolumn{3}{c}{$T_{\rm eff} (V_T - H)$} &
  \colhead{} &
  \multicolumn{3}{c}{$T_{\rm eff} (V_T - K_s)$} &
  \colhead{} &
  \multicolumn{3}{c}{$T_{\rm eff} (J - K_s)$} \nl
  \cline{2-4}
  \cline{6-8}
  \cline{10-12}
  \cline{14-16}
  \colhead{KIC\_ID} &
  \colhead{$T_{\rm eff}$} &
  \colhead{$\sigma_{\rm tot}$} &
  \colhead{$\sigma_{\rm ran}$} &
  \colhead{} &
  \colhead{$T_{\rm eff}$} &
  \colhead{$\sigma_{\rm tot}$} &
  \colhead{$\sigma_{\rm ran}$} &
  \colhead{} &
  \colhead{$T_{\rm eff}$} &
  \colhead{$\sigma_{\rm tot}$} &
  \colhead{$\sigma_{\rm ran}$} &
  \colhead{} &
  \colhead{$T_{\rm eff}$} &
  \colhead{$\sigma_{\rm tot}$} &
  \colhead{$\sigma_{\rm ran}$}
}
\startdata
1026309 & $4684$  & $102 $  & $  84 $ && $4607 $ & $  90 $ & $  72 $ && $ 4623$ & $ 79 $  & $  60 $ && $4469$ & $103$ & $ 96$ \nl
1160789 & $4997$  & $ 59 $  & $  48 $ && $4915 $ & $  47 $ & $  37 $ && $ 4951$ & $ 44 $  & $  33 $ && $4866$ & $122$ & $120$ \nl
1717271 & $4346$  & $ 63 $  & $  24 $ && $4279 $ & $  65 $ & $  27 $ && $ 4327$ & $ 57 $  & $  20 $ && $4267$ & $105$ & $ 95$ \nl
1718046 & $4811$  & $106 $  & $  81 $ && $4708 $ & $  86 $ & $  62 $ && $ 4754$ & $ 82 $  & $  57 $ && $4625$ & $112$ & $105$ \nl
1718401 & \nodata & \nodata & \nodata && \nodata & \nodata & \nodata && \nodata & \nodata & \nodata && $6531$ & $164$ & $161$ \nl
\enddata
\tablecomments{Only a portion of this table is shown here to demonstrate its form and content. A machine-readable version of the full table is available.}
\tablecomments{Effective temperatures presented here were computed at a fixed [Fe/H]$=-0.2$.}
\end{deluxetable*}

%% file: taba4.tex
\begin{deluxetable}{cccrrrrrr}
\tabletypesize{\scriptsize}
\tablewidth{0pt}
\tablecaption{Gravity Corrections\label{taba:logg}}
\tablehead{
  \colhead{} &
  \colhead{} &
  \colhead{$\log{g}$\tablenotemark{a}} &
  \multicolumn{6}{c}{$\log{g} {\rm (star)} - \log{g} {\rm (YREC)}$} \nl
  \cline{4-9}
  \colhead{$g - r$} &
  \colhead{$g - i$/$g - z$} &
  \colhead{(YREC)} &
  \colhead{$-0.5$} &
  \colhead{$-1.0$} &
  \colhead{$-1.5$} &
  \colhead{$-2.0$} &
  \colhead{$-2.5$} &
  \colhead{$-3.0$}
}
\startdata
\multicolumn{9}{c}{$\Delta T_{\rm eff}$ when $T_{\rm eff}$ are estimated from $g - r$} \nl
\hline
$0.500$ & \nodata & $4.57$ & $   0.0$ & $    0.0$ & $    0.0$ & $    0.0$ & $    0.0$ & $    0.0$ \nl 
$0.550$ & \nodata & $4.59$ & $   0.0$ & $    0.0$ & $    0.0$ & $    0.0$ & $    0.0$ & $    0.0$ \nl 
$0.600$ & \nodata & $4.61$ & $  -1.5$ & $   -2.2$ & $   -2.3$ & $   -2.0$ & $   -1.2$ & $    0.2$ \nl 
$0.650$ & \nodata & $4.62$ & $  -5.8$ & $   -8.1$ & $   -7.3$ & $   -4.5$ & $    0.2$ & $    7.2$ \nl 
$0.700$ & \nodata & $4.63$ & $ -11.2$ & $  -15.3$ & $  -12.9$ & $   -6.1$ & $    4.4$ & $   18.5$ \nl 
$0.750$ & \nodata & $4.64$ & $ -18.4$ & $  -26.6$ & $  -24.5$ & $  -13.8$ & $    3.8$ & $   25.7$ \nl 
$0.800$ & \nodata & $4.65$ & $ -28.2$ & $  -43.2$ & $  -44.3$ & $  -31.6$ & $   -7.3$ & $   22.8$ \nl 
$0.850$ & \nodata & $4.65$ & $ -36.2$ & $  -58.6$ & $  -65.1$ & $  -55.1$ & $  -29.0$ & $    6.6$ \nl 
$0.900$ & \nodata & $4.66$ & $ -41.2$ & $  -69.3$ & $  -81.0$ & $  -76.3$ & $  -54.2$ & $  -18.9$ \nl 
$0.950$ & \nodata & $4.66$ & $ -44.6$ & $  -77.2$ & $  -93.6$ & $  -94.7$ & $  -78.4$ & $  -45.1$ \nl 
$1.000$ & \nodata & $4.67$ & $ -47.1$ & $  -83.4$ & $ -103.8$ & $ -109.3$ & $  -97.9$ & $  -69.8$ \nl 
$1.050$ & \nodata & $4.67$ & $ -50.6$ & $  -88.9$ & $ -110.8$ & $ -118.8$ & $ -111.4$ & $  -88.2$ \nl 
$1.100$ & \nodata & $4.68$ & $ -51.5$ & $  -90.4$ & $ -111.8$ & $ -119.4$ & $ -114.0$ & $  -95.5$ \nl 
$1.150$ & \nodata & $4.69$ & $ -53.1$ & $  -90.2$ & $ -106.5$ & $ -109.9$ & $ -104.2$ & $  -88.6$ \nl 
$1.200$ & \nodata & $4.69$ & $ -48.4$ & $  -80.0$ & $  -87.8$ & $  -83.8$ & $  -75.8$ & $  -61.8$ \nl 
$1.250$ & \nodata & $4.70$ & $ -38.4$ & $  -56.8$ & $  -55.0$ & $  -44.8$ & $  -32.7$ & $  -18.7$ \nl
$1.300$ & \nodata & $4.72$ & $ -17.5$ & $  -25.9$ & $  -21.6$ & $  -13.0$ & $   -2.7$ & $    7.8$ \nl 
$1.350$ & \nodata & $4.73$ & $   0.0$ & $    0.0$ & $    0.0$ & $    0.0$ & $    0.0$ & $    0.0$ \nl
\hline
\multicolumn{9}{c}{$\Delta T_{\rm eff}$ from $g - i$} \nl
\hline
$0.500$ & $0.655$ & $4.57$ & $  0.0$ & $    0.0$ & $    0.0$ & $    0.0$ & $    0.0$ & $    0.0$ \nl
$0.550$ & $0.725$ & $4.59$ & $  0.0$ & $    0.0$ & $    0.0$ & $    0.0$ & $    0.0$ & $    0.0$ \nl
$0.600$ & $0.795$ & $4.61$ & $ -1.2$ & $   -1.8$ & $   -1.8$ & $   -1.4$ & $   -0.6$ & $    0.5$ \nl
$0.650$ & $0.865$ & $4.62$ & $ -5.2$ & $   -7.5$ & $   -6.5$ & $   -4.0$ & $    0.2$ & $    5.9$ \nl
$0.700$ & $0.934$ & $4.63$ & $-10.3$ & $  -14.6$ & $  -13.3$ & $   -8.3$ & $    0.3$ & $   12.0$ \nl
$0.750$ & $1.003$ & $4.64$ & $-16.9$ & $  -25.2$ & $  -25.8$ & $  -19.4$ & $   -7.0$ & $   10.3$ \nl
$0.800$ & $1.071$ & $4.65$ & $-24.7$ & $  -38.7$ & $  -43.5$ & $  -37.4$ & $  -22.5$ & $   -0.9$ \nl
$0.850$ & $1.138$ & $4.65$ & $-31.1$ & $  -50.8$ & $  -60.1$ & $  -57.1$ & $  -43.4$ & $  -21.5$ \nl
$0.900$ & $1.205$ & $4.66$ & $-34.9$ & $  -58.6$ & $  -71.4$ & $  -72.3$ & $  -61.9$ & $  -42.8$ \nl
$0.950$ & $1.272$ & $4.66$ & $-37.9$ & $  -65.1$ & $  -80.7$ & $  -85.2$ & $  -78.6$ & $  -63.0$ \nl
$1.000$ & $1.341$ & $4.67$ & $-41.3$ & $  -70.9$ & $  -88.3$ & $  -95.3$ & $  -92.0$ & $  -79.2$ \nl
$1.050$ & $1.411$ & $4.67$ & $-43.4$ & $  -74.3$ & $  -92.5$ & $ -101.1$ & $ -100.6$ & $  -90.4$ \nl
$1.100$ & $1.483$ & $4.68$ & $-43.8$ & $  -75.3$ & $  -93.6$ & $ -102.7$ & $ -103.9$ & $  -95.9$ \nl
$1.150$ & $1.559$ & $4.69$ & $-43.1$ & $  -73.7$ & $  -90.8$ & $  -98.7$ & $ -100.4$ & $  -94.2$ \nl
$1.200$ & $1.639$ & $4.69$ & $-38.3$ & $  -66.2$ & $  -80.6$ & $  -85.9$ & $  -86.7$ & $  -82.0$ \nl
$1.250$ & $1.724$ & $4.70$ & $-29.8$ & $  -52.7$ & $  -63.0$ & $  -63.2$ & $  -62.4$ & $  -58.5$ \nl
$1.300$ & $1.816$ & $4.72$ & $-16.1$ & $  -31.1$ & $  -38.2$ & $  -34.7$ & $  -34.0$ & $  -30.9$ \nl
$1.350$ & $1.920$ & $4.73$ & $-2.6 $ & $  -6.0 $ & $  -7.6 $ & $  -6.1 $ & $  -6.0 $ & $  -5.2 $ \nl
\hline
\multicolumn{9}{c}{$\Delta T_{\rm eff}$ from $g - z$} \nl
\hline 
$0.500$ & $0.708$ & $4.57$ & $  0.0$ & $    0.0$ & $    0.0$ & $    0.0$ & $    0.0$ & $    0.0$ \nl
$0.550$ & $0.795$ & $4.59$ & $  0.0$ & $    0.0$ & $    0.0$ & $    0.0$ & $    0.0$ & $    0.0$ \nl
$0.600$ & $0.881$ & $4.61$ & $ -0.8$ & $   -1.2$ & $   -1.1$ & $   -0.9$ & $   -0.4$ & $    0.4$ \nl
$0.650$ & $0.966$ & $4.62$ & $ -3.6$ & $   -5.1$ & $   -4.4$ & $   -2.5$ & $    0.3$ & $    4.4$ \nl
$0.700$ & $1.050$ & $4.63$ & $ -7.7$ & $  -11.2$ & $  -10.1$ & $   -6.3$ & $   -0.6$ & $    7.5$ \nl
$0.750$ & $1.133$ & $4.64$ & $-13.5$ & $  -20.6$ & $  -20.7$ & $  -15.6$ & $   -7.5$ & $    4.2$ \nl
$0.800$ & $1.215$ & $4.65$ & $-20.4$ & $  -32.5$ & $  -35.5$ & $  -30.7$ & $  -21.1$ & $   -6.5$ \nl
$0.850$ & $1.294$ & $4.65$ & $-25.5$ & $  -41.7$ & $  -48.3$ & $  -46.4$ & $  -37.3$ & $  -22.4$ \nl
$0.900$ & $1.373$ & $4.66$ & $-28.1$ & $  -47.0$ & $  -56.6$ & $  -58.3$ & $  -50.9$ & $  -37.5$ \nl
$0.950$ & $1.452$ & $4.66$ & $-30.2$ & $  -51.6$ & $  -63.9$ & $  -68.4$ & $  -63.3$ & $  -51.8$ \nl
$1.000$ & $1.533$ & $4.67$ & $-33.0$ & $  -56.5$ & $  -70.5$ & $  -76.7$ & $  -74.0$ & $  -64.3$ \nl
$1.050$ & $1.616$ & $4.67$ & $-34.8$ & $  -59.9$ & $  -74.9$ & $  -82.2$ & $  -81.7$ & $  -73.9$ \nl
$1.100$ & $1.703$ & $4.68$ & $-35.4$ & $  -61.3$ & $  -76.7$ & $  -84.6$ & $  -85.9$ & $  -80.0$ \nl
$1.150$ & $1.794$ & $4.69$ & $-34.6$ & $  -60.1$ & $  -75.3$ & $  -83.2$ & $  -85.5$ & $  -81.3$ \nl
$1.200$ & $1.891$ & $4.69$ & $-30.3$ & $  -53.7$ & $  -67.8$ & $  -74.9$ & $  -77.6$ & $  -75.2$ \nl
$1.250$ & $1.995$ & $4.70$ & $-22.7$ & $  -42.1$ & $  -54.2$ & $  -59.5$ & $  -61.8$ & $  -61.0$ \nl
$1.300$ & $2.110$ & $4.72$ & $-13.0$ & $  -25.9$ & $  -32.0$ & $  -40.3$ & $  -42.2$ & $  -41.6$ \nl
$1.350$ & $2.241$ & $4.73$ & $-2.6 $ & $  -5.3 $ & $  -6.1 $ & $  -8.9 $ & $  -9.3 $ & $  -9.2 $ \nl
\enddata
\tablecomments{In the original version of Table~4, gravity corrections had sign flip errors.
$\Delta T_{\rm eff}$ values in Table~4 should be added to dwarf-based $T_{\rm eff}$
estimates, if one wishes to infer $T_{\rm eff}$ for giants or subgiants.}
\tablenotetext{a}{The $\log{g}$ values in the base isochrone (Table~1).}
\end{deluxetable}

%% file: taba7.tex
\begin{deluxetable}{rccccccccccccc}
\tablewidth{0pt}
\tabletypesize{\scriptsize}
\tablecaption{Catalog with Revised $T_{\rm eff}$\label{taba:main}}
\tablehead{
  \colhead{} &
  \multicolumn{3}{c}{SDSS} &
  \colhead{} &
  \multicolumn{3}{c}{IRFM\tablenotemark{a}} &
  \colhead{} &
  \multicolumn{3}{c}{KIC} &
  \colhead{} &
  \colhead{} \nl
  \cline{2-4}
  \cline{6-8}
  \cline{10-12}
  \colhead{KIC\_ID} & 
  \colhead{$T_{\rm eff}$} &
  \colhead{$\sigma_{\rm tot}$} &
  \colhead{$\sigma_{\rm ran}$} &
  \colhead{} &
  \colhead{$T_{\rm eff}$} &
  \colhead{$\sigma_{\rm tot}$} &
  \colhead{$\sigma_{\rm ran}$} &
  \colhead{} &
  \colhead{$T_{\rm eff}$} &
  \colhead{$\log{g}$} &
  \colhead{[Fe/H]} &
  \colhead{$\Delta T_{\rm eff}$\tablenotemark{b}} &
  \colhead{Flag\tablenotemark{c}} \nl
  \colhead{} &
  \colhead{(K)} &
  \colhead{(K)} &
  \colhead{(K)} &
  \colhead{} &
  \colhead{(K)} &
  \colhead{(K)} &
  \colhead{(K)} &
  \colhead{} &
  \colhead{(K)} &
  \colhead{(dex)} &
  \colhead{(dex)} &
  \colhead{(K)} &
  \colhead{}
}
\startdata
757076 & $5137$ & $ 85$ & $55$ && $5150$ & $ 98$ & $ 94$ && $5174$ & $3.60$ & $-0.08$ & $  0$ & $0$ \nl
757099 & $5523$ & $ 97$ & $34$ && $5270$ & $110$ & $101$ && $5589$ & $3.82$ & $-0.21$ & $  0$ & $0$ \nl
757137 & $4724$ & $ 74$ & $42$ && $4536$ & $101$ & $ 99$ && $4879$ & $2.58$ & $-0.08$ & $-49$ & $0$ \nl
757218 & $4594$ & $ 79$ & $17$ && $4489$ & $ 90$ & $ 75$ && $4555$ & $2.28$ & $-0.12$ & $-67$ & $0$ \nl
757231 & $4861$ & $116$ & $64$ && $4974$ & $111$ & $ 89$ && $4825$ & $2.60$ & $-0.08$ & $-24$ & $0$ \nl
\enddata
\tablecomments{Only a portion of this table is shown here to demonstrate its
form and content. A machine-readable version of the full table is available.
In the original version of Table~7, there were sign flip errors
in the gravity corrections ($\Delta T_{\rm eff}$), which affect SDSS $T_{\rm
eff}$ for giants ($\log{g} \leq 3.5$) in the above table. Other columns are
unaffected. Effective temperatures presented here were computed at a fixed
[Fe/H]$=-0.2$.}
\tablenotetext{a}{IRFM $T_{\rm eff}$ estimates based on $J\, -\, K_s$.}
\tablenotetext{b}{$T_{\rm eff}$ correction for giants. The correction factor
has already been applied to the SDSS $T_{\rm eff}$ estimate in the second column
of the above table.}
\tablenotetext{c}{Quality flag indicating stars with unusually discrepant SDSS
$T_{\rm eff}$ estimates (Flag $=1$). See Section~4.2.4 for details.}
\end{deluxetable}

%% file: taba8.tex
\begin{deluxetable}{ccrrrrrrrrrrrrrrrrrrr}
\tablewidth{0pt}
\tabletypesize{\scriptsize}
\tablecaption{Statistical Properties of $T_{\rm eff}$\label{taba:stat}}
\tablehead{
\colhead{$\langle T_{\rm eff} \rangle$} &
\colhead{} &
\colhead{} &
\multicolumn{3}{c}{IRFM $-$ KIC\tablenotemark{a}} &
\colhead{} &
\multicolumn{3}{c}{SDSS $-$ KIC\tablenotemark{a}} &
\colhead{} &
\multicolumn{3}{c}{SDSS $-$ IRFM\tablenotemark{a}} &
\colhead{} &
\multicolumn{3}{c}{$T_{\rm eff}$(color) $-$ $T_{\rm eff}$($griz$)} &
\colhead{} &
\multicolumn{2}{c}{SDSS} \nl
\cline{4-6} \cline{8-10} \cline{12-14} \cline{16-18} \cline{20-21}
\colhead{(KIC)} &
\colhead{$\langle (g\, -\, r)_0 \rangle$} &
\colhead{${\rm N_{stars}}$} &
\colhead{$\Delta T_{\rm eff}$} &
\colhead{$\sigma$} &
\colhead{$\sigma_{\rm prop}$} &
\colhead{} &
\colhead{$\Delta T_{\rm eff}$} &
\colhead{$\sigma$} &
\colhead{$\sigma_{\rm prop}$} &
\colhead{} &
\colhead{$\Delta T_{\rm eff}$} &
\colhead{$\sigma$} &
\colhead{$\sigma_{\rm prop}$} &
\colhead{} &
\colhead{$g\, -\, r$} &
\colhead{$g\, -\, i$} &
\colhead{$g\, -\, z$} &
\colhead{} &
\colhead{$\sigma_{griz}$\tablenotemark{b}} &
\colhead{$\sigma_{\rm prop}$\tablenotemark{c}}
} 
\startdata
\multicolumn{21}{c}{giants (KIC $\log{g} \leq 3.5$)} \nl
\hline
$5292$ & $0.51$ & $  35$ & $246$ & $204$ & $122$ && $216$ & $38$  & $35$ && $-62$ & $198$ & $132$ && $ -5$ & $-4$ & $29$ && $34$ & $29$ \nl
$5184$ & $0.56$ & $ 175$ & $167$ & $111$ & $112$ && $214$ & $41$  & $35$ && $ 30$ & $132$ & $121$ && $-31$ & $-1$ & $37$ && $39$ & $27$ \nl
$5086$ & $0.60$ & $ 676$ & $159$ & $100$ & $108$ && $215$ & $45$  & $34$ && $ 36$ & $104$ & $117$ && $-32$ & $ 0$ & $35$ && $37$ & $25$ \nl
$4995$ & $0.65$ & $2098$ & $135$ & $ 99$ & $105$ && $208$ & $45$  & $33$ && $ 50$ & $106$ & $113$ && $-38$ & $ 1$ & $36$ && $38$ & $23$ \nl
$4897$ & $0.69$ & $3376$ & $129$ & $ 96$ & $101$ && $207$ & $44$  & $32$ && $ 52$ & $103$ & $110$ && $-38$ & $ 1$ & $35$ && $37$ & $21$ \nl
$4800$ & $0.74$ & $4316$ & $124$ & $ 91$ & $ 98$ && $202$ & $46$  & $30$ && $ 48$ & $ 99$ & $105$ && $-38$ & $ 1$ & $36$ && $36$ & $19$ \nl
$4702$ & $0.79$ & $3435$ & $118$ & $ 91$ & $ 94$ && $192$ & $51$  & $27$ && $ 39$ & $101$ & $101$ && $-37$ & $ 0$ & $37$ && $34$ & $17$ \nl
$4599$ & $0.85$ & $3002$ & $110$ & $ 95$ & $ 91$ && $178$ & $56$  & $24$ && $ 36$ & $108$ & $ 96$ && $-30$ & $-2$ & $34$ && $28$ & $16$ \nl
$4509$ & $0.91$ & $1148$ & $ 71$ & $106$ & $ 87$ && $151$ & $53$  & $20$ && $ 56$ & $118$ & $ 91$ && $-25$ & $-3$ & $32$ && $23$ & $14$ \nl
$4401$ & $0.97$ & $ 930$ & $ 58$ & $ 97$ & $ 84$ && $148$ & $46$  & $17$ && $ 75$ & $109$ & $ 87$ && $-19$ & $-5$ & $29$ && $18$ & $13$ \nl
$4307$ & $1.03$ & $ 861$ & $ 64$ & $ 80$ & $ 81$ && $138$ & $41$  & $16$ && $ 62$ & $ 88$ & $ 84$ && $-16$ & $-6$ & $27$ && $16$ & $12$ \nl
$4202$ & $1.10$ & $ 665$ & $ 97$ & $ 53$ & $ 79$ && $123$ & $32$  & $14$ && $ 20$ & $ 60$ & $ 81$ && $ -3$ & $-7$ & $19$ && $13$ & $12$ \nl
$4105$ & $1.20$ & $ 631$ & $169$ & $ 29$ & $ 80$ && $103$ & $28$  & $14$ && $-60$ & $ 33$ & $ 83$ && $ 22$ & $-7$ & $ 3$ && $15$ & $11$ \nl
\enddata
\tablecomments{We only present statistical properties for giants (KIC $\log{g}
\leq 3.5$) in the revised Table~8. Results are the same for dwarfs (KIC
$\log{g} > 3.5$) in the original Table~8.
Statistical properties derived from the full long-cadence
sample, after applying the hot $T_{\rm eff}$ corrections. No metallicity and
binary corrections were applied.}
\tablenotetext{a}{Weighted mean difference ($T_{\rm eff}$), weighted standard
deviation ($\sigma$), and the expected standard deviation propagated from random errors
($\sigma_{\rm prop}$).}
\tablenotetext{b}{Median of the standard deviation, which is derived from individual
$T_{\rm eff}$ estimates in $g\, -\, r$, $g\, -\, i$, and $g\, -\, z$ for each star.}
\tablenotetext{c}{Median of the standard deviation, which is propagated from photometric
errors in $griz$ for each star.}
\end{deluxetable}